%% file: ms.tex
\documentclass[times,astrobib,amssymb,usenatbib]{mn2e}
\usepackage{times} 
\usepackage{latexsym,amsmath,amssymb,amsfonts} 
\usepackage{graphicx} 
\usepackage{fancyhdr}  
\usepackage{natbib}  
\usepackage{supertabular}
\usepackage{longtable}
\usepackage{url}
\usepackage{dcolumn}
\usepackage{morefloats}
\usepackage{hyperref}  
\begin{document}
\newcommand{\sqcm}{cm$^{-2}$}  
\newcommand{\lya}{Ly$\alpha$}
\newcommand{\lyb}{Ly$\beta$}
\newcommand{\lyg}{Ly$\gamma$}
\newcommand{\lyd}{Ly$\delta$}
\newcommand{\HeI}{\mbox{He\,{\sc i}}}
\newcommand{\HeII}{\mbox{He\,{\sc ii}}}
\newcommand{\HI}{\mbox{H\,{\sc i}}}
\newcommand{\HII}{\mbox{H\,{\sc ii}}}
\newcommand{\HH}{\mbox{H{$_2$}}}
\newcommand{\hh}{\mbox{\tiny H{$_2$}}} 
\newcommand{\hi}{\mbox{\tiny H{\sc i}}} 
\newcommand{\OI}{\mbox{O\,{\sc i}}}
\newcommand{\OII}{\mbox{O\,{\sc ii}}}
\newcommand{\OIII}{\mbox{O\,{\sc iii}}}
\newcommand{\OIV}{\mbox{O\,{\sc iv}}}
\newcommand{\OV}{\mbox{O\,{\sc v}}}
\newcommand{\OVI}{\mbox{O\,{\sc vi}}}
\newcommand{\OVII}{\mbox{O\,{\sc vii}}}
\newcommand{\OVIII}{\mbox{O\,{\sc viii}}}
\newcommand{\CaVIII}{\mbox{Ca\,{\sc viii}}}
\newcommand{\CaVII}{\mbox{Ca\,{\sc vii}}}
\newcommand{\CaVI}{\mbox{Ca\,{\sc vi}}}
\newcommand{\CaV}{\mbox{Ca\,{\sc v}}}
\newcommand{\CIV}{\mbox{C\,{\sc iv}}}
\newcommand{\CV}{\mbox{C\,{\sc v}}}
\newcommand{\CVI}{\mbox{C\,{\sc vi}}}
\newcommand{\CII}{\mbox{C\,{\sc ii}}}
\newcommand{\CI}{\mbox{C\,{\sc i}}}
\newcommand{\CIIs}{\mbox{C\,{\sc ii}}$^\ast$}
\newcommand{\CIII}{\mbox{C\,{\sc iii}}}
\newcommand{\SiI}{\mbox{Si\,{\sc i}}}
\newcommand{\SiII}{\mbox{Si\,{\sc ii}}}
\newcommand{\SiIII}{\mbox{Si\,{\sc iii}}}
\newcommand{\SiIV}{\mbox{Si\,{\sc iv}}}
\newcommand{\SiXII}{\mbox{Si\,{\sc xii}}}
\newcommand{\SV}{\mbox{S\,{\sc v}}}
\newcommand{\SIV}{\mbox{S\,{\sc iv}}}
\newcommand{\SIII}{\mbox{S\,{\sc iii}}}
\newcommand{\SII}{\mbox{S\,{\sc ii}}}
\newcommand{\SI}{\mbox{S\,{\sc i}}}
\newcommand{\ClI}{\mbox{Cl\,{\sc i}}}
\newcommand{\ArI}{\mbox{Ar\,{\sc i}}}
\newcommand{\NI}{\mbox{N\,{\sc i}}}
\newcommand{\NII}{\mbox{N\,{\sc ii}}}
\newcommand{\NIII}{\mbox{N\,{\sc iii}}}
\newcommand{\NIV}{\mbox{N\,{\sc iv}}}
\newcommand{\NV}{\mbox{N\,{\sc v}}}
\newcommand{\PV}{\mbox{P\,{\sc v}}}
\newcommand{\PII}{\mbox{P\,{\sc ii}}}
\newcommand{\PIII}{\mbox{P\,{\sc iii}}}
\newcommand{\PIV}{\mbox{P\,{\sc iv}}}
\newcommand{\NeVIII}{\mbox{Ne\,{\sc viii}}}
\newcommand{\ArVIII}{\mbox{Ar\,{\sc viii}}}
\newcommand{\NeV}{\mbox{Ne\,{\sc v}}}
\newcommand{\NeVI}{\mbox{Ne\,{\sc vi}}}
\newcommand{\NeX}{\mbox{Ne\,{\sc x}}} 
\newcommand{\NaIX}{\mbox{Na\,{\sc ix}}} 
\newcommand{\MgII}{\mbox{Mg\,{\sc ii}}}
\newcommand{\FeII}{\mbox{Fe\,{\sc ii}}}
\newcommand{\MgX}{\mbox{Mg\,{\sc x}}}
\newcommand{\AlXI}{\mbox{Al\,{\sc xi}}}
\newcommand{\FeIII}{\mbox{Fe\,{\sc iii}}}
\newcommand{\NaI}{\mbox{Na\,{\sc i}}}
\newcommand{\CaII}{\mbox{Ca\,{\sc ii}}}
\newcommand{\zabs}{$z_{\rm abs}$}
\newcommand{\zmin}{$z_{\rm min}$}
\newcommand{\zmax}{$z_{\rm max}$}
\newcommand{\zqso}{$z_{\rm qso}$}
\newcommand{\subHe}{_{\it HeII}}
\newcommand{\subH}{_{\it HI}}
\newcommand{\subHLy}{_{\it H Ly}}
\newcommand{\degree}{\ensuremath{^\circ}}
\newcommand{\lapp}{\mbox{\raisebox{-0.3em}{$\stackrel{\textstyle <}{\sim}$}}}
\newcommand{\gapp}{\mbox{\raisebox{-0.3em}{$\stackrel{\textstyle >}{\sim}$}}}
\newcommand{\be}{\begin{equation}}
\newcommand{\en}{\end{equation}}
\newcommand{\di}{\displaystyle}
\def\tworule{\noalign{\medskip\hrule\smallskip\hrule\medskip}} 
\def\onerule{\noalign{\medskip\hrule\medskip}} 
\def\bl{\par\vskip 12pt\noindent}
\def\bll{\par\vskip 24pt\noindent}
\def\blll{\par\vskip 36pt\noindent}
\def\rot{\mathop{\rm rot}\nolimits}
\def\alf{$\alpha$}
\def\refff{\leftskip20pt\parindent-20pt\parskip4pt}
\def\zabs{$z_{\rm abs}$}
\def\zqso{$z_{\rm qso}$}
\def\zem{$z_{\rm em}$}
\def\mgii{Mg\,{\sc ii}~} 
\def\feiia{Fe\,{\sc ii}$\lambda$2600}
\def\mgia{Mg\,{\sc i}$\lambda$2852}
\def\mgiia{Mg\,{\sc ii}$\lambda$2796}
\def\mgiib{Mg\,{\sc ii}$\lambda$2803}
\def\mgiiab{Mg\,{\sc ii}$\lambda\lambda$2796,2803}
\def\wobs{$w_{\rm obs}$}
\def\kms{km~s$^{-1}$}
\def\bnt{$b_{\rm nt}$}
\def\fosc{$f_{\rm osc}$}
\def\chisq{$\chi^{2}$}
\def\dtype{$\delta_{\rm type}$}  
\title[\HH\ absorbers at low-$z$]{A $HST/$COS survey of molecular hydrogen in DLAs \& sub-DLAs at $z < 1$: Molecular fraction and excitation temperature}        
\author[S. Muzahid et al.]
{
\parbox{\textwidth}{ 
S. Muzahid$^{1}$\thanks{E-mail: sowgatm@gmail.com},     
R. Srianand$^{2}$, and 
J. Charlton$^{1}$   
} 
\vspace*{4pt}\\  
$^{1}$ The Pennsylvania State University, 525 Davey Lab, University Park, 
State College, PA 16802, USA \\ 
$^{2}$ Inter-University Centre for Astronomy and Astrophysics, Post Bag 4, 
Ganeshkhind, Pune 411007, India \\}   
\date{Accepted. Received; in original form }
\maketitle
\label{firstpage}

\begin {abstract}  
We present the results of a systematic search for molecular hydrogen (\HH) in low redshift \linebreak ($ 0.05 \lesssim z \lesssim 0.7$) DLAs and sub-DLAs with $N(\HI) \gtrsim 10^{19.0}$~cm$^{-2}$, in the archival $HST/$COS spectra. Our core sample is comprised of 27 systems with a median $\log N(\HI) = 19.6$. On the average, our survey is sensitive down to $\log~N(\HH) = 14.4$ corresponding to a molecular fraction of $\log f_{\hh} = -4.9$ at the median $N(\HI)$. \HH\ is detected in 10 cases (3/5 DLAs and 7/22 sub-DLAs) down to this $f_{\hh}$ limit. The \HH\ detection rate of $50^{+25}_{-12}$ percent seen in our sample, is a factor of $\gtrsim 2$ higher than that of the high-$z$ sample of \citet{Noterdaeme08}, for systems with  $N(\HH) > 10^{14.4}$~cm$^{-2}$. In spite of having $N(\HI)$ values typically lower by a factor of 10, low-$z$ \HH\ systems show molecular fractions ($\log f_{\hh}=-1.93\pm0.63$) that are comparable to the high-$z$ sample. The rotational excitation temperatures ($T_{01} = 133\pm55$~K), as measured in our low-$z$ sample, are typically consistent with high-$z$ measurements. Simple photoionization models favor a radiation field much weaker than the mean Galactic ISM field for a particle density in the range 10--100 cm$^{-3}$. The impact parameters of the identified host-galaxy candidates are  in the range $10 \lesssim \rho$ (kpc)  $\lesssim 80$. We, therefore, conjecture that the low-$z$ \HH\ bearing gas is not related to star-forming disks but stems from self-shielded, tidally stripped or ejected disk-material in the extended halo. 
\end {abstract}
\begin{keywords} 
galaxies: ISM -- galaxies: haloes -- quasar: absorption line   
\end{keywords}
\section{Introduction}

Damped \lya\ absorbers (DLAs) and sub-DLAs seen in QSO spectra are characterized by very high neutral hydrogen column densities: $N(\HI) \geqslant 1\times10^{19}$ cm$^{-2}$ for sub-DLAs and $\geqslant 2\times10^{20}$ cm$^{-2}$ for DLAs \citep[see][for a review]{Wolfe05}. Whereas the ionization correction can be appreciable in sub-DLAs \citep[]{Peroux03a,Peroux03b}, DLAs are predominantly neutral and form the major reservoirs of neutral-gas at high redshifts \citep[]{Prochaska09a,Noterdaeme09dla,Noterdaeme12}. In general, DLAs/sub-DLAs show higher metallicities compared to the \lya\ forest systems \citep[]{Schaye03,Wolfe05,Kulkarni05}. An increase in the cosmic mean metallicity of DLAs/sub-DLAs with cosmic time has been reported \citep[see e.g.][]{Prochaska03Zz,Kulkarni07,Rafelski12,Som13}. Moreover, DLA metallicities are found to be correlated with the velocity spread ($\Delta v_{90}$) of neutral or singly ionized metal lines \citep[]{Wolfe98,Ledoux06a}, which has been interpreted as the mass-metallicity relation seen in galaxies \citep[]{Tremonti04}. All these suggest that the DLAs/sub-DLAs are located in over-dense regions where star formation activity takes place \citep[see e.g.][]{Pettini97}. Direct detections of \lya$/$H$\alpha$$/$[\OIII] emission lines in a handful of DLAs \citep[]{Moller04,Fynbo10,Peroux11,Noterdaeme12AA,Jorgenson14dla} made such an idea more compelling. Absorption studies of DLA/sub-DLA systems, thus, provide a unique means to probe galaxy neighbourhoods over a wide range of redshift (0~$<z<$~5) without the biases introduced in magnitude limited sample of galaxies studied in emission.  

The connection between DLAs/sub-DLAs and galaxies can be firmly established by directly detecting galaxies at the same redshifts. Alternatively, such a connection can be indirectly established by showing that the physical conditions in the absorbing gas are consistent with those seen in the diffuse interstellar medium (ISM). Except the previously mentioned few cases, direct detections of candidate galaxies hosting DLAs or sub-DLAs at $z > 1.7$ have not been very successful \citep[]{Kulkarni06,Christensen07}. Thus, our understanding of the physical conditions in high-$z$ DLAs/sub-DLAs primarily relies on the optical absorption-line spectroscopy using low-ionization metal lines and, in a few cases, \HH, HD, and CO molecular absorption \citep[e.g.][]{Ledoux03,Srianand05,Srianand08,Noterdaeme08,Noterdaeme08hd,Noterdaeme09co,Noterdaeme10co,Tumlinson10,Guimaraes12,Vasquez14}. At low redshift ($z<1$) it is relatively easy to detect candidate host-galaxies in most cases \citep{Rao11}. 

Molecular hydrogen (\HH) is the most abundant molecule in the universe which also acts as the most important molecular coolant. Formation of \HH\ is expected on the surface of dust grains if the gas is cool, dense and mostly neutral \citep[]{Gould63,Hollenbach71a}. In the case of warm and dust-free gas, \HH\ can form via H$^{-}$ ions \citep[]{Jenkins97}. Photo-dissociation of H$_{\rm 2}$ takes place in the energy range 11.1--13.6 eV via Lyman- and Warner-band absorption. Therefore, information on the kinetic and rotational excitation temperatures, the particle density, and the radiation field can be derived from good quality data when \HH\ is detected. For example, at high redshift, DLA subcomponents in which H$_{\rm 2}$ absorption is detected have typical kinetic temperature of $T=$~153$\pm$78 K and density of $n_{\rm H}=$ 10--200 cm$^{-3}$ \citep[]{Srianand05}. The typical inferred radiation field in the H$_2$ bearing components is of the order of the mean UV radiation field in the Galactic ISM. Systematic surveys at high redshift \citep[i.e. $z > 1.7$,][]{Ledoux03,Noterdaeme08} have shown that the \HH\ is detected in 10--20\% of DLAs/sub-DLAs \citep[but see][]{Jorgenson14H2}. \HH\ content is typically low and is found to be correlated with the metallicity and dust depletion, in those studies. The frequent detections of molecular hydrogen towards dusty and high-metallicity regions hints at its connection to star-forming regions \citep[see also][]{Petitjean06}. 

In spite of having enormous diagnostic potential to probe physical conditions in different gas phases of galaxies in a luminosity unbiased way, molecular hydrogen has not been well studied at low-$z$. This is primarily because the atmospheric cutoff of light below 3000 \AA\ makes ground-based observations of \HH\ impossible for redshift $z<1.7$. In addition, due to the unavailability of high resolution UV sensitive spectrographs in the past, detecting molecular hydrogen in absorption was not possible beyond the Galactic disk/halo \citep[]{Savage77,Gillmon06} and the Magellanic Clouds \citep[]{Tumlinson02,Welty12}. The only known \HH\ system at $z < 1.5$ (i.e. \zabs\ = 1.15 toward HE~0515--4414) was reported by \citet{Reimers03} in $HST/$STIS data. The high throughput of $HST/$COS FUV gratings now gives us a window of opportunity to explore molecular hydrogen in low-$z$ DLAs/sub-DLAs. The numerous UV absorption lines of \HH\ from different rotational ($J$) levels are now accessible for absorbers with redshift as low as $z \sim 0.05$ in $HST/$COS spectra \citep[see e.g.][]{Crighton13,Oliveira14,Srianand14}. Here we present the results of our systematic search for low-$z$ molecular hydrogen in DLAs/sub-DLAs that are detected in the medium resolution COS spectra from the HST archive.  

This article is organized as follows. In Section~\ref{sec:obs} we discuss the observations and reduction of the spectra we have used to search for \HH\ systems. In Section~\ref{sec:search} we describe our search technique, final data sample, and the absorption line measurements. Detailed analysis of the sample is presented in Section~\ref{sec:ana}. In Section~\ref{sec:diss} we discuss our results. The important findings are summarized in Section~\ref{sec:summ}. Throughout this work we assume a flat $\Lambda$CDM cosmology with \linebreak $H_{0} = 70$ \kms~Mpc$^{-1}$, $\Omega_{\rm M} = 0.3$, and $\Omega_{\Lambda} = 0.7$. The wavelengths of \HH\ transitions are taken from \citet{Bailly10}.

\input{tab1}

\begin{figure*} 
\centerline{\hbox{ 
\centerline{\vbox{ 
\includegraphics[height=5.6cm,width=8.6cm,angle=00]{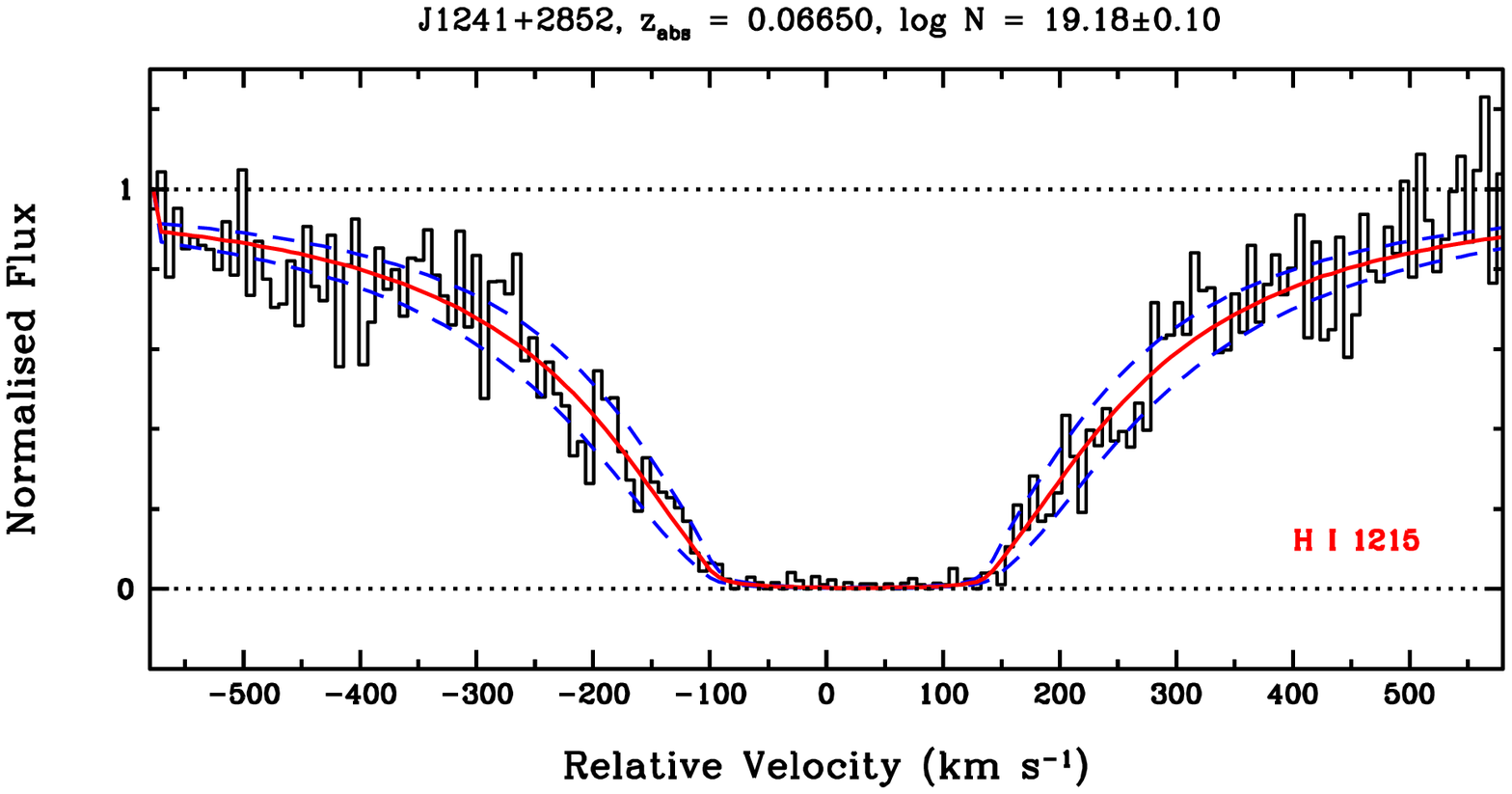}    
\includegraphics[height=5.6cm,width=8.6cm,angle=00]{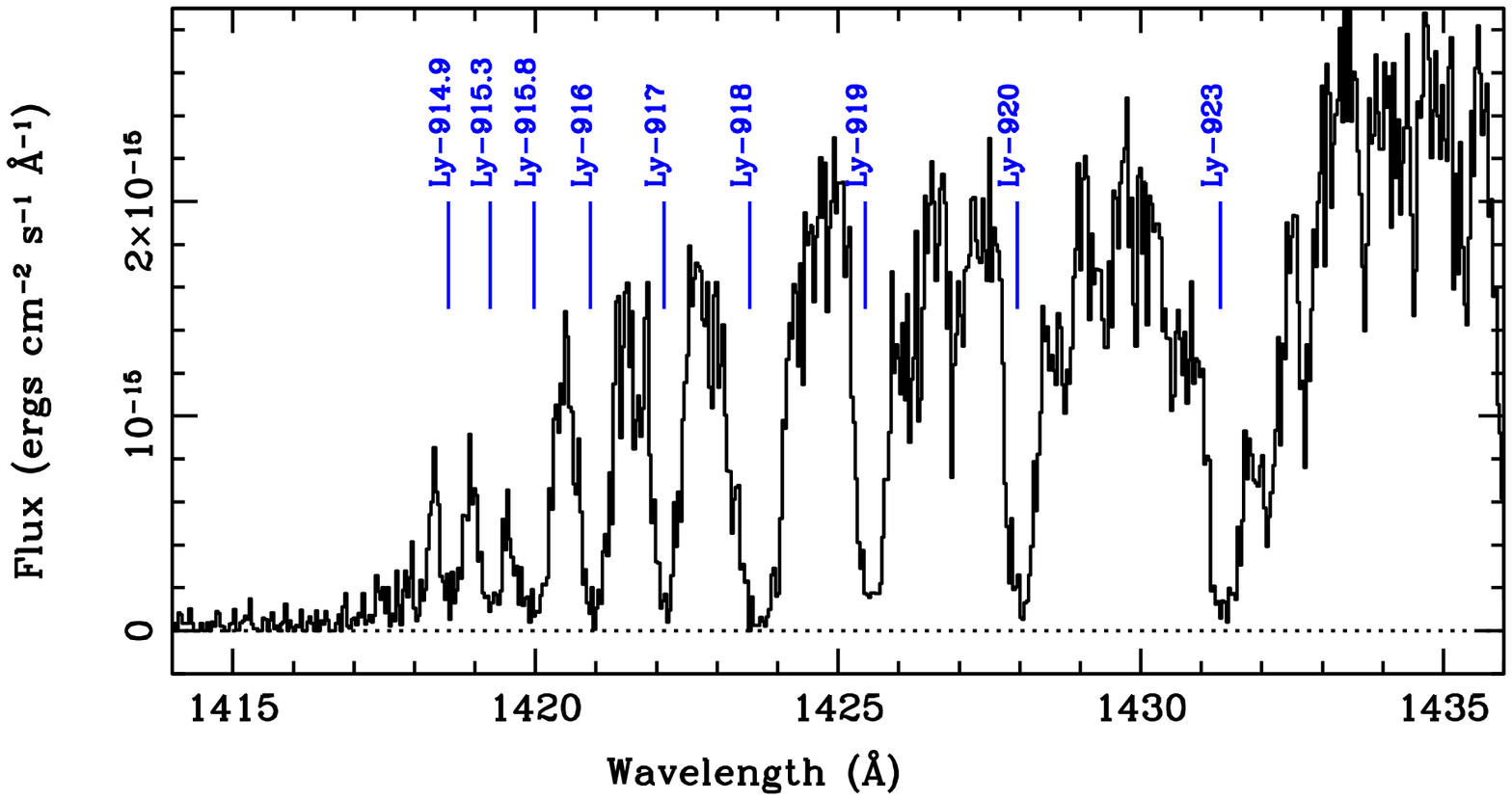}    
}}
}}
\caption{Left: Sub-damped \lya\ absorption from the \zabs\ = 0.06650 towards J~1241+2852 system. 
The solid (red) and dashed (blue) curves represent best fitting Voigt profile to the data 
(black histogram) and its 1$\sigma$ uncertainty, respectively.   
Fits for other systems are shown in Appendix-\ref{HIfits1} \& \ref{HIfits2}. 
Right: Strong Lyman limit break from the 
\zabs\ = 0.55048 towards Q~1241+176 system. The \lya\ is not covered in COS whereas the \lyb\ 
falls in the gap of the existing spectrum. Both the systems show \HH\ absorption.   
}       
\label{exampleHI}   
\end{figure*} 

\begin{figure*} 
\centerline{\vbox{ 
\centerline{\hbox{ 
\includegraphics[height=8.0cm,width=8.5cm,angle=00]{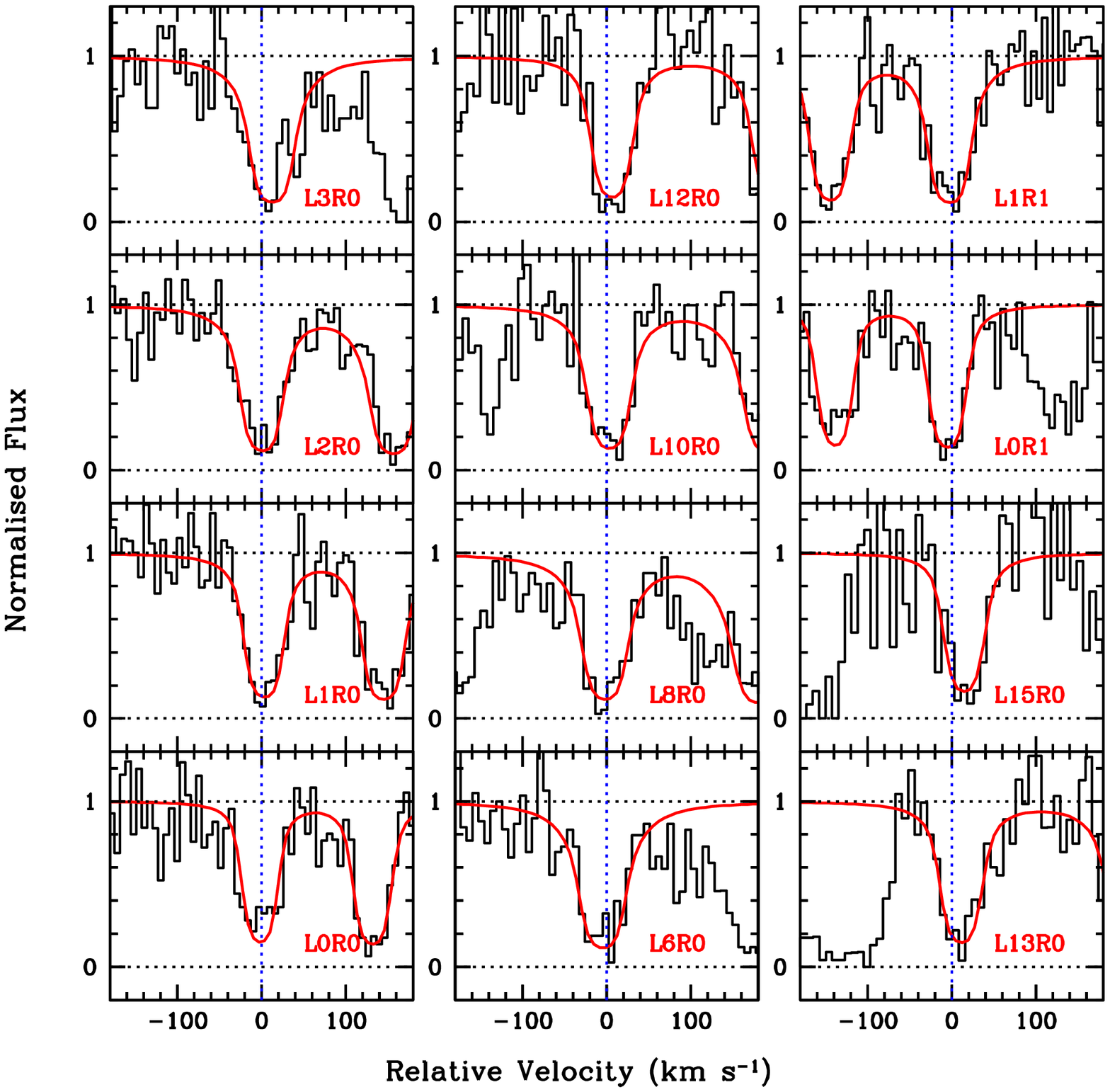} 
\includegraphics[height=8.0cm,width=8.5cm,angle=00]{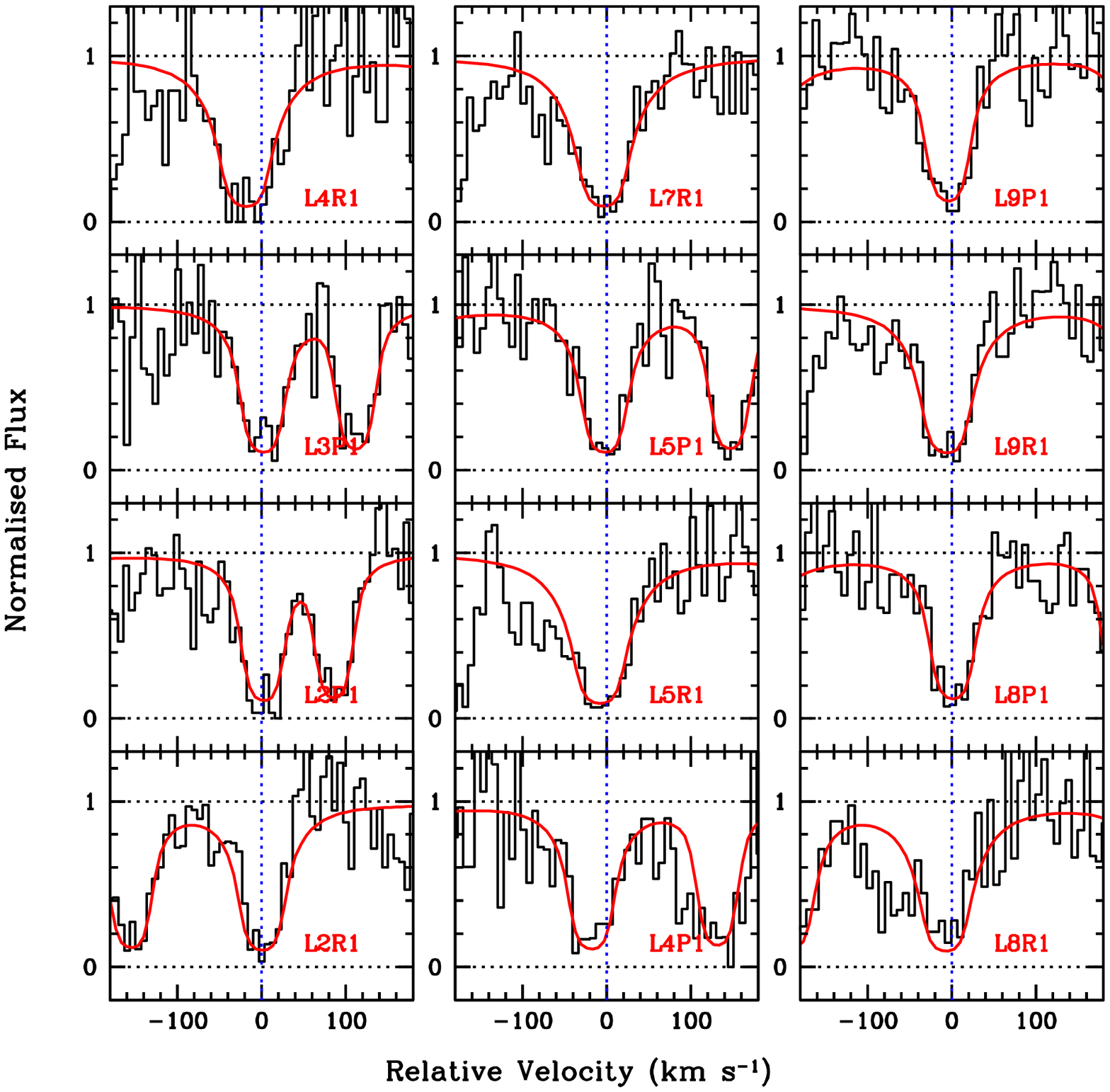} 
}}
\vskip-0.5cm 
\centerline{\hbox{ 
\includegraphics[height=8.0cm,width=8.5cm,angle=00]{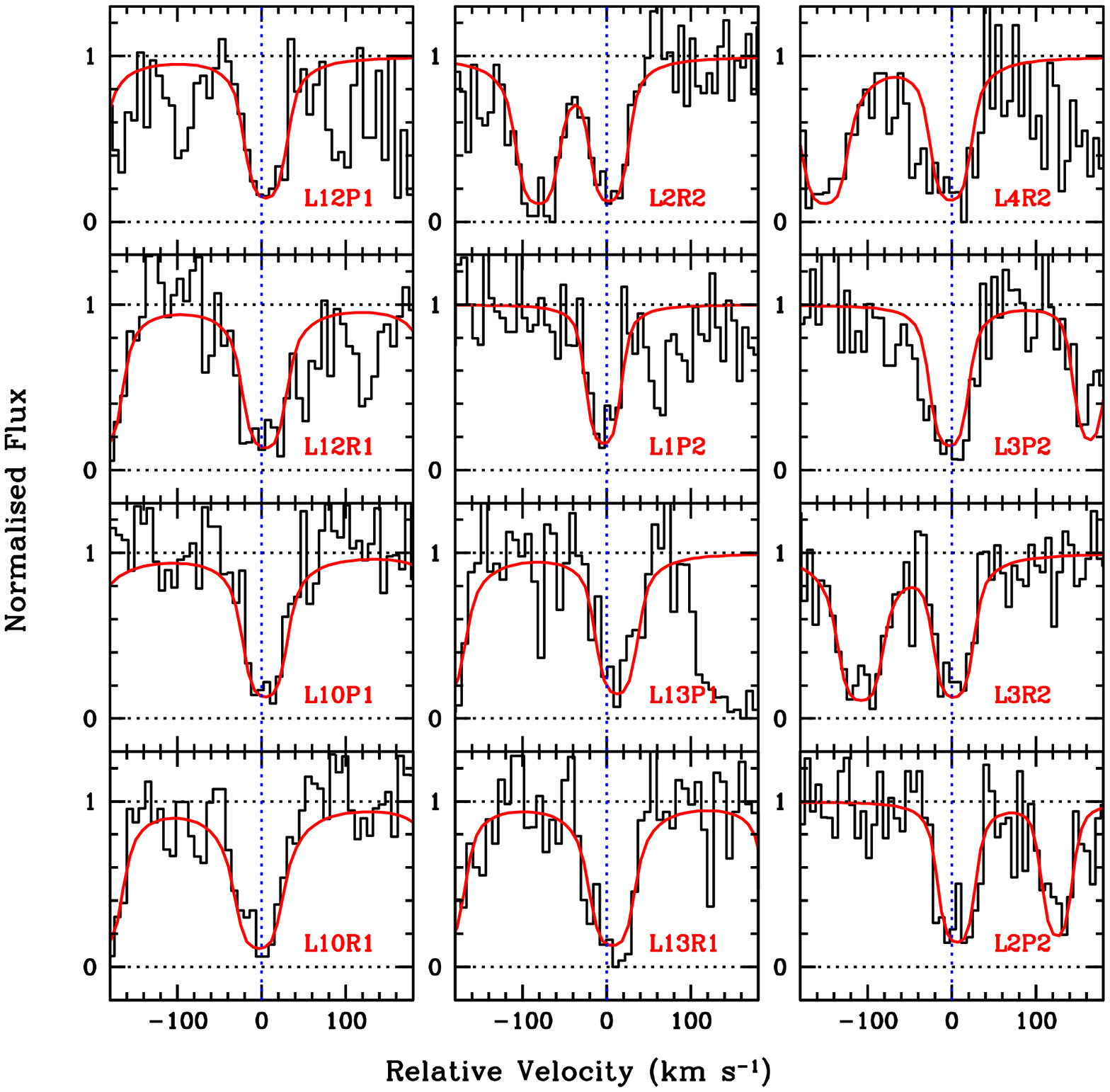} 
\includegraphics[height=8.0cm,width=8.5cm,angle=00]{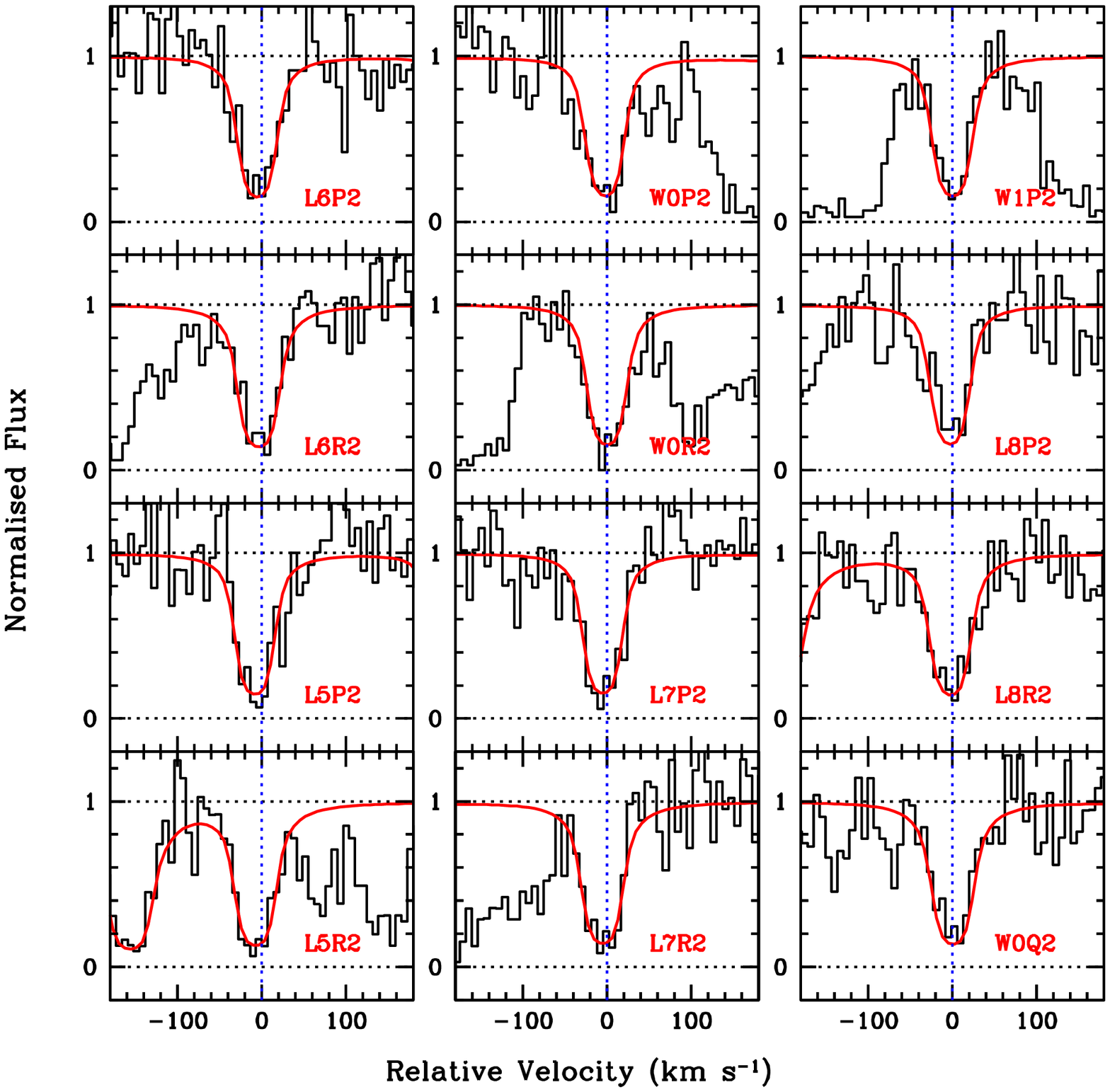} 
}}
\vskip-0.5cm 
\centerline{\hbox{ 
\includegraphics[height=8.0cm,width=8.5cm,angle=00]{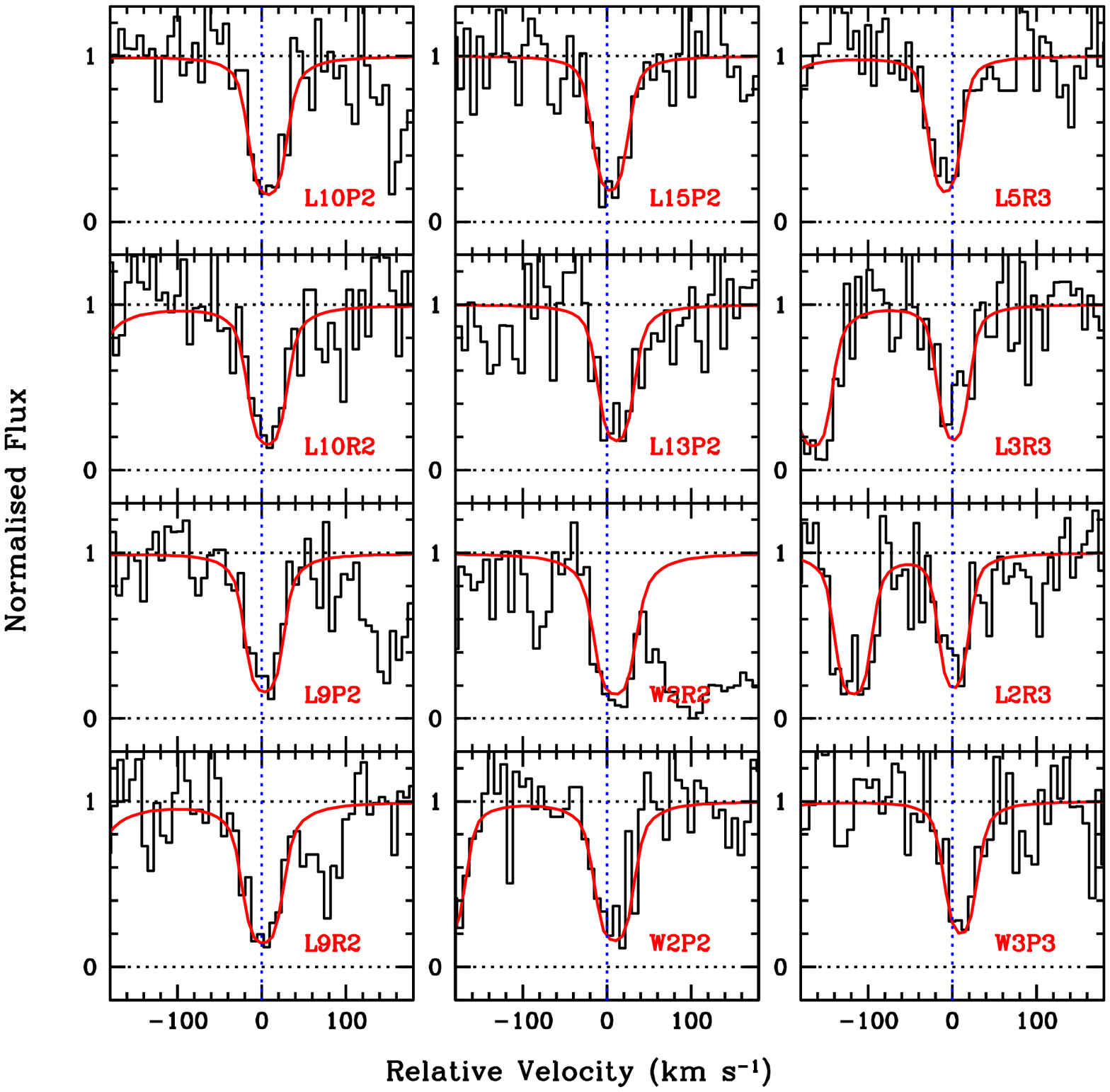} 
\includegraphics[height=8.0cm,width=8.5cm,angle=00]{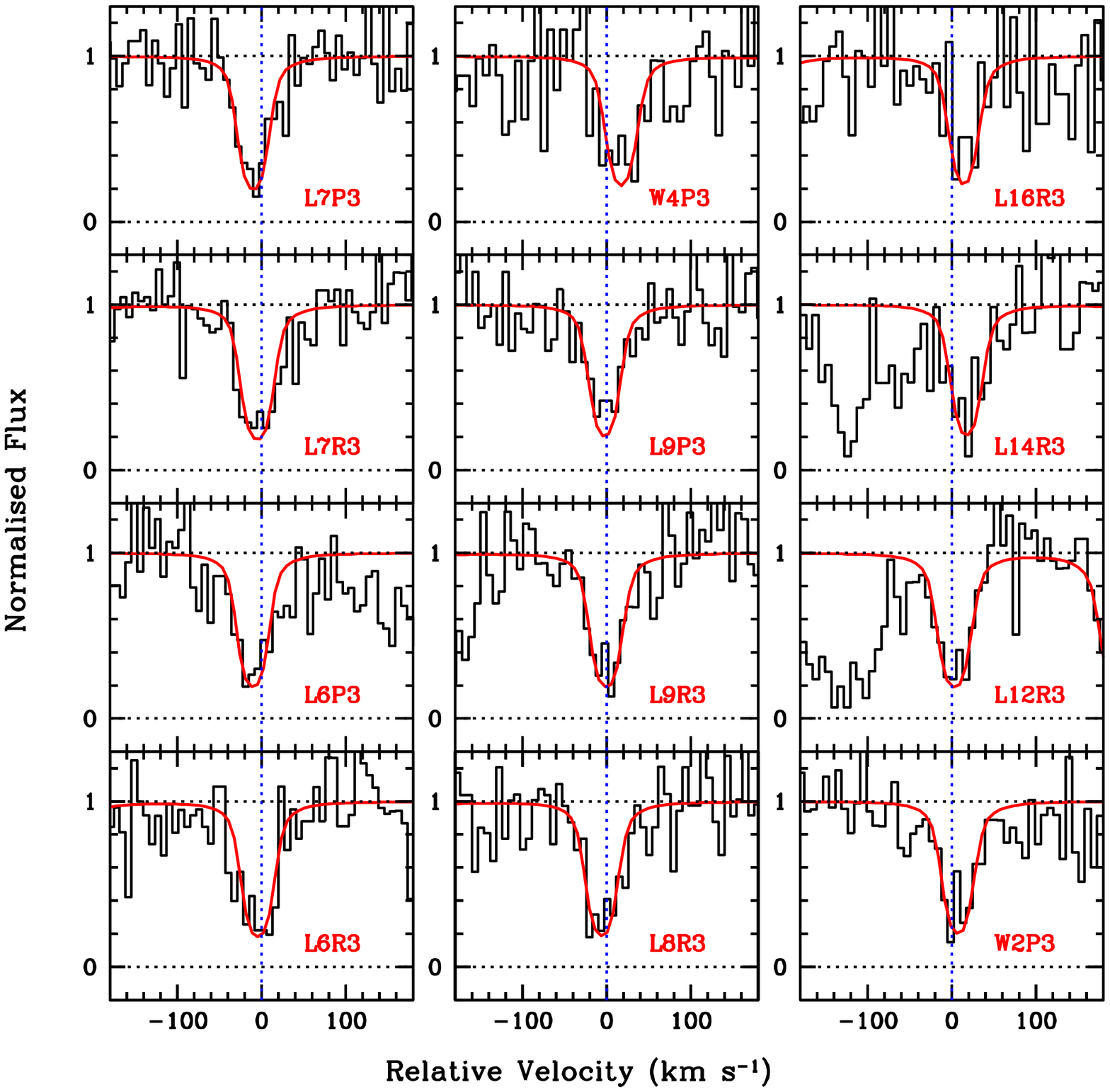} 
}}
}}
\caption{Numerous molecular hydrogen absorption lines from different $J$ levels from 
the \zabs\ = 0.24788 towards J~0925+4004 system. The (red) smooth curves are the best 
fitting Voigt profiles to the data (in black histogram).        
}    
\label{exampleH2}  
\end{figure*} 

\section{Observations and Data Reduction}  
\label{sec:obs}

We have searched for DLAs and sub-DLAs in nearly 400 far-ultraviolet (FUV) spectra of intermediate redshift quasars, observed with the $HST/$COS, that were available in the public $HST$ archive before March, 2014. The properties of COS and its in-flight operations are discussed by \citet{Osterman11} and \citet{Green12}. All the COS spectra were obtained using medium resolution ($R\sim$~20,000) FUV COS gratings (G130M and/or G160M). The data were retrieved from the $HST$ archive and reduced using the {\sc calcos} pipeline software. The pipeline reduced data (so called ``$x1d$" files) were flux calibrated. To increase the spectral signal-to-noise ratio (SNR), individual G130M and G160M integrations were aligned and coadded using the IDL code (``$coadd\_x1d$") developed by \citet{Danforth10}\footnote{http://casa.colorado.edu/∼danforth/science/cos/costools.html}. The exposures were weighted by the integration time while coadding in flux units. Since our data comes from different observing programs, the final data sample show a range in spectral SNR (e.g. $\sim$~5 to 25 per resolution element). To take care of this SNR differences, in the next section, we have defined the ``\HH\ sensitivity" limit for each detected DLA and sub-DLA. Depending on the spectral SNR, the ``\HH\ sensitivity" limit varies between $10^{13.9}-10^{15.6}$ \sqcm\ with a median value of 10$^{14.4}$ \sqcm. Therefore, about half of the spectra in our sample are sensitive enough to detect \HH\ with $N(\HH) >$~10$^{14.4}$ \sqcm. All the low-$z$ \HH\ bearing systems show $N(\HH)$ higher than this median value.     

As the COS FUV spectra are significantly oversampled (i.e. six raw pixels per resolution element), we binned the data by three pixels. This further improves SNR per pixel. All our measurements and analyses were, subsequently, performed on the binned data.  Measurements are, however, found to be fairly independent of binning. Continuum normalization was done by fitting the line-free regions with a smooth lower-order polynomial.

\section{Search technique, data sample, and absorption line measurements}     
\label{sec:search}  

First of all, we do not perform a blind search for \HH\ systems in the spectra. Instead, we search for DLAs/sub-DLAs first and then look for \HH\ absorption from the identified DLAs/sub-DLAs. Identifying DLAs/sub-DLAs is fairly straightforward in intermediate redshift QSO spectra as they produce distinct damping wings in the \lya\ absorption and as there is no \lya\ forest crowding at low-$z$ (see e.g. Fig.~\ref{exampleHI}). Note that, \lya\ absorption will only be covered by the COS spectra for systems with 0~$<$~\zabs~$<$~0.47, provided both G130M and G160M data are available. With the  G130M data alone, \lya\ can be covered up to 0~$<$~\zabs~$\lesssim$~0.19. Therefore, in order to account for the systems for which \lya\ absorption is not covered, we searched for systems that produce a strong Lyman limit ($\lambda_{\rm rest} = 912$~\AA) break and/or show strong Lyman series lines as shown in the right panel of  Fig.~\ref{exampleHI}. Using this conservative approach we made sure that none of the systems with $N(\HI) \gtrsim 10^{19}$~cm$^{-2}$ are missed. In total we have found 33 systems with $\log N(\HI) \gtrsim$~19.0 as listed by increasing absorption redshift (\zabs) in Table~\ref{sample1} \& \ref{sample2}. Our core sample, for which Lyman- and Warner-band absorption lines of \HH\ are covered by the COS spectra, is comprised of the 27 systems that are listed in Table~\ref{sample1}. In Table~\ref{sample2} we have listed six other systems with $\log N(\HI) \gtrsim$~19.0 for which \HH\ information is not available in the existing spectra. Note that for 20 of these 27 systems, $N(\HI)$ are constrained from the \lya\ profiles. For the other 7 systems, $N(\HI)$ are constrained from \lyb\ and/or higher order Lyman series lines and thus $N(\HI)$ values should be taken with caution. For these systems we adopted a minimum number of components approach to fit all the unblended Lyman series lines simultaneously. Using low resolution spectra covering \lya\ absorption, 5 out of the 7 systems were reported to have large $N(\HI)$ values (i.e. $\log N(\HI)>$~18.5, see the footnote of Table~\ref{sample1}). Note that for all the systems listed in Table~\ref{sample2}, $N(\HI)$ is constrained from \lya\ profiles.        

For every identified DLAs/sub-DLAs in Table~\ref{sample1}, we have searched for the Lyman- and Warner-band absorption lines from the molecular hydrogen. Numerous absorption lines from different rotational levels (starting from $J$~=~0) have been searched one-by-one. In total we have found 10 systems where \HH\ absorption lines are detected. Three of these 10 systems, i.e. \zabs\ = 0.55733 towards Q~0107--0232, \citep{Crighton13}; \zabs\ = 0.18562 towards B~0120--28, \citep[]{Oliveira14}; and \zabs\ = 0.09630 towards J~1619+3342, \citep[]{Srianand14} were known to have \HH\ before. Here we report seven new \HH\ detections.  

For each of the seven cases with new \HH\ detections, we choose a set of clean (free from severe blending) \HH\ lines from different $J$ levels for Voigt profile fitting. We fit all these clean lines simultaneously by keeping the Doppler-parameter ($b$) tied. In addition, we constrain the column density to be same for all the transitions from a given $J$ level and allow it to be different for different $J$ levels. The COS wavelength calibration is known to have uncertainties at the level of 10 -- 15 \kms\ \citep[]{Savage11,Meiring13}. Therefore, we did allow the redshift of individual transitions to be different. In majority of the cases lines are aligned within $\pm5$~\kms. However, occasionally, the velocity offset can be up to 20 \kms. Such extreme offsets are noticed only when the lines fall at the edge of the spectra. As an example, in Fig.~\ref{exampleH2} we show our \HH\ fits for the \zabs\ = 0.24788 system towards J~0925+4004. All the \HH\ lines, from different $J$ levels, that are used for fitting are shown in the plot. Voigt profile fits to \HH\ absorption from all other newly reported systems are shown in Appendix-\ref{H2fits}. The line spread function (LSF) of the COS spectrograph is not a Gaussian. A characterization of the non-Gaussian LSF for COS can be found in \citet{Ghavamian09} and \citet{Kriss11}. For our Voigt profile fit analysis we adopt the latest LSF given by \citet{Kriss11}. Interpolated LSFs at the line center were convolved with the model Voigt profile while fitting an absorption line using the {\sc vpfit}\footnote{http://www.ast.cam.ac.uk/ rfc/vpfit.html} software. 

\input{tab2}

In the case of non-detections we estimate 3$\sigma$ upper limits on $N(\HH)$, based on the observed error in the continuum, using the following steps: (a) First we identify the expected wavelength ranges of \HH\ transitions from $J=0$ and $J=1$ rotational levels that are uncontaminated by absorption from other systems. (b)  We derive upper limits on $N(\HH)$ for each of those transitions using the prescription by \citet{Hellsten98}, assuming a $b$-parameter of 10~\kms\ (close to the median $b(\HH)$ in our sample) for the unresolved \HH\ line. (c) We adopt the best limiting \HH\ column densities we get for $J=0$ and $J=1$ levels. (d) Finally, we sum the $N(\HH)$ limits that we get from  $J = 0$ and $J = 1$ levels. The estimated upper limits are given in column 8 of Table~\ref{sample1}. In column 7 of Table~\ref{sample1} we present the \HH\ detection sensitivity, for each spectrum in our sample, at 3$\sigma$ level. For the systems in which \HH\ is not detected, the \HH\ detection sensitivity is the same as the 3$\sigma$ upper limits on $N(\HH)$ presented in column 8 of Table~\ref{sample1}. For the systems with \HH\ detections, we estimate 3$\sigma$ upper limits on $N(\HH)$, as described above, from the SNR in the unabsorbed pixels near to the absorption. About half of the spectra in the core sample are sensitive down to $\log N(\HH) = $~14.4, which we denote as the ``median sensitivity" limit for our sample.

\input{tab3}

In Table~\ref{H2tab} we provide \HH\ column densities for different $J$ levels, $N(\HH, J)$, rotational excitation temperatures, $T_{ij}$, Doppler parameters, $b(\HH)$, and average metallicities for the 10 \HH\ bearing systems. All the newly reported \HH\ systems, except one, are fitted with a single component Voigt profile (see column 2 of Table~\ref{H2tab}). Apart from the strongest component at 0~\kms, the \zabs\ = 0.10115 towards Q~0439--433 system shows an additional weak, tentative component at $-65$~\kms. In general, the $b(\HH)$ values obtained in our single component Voigt profile fit, as listed in column 8 of Table~\ref{H2tab}, are slightly higher than what is typically measured in high-$z$ DLAs. This could indicate the presence of multiple narrow components. Indeed, in cases where we have access to high resolution optical spectra (e.g. \zabs\ = 0.10115 system towards Q~0439--433), the \NaI\ and \CaII\ absorption lines show multiple narrow components. However, from high-$z$ \HH\ systems studied with high resolution, it is known that all the metal components need not produce \HH\ absorption and observed \HH\ need not necessarily be associated with the strongest metal absorption \citep[see e.g.][]{Rodriguez06}. Additionally, because of the wavelength scale uncertainty and relatively lower spectral resolution of COS we do not try to fit \HH\ lines with multiple components with $b$-parameters much less than the spectral resolution. This could lead to underestimation of $N(\HH)$ measurements and associated errors.       

\section{Analysis} 
\label{sec:ana}      

\subsection{Detection Rate}  
\label{subsec:inc}

As mentioned before, there are a total of 27 systems (5 DLAs and 22 sub-DLAs) in our core sample for which the COS spectra cover the expected wavelength range of \HH\ transitions. Ten out of these 27 systems show absorption lines from molecular hydrogen. Only three of these systems are DLAs and the remaining ones are sub-DLAs. The majority of these systems show \HH\ absorption from $J\leqslant3$ rotational levels. None of them show higher (i.e. $J\geqslant4$) rotational level transitions. The \zabs\ = 0.22711 system towards J~1342--0053 is detected via $J=1$ level transitions alone. Furthermore, for two other systems, i.e. \zabs\ = 0.06650 towards J~1241+2852 and \zabs\ = 0.16375 towards Q~0850+440, we do not detect any $J = 3$ level transitions. Note that \citet{Oliveira14} did not report any $J \geqslant 2$ transitions for the \zabs\ = 0.18495 \HH\ component towards B~0120--28. We have verified that  $J \geqslant 2$ transitions in this component are not detected.   

\begin{figure} 
\centerline{\vbox{ 
\centerline{\hbox{ 
\includegraphics[height=6.4cm,width=8.6cm,angle=00]{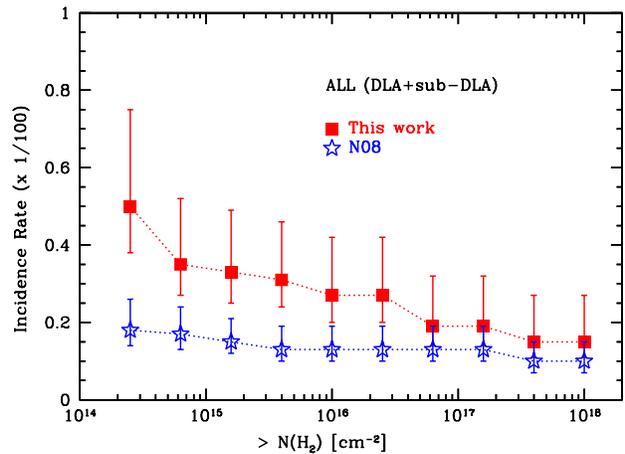}    
}}
}}
\caption{Detection rates of molecular hydrogen for our low-$z$ sample (red squares) and the high-$z$  
sample of N08 (blue stars) against different threshold $N(\HH)$ values. Errorbars represent Gaussian 
1$\sigma$ confidence intervals computed using tables of \citet{Gehrels86} assuming 
a Poisson distribution.}               
\label{IncRate}  
\end{figure}

At high redshift (1.8~$<$~\zabs~$<$4.2) \citet[hereafter N08]{Noterdaeme08} have conducted a survey of molecular hydrogen in DLAs and strong sub-DLAs (with $\log N(\HI) >$~20.0) using high resolution spectra obtained with the Very Large Telescope (VLT) Ultraviolet and Visual Echelle Spectrograph (UVES). The spectral resolution ($R\sim$~45,000) and the typical SNR of their optical spectra are higher than our present survey. They have found only 13 systems with \HH\ detections in a sample of 77 DLAs/sub-DLAs. Therefore, it appears that the \HH\ detection is much more frequent at low-$z$ compared to high-$z$. We quantify this in Fig.~\ref{IncRate}, where we have compared the \HH\ detection rates in the high and low redshift samples for different $N(\HH)$ threshold values. In order to estimate detection rate for a given threshold value, we use {\sl only} spectra that are sensitive to detect \HH\ down to the threshold $N(\HH)$. The detection rate is then derived by taking the ratio of number of systems in which \HH\ is detected with $N(\HH)$ higher than the threshold value to the total number of systems detected in those spectra. As mentioned before, the ``median sensitivity" of our sample is $\log N(\HH) = 14.4$ whereas for the high-$z$ sample it is $\log N(\HH) = 14.2$. Note that for the 13 high-$z$ \HH\ systems of N08 sample we have derived the spectral sensitivity for \HH\ detection at 3$\sigma$ level in the same fashion as described in the previous section.       

It is apparent from Fig.~\ref{IncRate} that the \HH\ detection rate is higher at low-$z$ for the whole range of $N(\HH)$ threshold values up to $\log N(\HH) = 16.5$. The difference is increasingly higher for lower $N(\HH)$ thresholds. For a threshold column density of $\log N(\HH) = 14.4$, the \HH\ detection rate at low-$z$ is $50^{+25}_{-12}$ percent, whereas it is only $18^{+8}_{-4}$ percent at high-$z$. Therefore, the \HH\ detection rate is $\gtrsim 2$ times higher at low-$z$ with a $\sim2\sigma$ confidence. Recently, in a blind survey for \HH\ in DLAs, observed with mostly low spectral resolution ($\sim$~71 \kms) data, \citet[]{Jorgenson14H2} have found a paucity of strong \HH\ systems (e.g. $\sim1$ percent detection rate for $\log N(\HH) > 18$). For a $N(\HH)$ threshold as high as $10^{18}$~cm$^{-2}$, the \HH\ detection rate in our low-$z$ sample (i.e. $15^{+12}_{-4}$ percent) is higher than that found by \citet[]{Jorgenson14H2} but consistent with the sample of N08 (i.e. $10^{+5}_{-3}$ percent). In passing, we wish to point out that using high redshift ($z > 2$) QSO spectra from the Sloan Digital Sky Survey (SDSS), \citet{Balashev14} have found an upper limit in detection rate of $\sim 7$ percent for strong \HH\ systems with $\log N(\HH) > 19.0$. For our sample, we find a detection rate of $7^{+10}_{-2}$ percent for a similar $N(\HH)$ threshold.  

\begin{figure*} 
\centerline{\hbox{ 
\centerline{\vbox{ 
\includegraphics[height=5.6cm,width=8.6cm,angle=00]{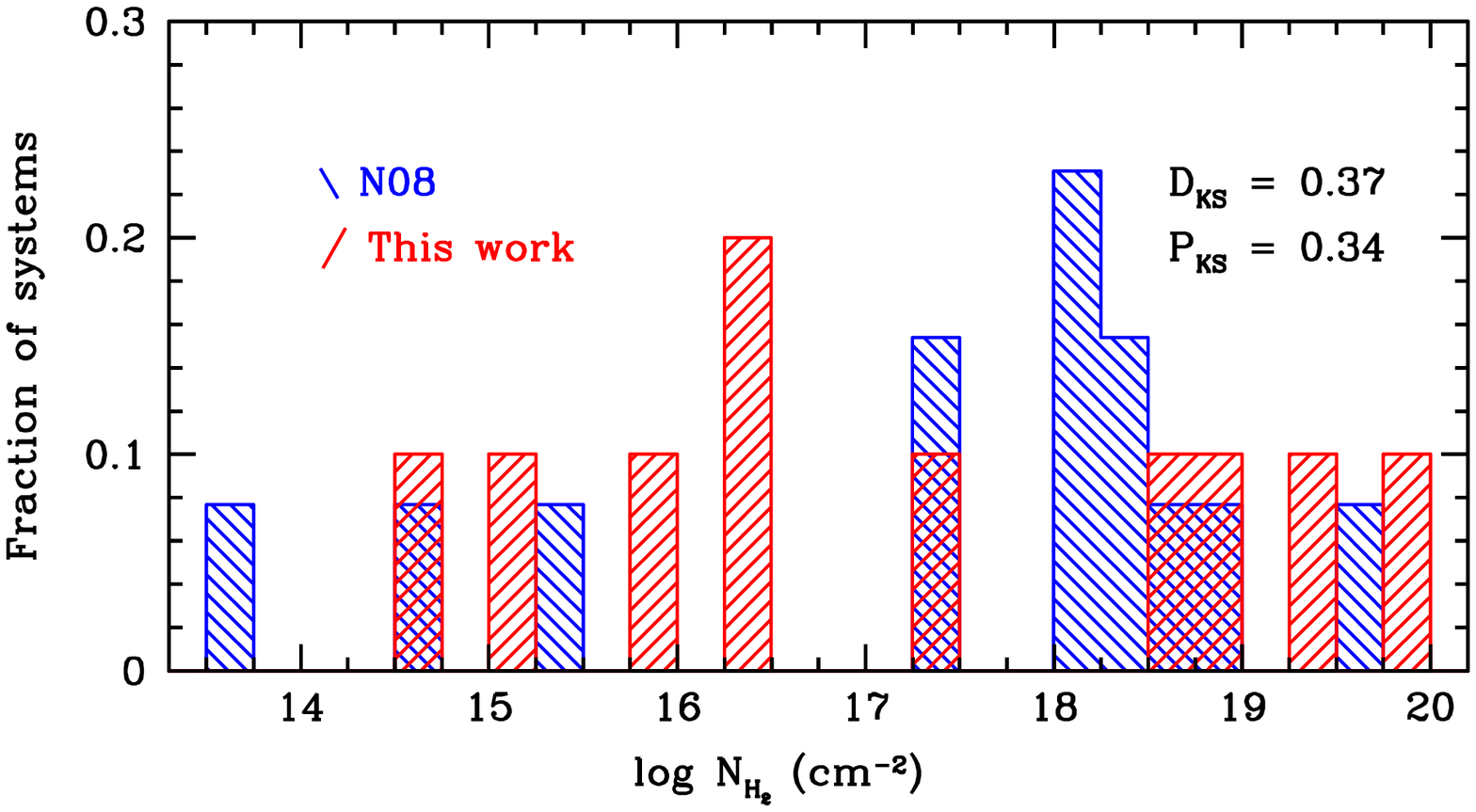}    
\includegraphics[height=5.6cm,width=8.6cm,angle=00]{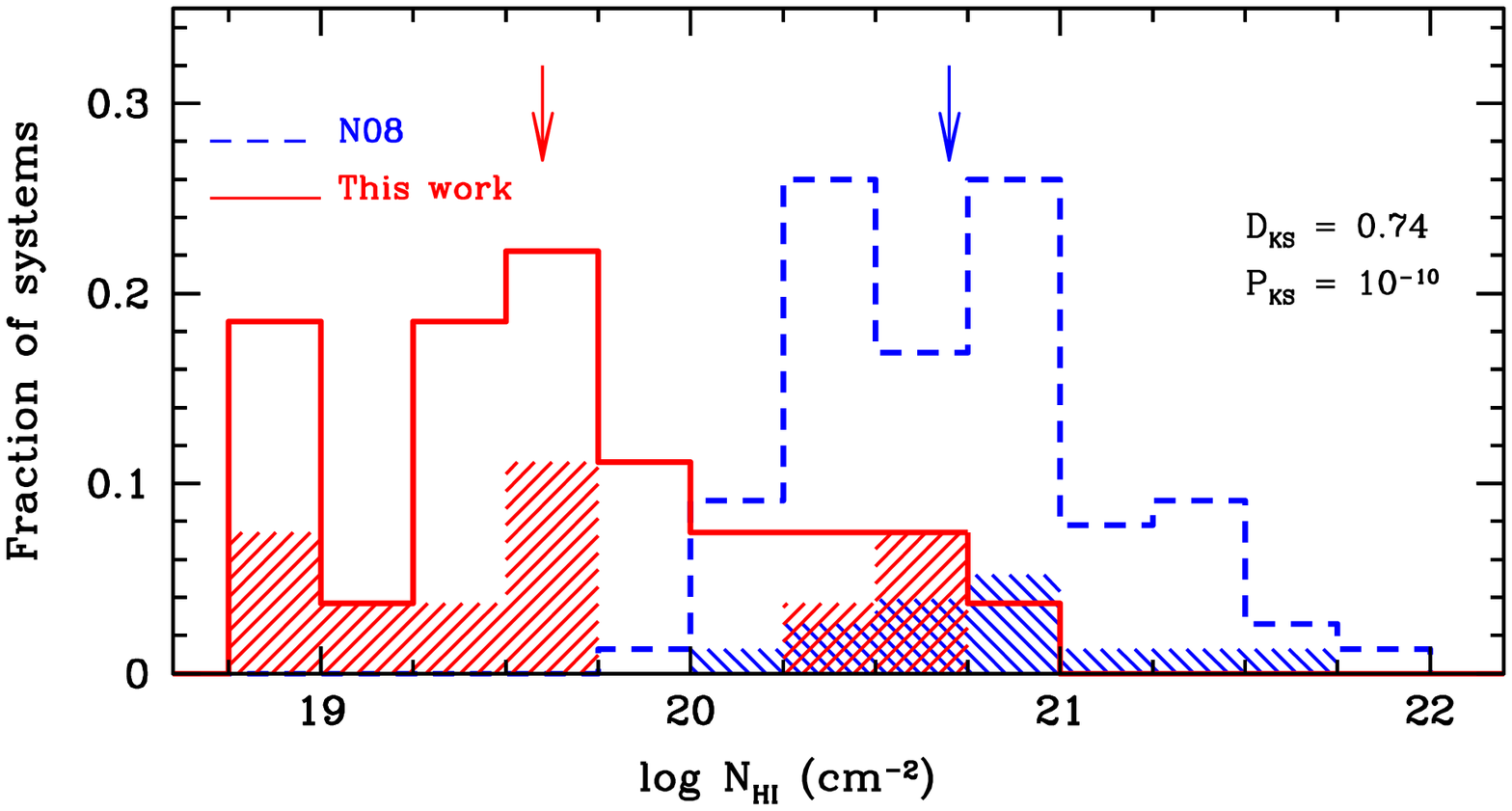}    
}}
}}
\caption{Left: The $N(\HH)$ distributions for the low-$z$ sample (red 45$^{\circ}$ hatched histogram) 
and the high-$z$ sample of N08 (blue 135$^{\circ}$ hatched histogram).  
Right: The $N(\HI)$ distributions of our core sample (red solid histogram) and the sample of 
N08 (blue dashed histogram). The arrows mark the median values. The (red) 45$^{\circ}$ and  
(blue) 135$^{\circ}$ hatched histograms show the $N(\HI)$ distributions for the systems with 
\HH\ detections in low- and high-$z$ samples respectively.}        
\label{Ndist}  
\end{figure*}   

\subsection{Column density distributions}   
\label{subsec:colden}

The total \HH\ column density of a given system with detected \HH, as listed in column 8 of Table~\ref{sample1}, is derived by summing the column densities measured in different $J$ levels as listed in columns 3, 4, 5, and 6 of Table~\ref{H2tab}. For multicomponent systems, component column densities are further summed up to get the total $N(\HH)$. The total \HH\ column density distribution for our low-$z$ sample is compared with that of the high-$z$ sample of N08 in the left panel of Fig.~\ref{Ndist}. We do not find any significant difference between them. A two-sided Kolmogorov-Smirnov (KS) test suggests that the maximum deviation between the two cumulative distribution functions is $D_{\rm KS} = 0.37$ with a probability of $P_{\rm KS} = 0.34$\footnote{$P_{\rm KS}$ is the probability of finding this $D_{\rm KS}$ value or lower by chance.}.

In the right panel of Fig.~\ref{Ndist} the \HI\ column density distributions of our core sample (total 27 systems) and the sample of N08 (total 77 systems) are shown. It is clearly evident that the two distributions are significantly different. A two-sided KS-test also supports this with a $D_{\rm KS}$ = 0.74 and a $P_{\rm KS} \sim 10^{-10}$. The median value of $\log N(\HI)=$~19.6 for our low-$z$ sample is an order of magnitude lower than that of the high-$z$ sample. In the low-$z$ sample only 7 systems show $\log N(\HI)>$~20.0, whereas all 77 high-$z$ systems have $\log N(\HI)>$~20.0. Therefore, in order to make a more realistic comparison between properties of \HH\ absorbers at high- and low-$z$, it is important to increase the sample sizes of the sub-DLAs at high-$z$ and the DLAs at low-$z$.

The hatched histograms in the right panel of Fig.~\ref{Ndist} show the $N(\HI)$ distributions for the systems in which \HH\ is detected. It is interesting to note that there is no preference for molecular hydrogen to originate from higher (or lower) $N(\HI)$ systems. A two-sided KS-test suggests that the $N(\HI)$ distributions for systems with and without \HH\ detections are very similar (e.g. $D_{\rm KS} = 0.18$ and $P_{\rm KS} = 0.97$). Moreover, we do not find any significant correlation between $N(\HI)$ and $N(\HH)$. N08 have found the same for their high-$z$ sample. As noted earlier, a significant fraction (i.e. 7 out of 10) of low-$z$ \HH\ systems show $\log N(\HI) <$~20. Detection of \HH\ at lower $N(\HI)$ and the enhanced detection rate could possibly suggest that the low-$z$ sub-DLAs have (a) higher metallicities and/or dust, (b) higher densities, and/or (c) weaker radiation field. We discuss these issues in Section~\ref{sec:diss}.  

\subsection{The molecular fraction, $f_{\hh}$}   
\label{subsec:fH2} 

\begin{figure*} 
\centerline{\vbox{ 
\centerline{\hbox{ 
\includegraphics[height=7.5cm,width=7.5cm,angle=00]{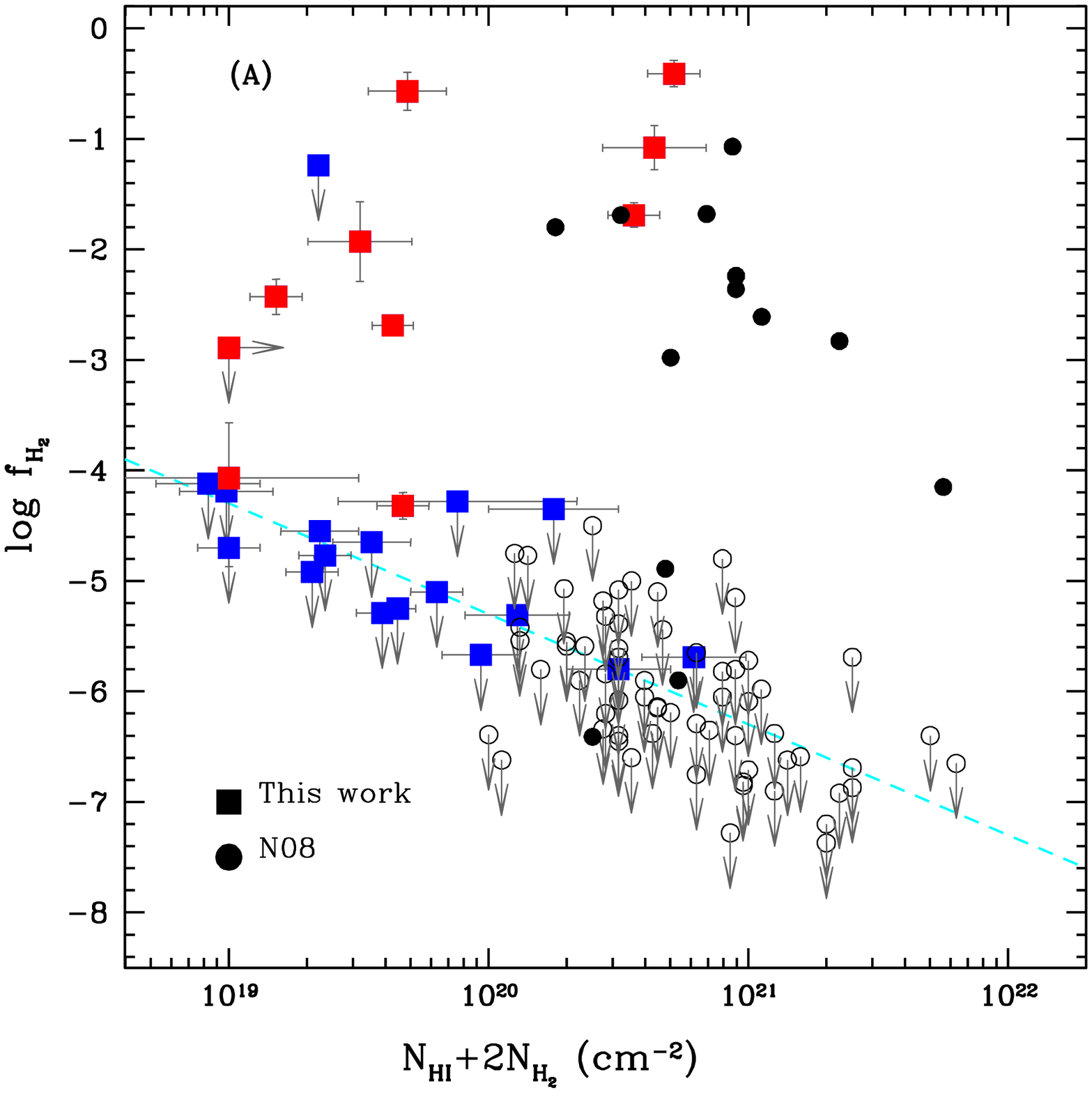}    
\includegraphics[height=7.5cm,width=7.5cm,angle=00]{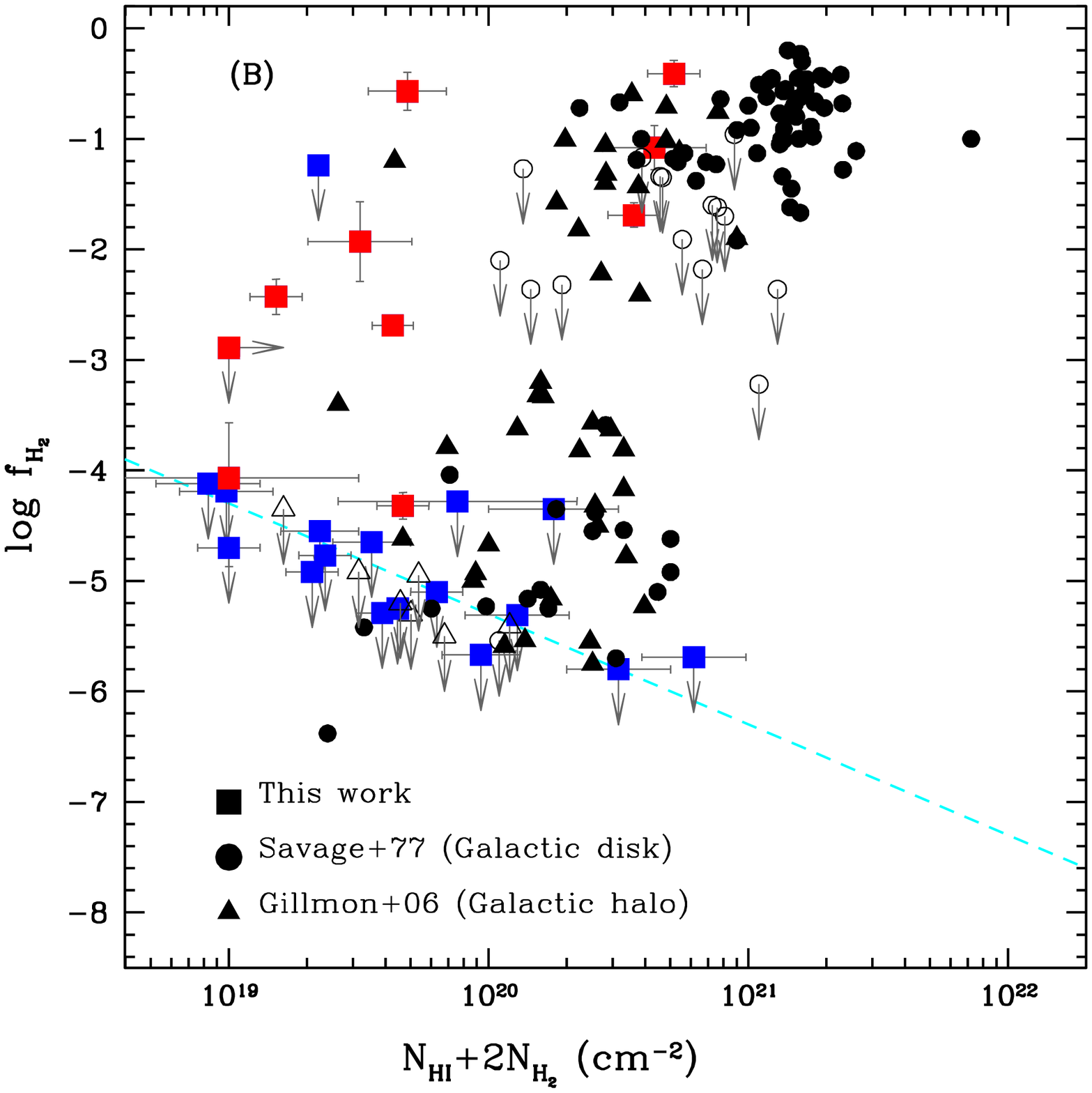}  
}}  
\centerline{\hbox{ 
\includegraphics[height=7.5cm,width=7.5cm,angle=00]{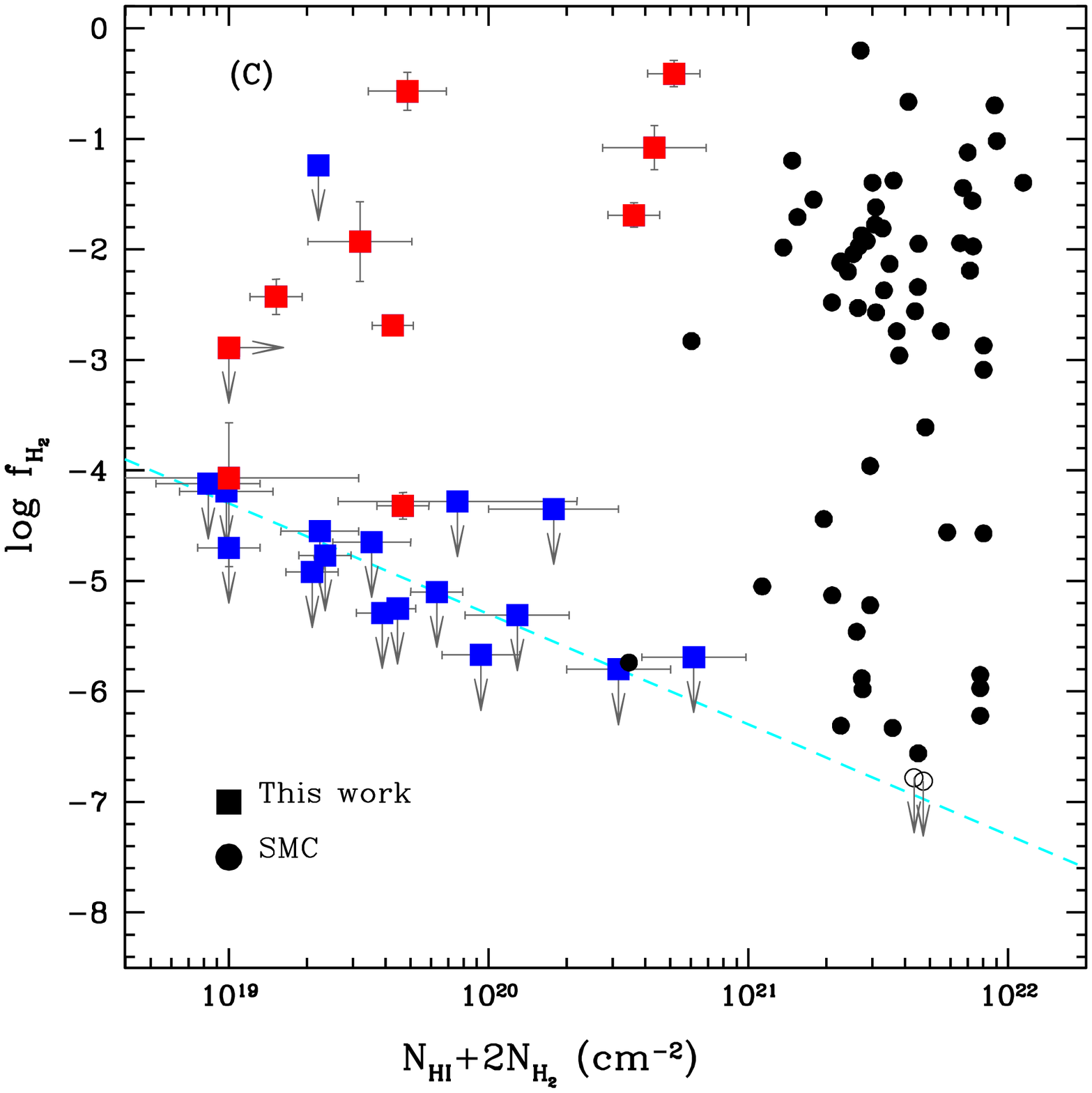}    
\includegraphics[height=7.5cm,width=7.5cm,angle=00]{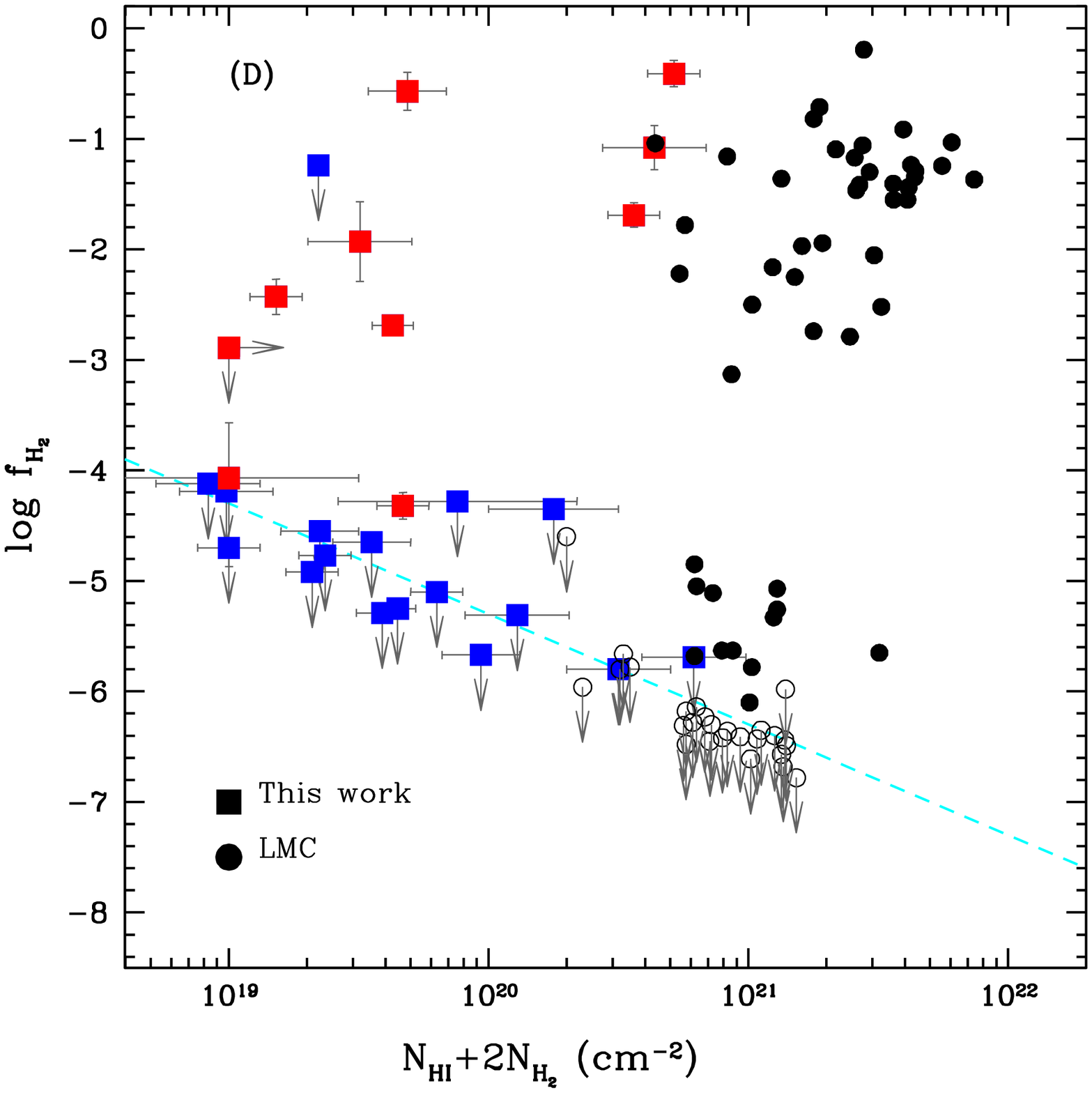}    
}}
}} 
\caption{The molecular fraction against the total hydrogen (neutral atomic + molecular)  
column density, $N(\HI+\HH)$. In each 
panel squares represent our low-$z$ sample (with red squares corresponding to \HH\ detections), 
whereas circles/triangles represent other samples as indicated in the panel (with filled ones 
corresponding to \HH\ detections). The dashed line corresponds to our ``median sensitivity" 
limit i.e. $\log N(\HH) = 14.4$.}       
\label{fH2}  
\end{figure*}

The molecular fraction, $f_{\hh}$, of a DLA/sub-DLA system is defined as:  
\begin{equation}  
f_{\hh} = \cfrac{2N(\HH)}{N(\HI)+2N(\HH)}~.       
\end{equation}  
At the median $N(\HI)$, the ``median sensitivity" limit of our survey (i.e. $\log N(\HH) = 14.4$) corresponds to a $f_{\hh} = 10^{-4.9}$. We, therefore, denote this as our median $f_{\hh}$ sensitivity limit.   
In four different panels of Fig.~\ref{fH2} we have plotted $f_{\hh}$ against total hydrogen column density, $N_{\rm H} = N(\HI)$+2$N(\HH)$. We note that for sub-DLAs this is a lower limit in $N_{\rm H}$ as we are ignoring the ionized fraction.

In panel-(A) we have compared low- and high-$z$ samples in the $f_{\hh}-N_{\rm H}$ plane. As already noted, the low-$z$ systems have lower $N_{\rm H}$ values (always $N_{\rm H} < 10^{20.7} {\rm cm^{-2}}$) compared to high-$z$. There are a few detections with very low molecular fractions $(f_{\hh} \lesssim 10^{-4.9}$) in the high-$z$ sample. Such low $f_{\hh}$ systems are not present in our sample. We perform a two-sample survival analysis, including censored data points (upper limits), in order to investigate if $f_{\hh}$ distributions at high-$z$ and low-$z$ are drawn from the same parent populations\footnote{We have used the ``{\sl survdiff}" function under the ``{\sl survival}" package in {\sc r} (http://www.r-project.org/ ).}. The log-rank test rejects the null hypothesis with a confidence of 98.7\% (see Table~\ref{tab:logrank}), suggesting only a mild difference. When we consider only systems with $\log f_{\hh} > -4.9$, the log-rank test rejects the null hypothesis with a confidence of only 56\%, implying a lack of any statistically significant difference in the $f_{\hh}$ distribution. We point out that the mild difference suggested by the log-rank test for the entire sample is merely due to the fact that the \HH\ detection rate is considerably higher at low-$z$ than at high-$z$.                

\begin{table}  
\begin{center}  
\caption{Results of log-rank tests between $f_{\hh}$ distributions.}           
\begin{tabular}{ccccc}  
\hline \hline 
Sample-1  &  Sample-2    & $> \log f_{\hh}^{a}$  &   \chisq$^{b}$  &  $P_{\rm log-rank}^{c}$  \\        
\hline  
Low-$z$   &  High-$z$    &  $-8.0$   &   6.2     &  $0.013$    \\ 
Low-$z$   &  High-$z$    &  $-4.9$   &   0.6     &  $0.444$    \\  
Low-$z$   &  ISM (Halo)  &  $-4.9$   &   6.7     &  $0.009$    \\ 
Low-$z$   &  ISM (Disk)  &  $-4.9$   &   14.8    &  $10^{-4}$  \\  
Low-$z$   &  SMC         &  $-4.9$   &   15.2    &  $10^{-4}$  \\  
Low-$z$   &  LMC         &  $-4.9$   &   15.2    &  $10^{-4}$  \\  
\hline \hline 
\end{tabular}        
\label{tab:logrank}	
~\\ ~\\  
Notes -- 
$^{a}$Threshold $\log f_{\hh}$.  
$^{b}$The \chisq\ statistic for a test of equality. \\  
$^{c}$Probability of rejecting null hypothesis by chance.       
\end{center}                                                   
\end{table}      

In panel-(B) of Fig.~\ref{fH2} we compare our sample with that of the \citet[][i.e. the Galactic disk]{Savage77} and \citet[][i.e. the Galactic halo]{Gillmon06} samples. The Galactic samples clearly show a transition near $\log N_{\rm H}=$~20.7, above which all systems have molecular fraction $\log f_{\hh} \gtrsim -1$. This transition, leading to a bimodality in the $f_{\hh}$ distribution, is generally identified as the threshold above which the \HH\ molecule gets completely self-shielded from interstellar radiation \citep[]{Savage77}. Such a transition is not apparent in the low-$z$ and/or high-$z$ DLA/sub-DLA sample. This is what is expected since they trace a wide variety of environments (e.g. different metallicities, dust depletion, radiation field etc.) unlike the sightlines passing through a single galaxy. A log-rank test of the $f_{\hh}$ distributions between our low-$z$ sample and the Galactic disk sample of \citet{Savage77}, for systems with $\log f_{\hh} > -4.9$, indicates that they are significantly different (e.g. rejects the null-hypothesis with a confidence of $>$~99.99\%). This is also true when we compare low-$z$ $f_{\hh}$ distribution with that of the Galactic halo \citep{Gillmon06} sample, albeit with lower confidence (e.g. 99.1\% confidence, see Table~\ref{tab:logrank}).

\begin{figure*} 
\centerline{\vbox{
\centerline{\hbox{ 
\includegraphics[height=8.4cm,width=8.4cm,angle=00]{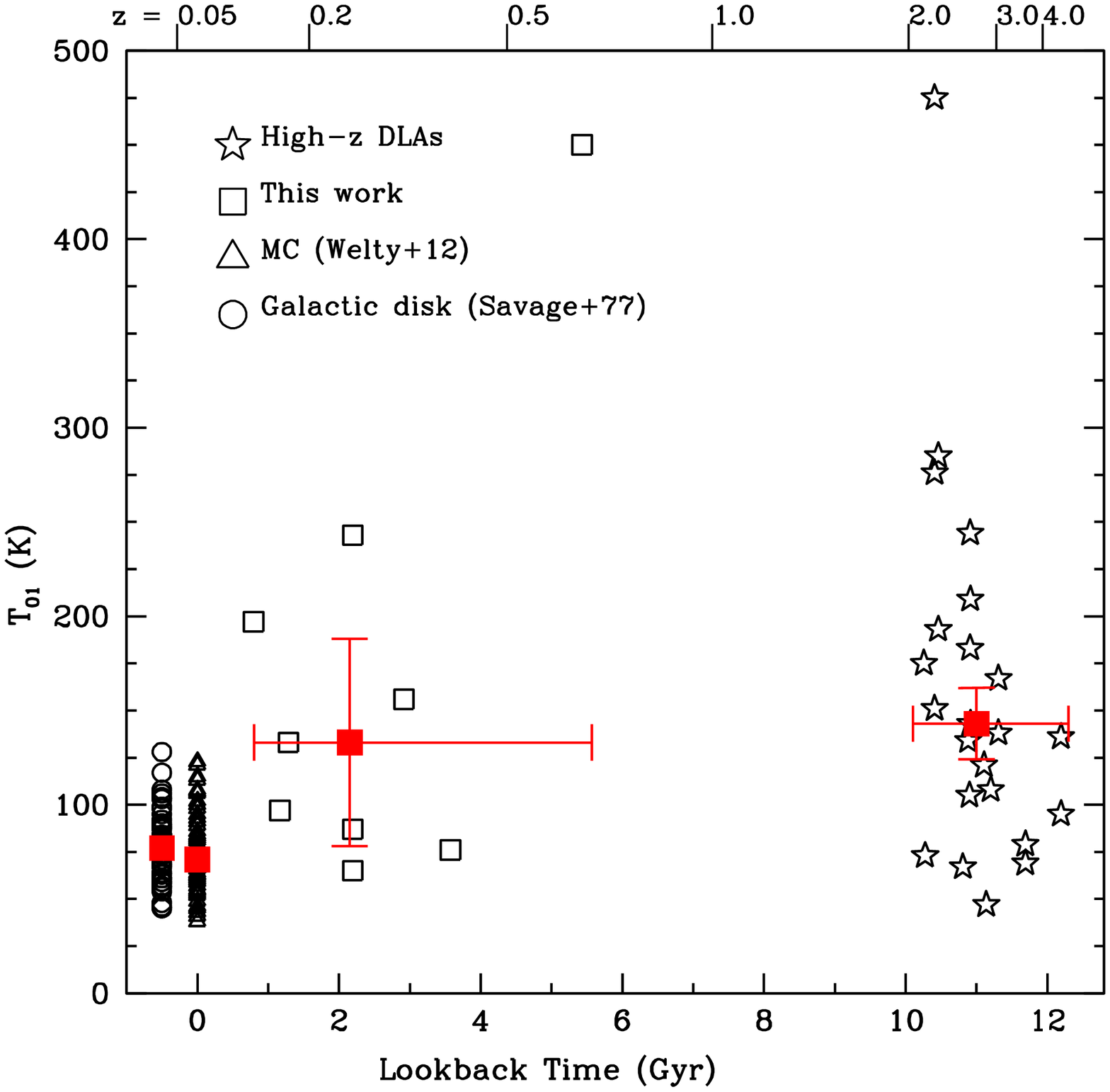}    
\includegraphics[height=8.4cm,width=8.4cm,angle=00]{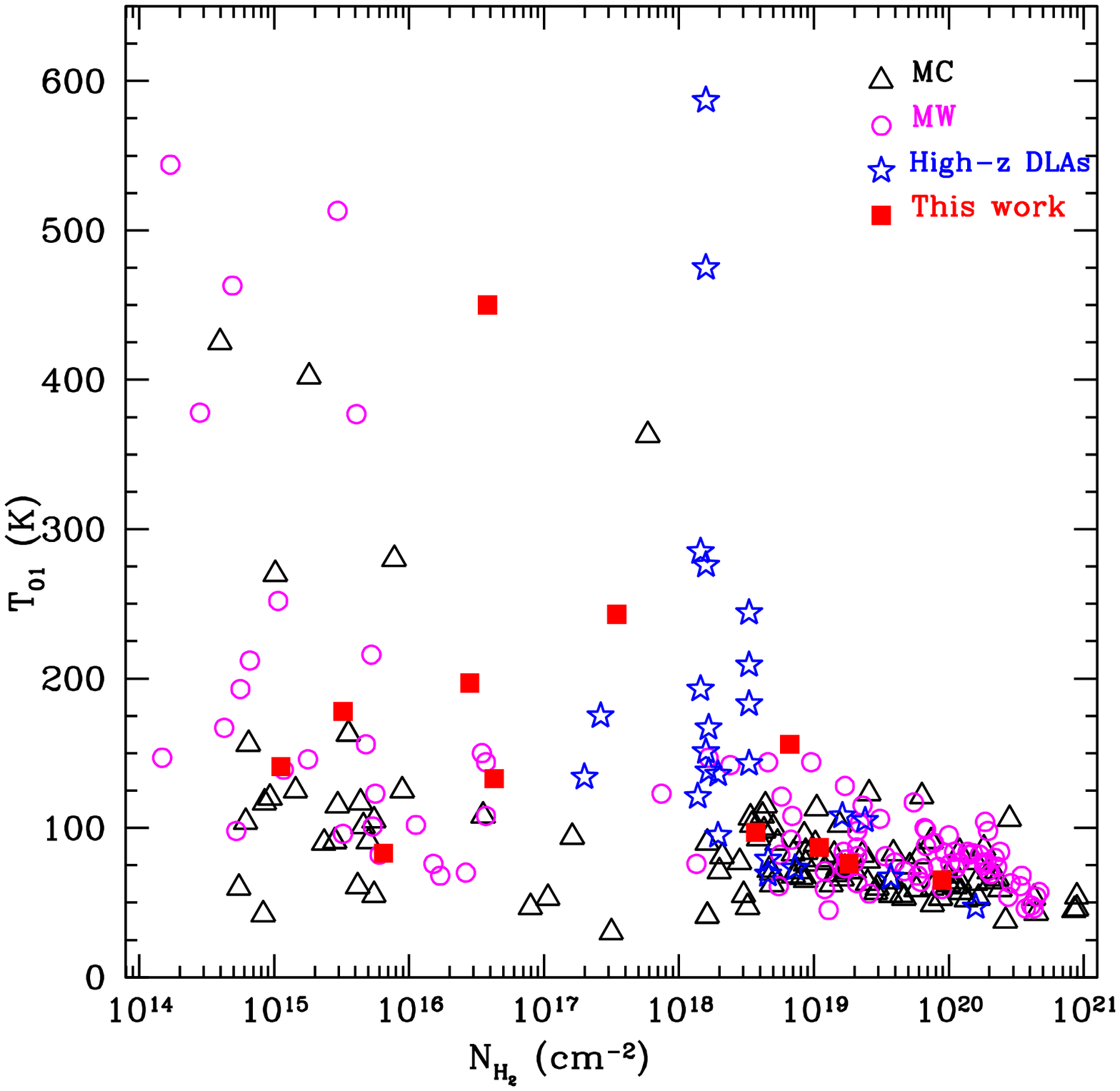}    
}}
}}
\caption{Left: Cosmic evolution of the rotational excitation temperature, $T_{01}$. Only components    
with total $\log N(\HH) > 16.5$ are used. The (red) filled squares represent the median values in 
each sample. Error-bars along y-axis are the 1$\sigma$ Gaussian errors on the median value computed 
from bootstrapping. Error-bars along x-axis are just to show the spread in look-back time for 
high- and low-$z$ samples. The Galactic measurements are shifted to the left for clarity.      
Here we have excluded the temperature measurements in the \zabs\ = 0.55715 component towards 
Q~0107$-$0232.                         
Right: $T_{01}$ against \HH\ column density for different samples plotted with different 
symbols. The (magenta) open circles (Milky Way, MW) are from \citet{Savage77} and \citet{Gillmon06}. 
The (black) open triangles (Magellanic Clouds, MC) are from \citet{Tumlinson02} and \citet{Welty12}. 
The (blue) stars are from the ``high-$z$ DLAs" as defined in the text. The data points plotted 
as (red) squares are from this study.}             
\label{fig:T01}  
\end{figure*}   

In panel-(C) and (D) of Fig.~\ref{fH2}, the low-$z$ sample is compared with the SMC and LMC, respectively. Here we have taken both \citet{Tumlinson02} and \citet{Welty12} measurements into account. For lines of sight that are common in both the studies, we used updated measurements from \citet{Welty12}. It is clear that our low-$z$ sample and the Magellanic Cloud systems have only marginal overlap, with the latter having systematically higher $N_{\rm H}$ values. Unlike the Galactic lines of sight, the transition in $f_{\hh}$ due to self-shielding is not very prominent for SMC systems. LMC systems, however, show such transition albeit at higher $N_{\rm H}$ values (i.e. at $\log N_{\rm H}$ = 21.3). A log-rank test of  $f_{\hh}$ distributions between our low-$z$ sample and the SMC$/$LMC, for systems with $\log f_{\hh} > -4.9$, suggests that they are drawn from significantly different parent populations (e.g. rejects the null-hypothesis with a probability of $>$~99.99\%, see Table~\ref{tab:logrank}). 

It is clearly evident from Fig.~\ref{fH2} that our low-$z$ \HH\ systems predominantly populate the upper-left corner of the $f_{\hh} - N_{\rm H}$ plane (i.e. $f_{\hh} \gtrsim -4.5$ and $\log N_{\rm H} \lesssim 20.7$). This is a unique region in the parameter-space. Only one of the Magellanic Cloud systems, three high-$z$ systems and a few (8/72) Galactic disk systems, with detected \HH, fall in this region. The key reason for this observation is that our low-$z$ sample is primarily comprised of sub-DLAs. The only possible exception is the Galactic halo sample. We would like to emphasize here that a significantly large fraction (e.g. $25/40 \sim$~63\%) of the Galactic halo systems, in which \HH\ is detected, have $f_{\hh}$ and $N_{\rm H}$ in the range similar to as seen for our low-$z$ sample. This perhaps indicates that the origin(s) and the physical conditions of molecular gas in low-$z$ DLAs/sub-DLAs are similar to that of the Milky Way halo gas. 

It is interesting to note that even with systematically lower $N_{\rm H}$ values, many low-$z$ systems show a considerably large molecular fraction. We discuss possible reasons for that in view of simple photoionization models in Section~\ref{PImod}. The Galactic disk/Magellanic Clouds/high-$z$ samples are primarily composed of DLAs with $N_{\rm H} >10^{21}$~cm$^{-2}$. Therefore, we point out that for a realistic comparison of $f_{\hh}$ distributions, as presented in  Table~\ref{tab:logrank}, more observations of molecular hydrogen in DLAs at low-$z$ are extremely important. Additionally, more observations of \HH\ in sub-DLAs at high-$z$ are crucial for comparing with our current low-$z$ sample.

\subsection{The excitation temperature, $T_{01}$} 
\label{subsec:T01}

The rotational excitation temperature, $T_{01}$, for $J=$~0 to $J=$~1 rotational levels 
can be expressed as:   
\begin{equation}  
\cfrac{N(J=1)}{N(J=0)} = \cfrac{g_{1}}{g_{0}}~{\rm exp}(-170.5/T_{01})~.     
\end{equation}  
Here $g_{0}$, $g_{1}$ are the statistical weights for $J=$~0 and $J=$~1 rotational levels respectively. When \HH\ is sufficiently self-shielded (e.g. $\log N(\HH) >$~16.5) from photo-dissociating photons, collisional processes dominate the level populations, then $T_{01}$ represents the kinetic temperature of the absorbing gas \citep[]{Snow00,Roy06}. The excitation temperatures ($T_{01}$) measured from \HH\ absorption in our sample are summarized in column 9 of Table~\ref{H2tab}. The median value of $T_{01}$ in our sample is $T_{01}=$~133$\pm$75~K for the \HH\ components with a total $N(\HH)> 10^{16.5}$ cm$^{-2}$. The large scatter results from the large $T_{01}$ value (i.e. 997~K) measured in one of the components in \zabs\ = 0.55733 towards Q~0107$-$0232 system. The median value becomes $T_{01}=$~133$\pm$55~K when we exclude the outlier. Note that the errors in the median $T_{01}$ values in this work are estimated from bootstrapping unless specified.       

In the left panel of Fig.~\ref{fig:T01} we show the evolution of $T_{01}$ over the last 12 Gyr of cosmic time. Measurements from different samples at different cosmic time (or redshift) are shown in different symbols. The local measurements are taken from \citet[the Galactic disk]{Savage77} and \citet[the Magellanic Clouds]{Welty12}, whereas the high-$z$ measurements are from \cite{Petitjean02}, \cite{Srianand05,Srianand08}, \cite{Ledoux06b}, \cite{Noterdaeme07,Noterdaeme08,Noterdaeme10co}, and \cite{Guimaraes12}. Hereafter, we will refer this ensemble of systems as ``high-$z$ DLAs". 
The median $T_{01}$ in our low-$z$ sample (i.e. 133$\pm$55~K) is very similar to that of the ``high-$z$ DLAs" (i.e. 143$\pm$19~K). The local measurements of $T_{01}$, i.e. 77$\pm$02~K in the Galactic disk, 115$\pm$13~K in the Galactic halo, and 71$\pm$03~K in the Magellanic Clouds, are slightly lower but roughly consistent with the low-$z$ measurements within the large scatter.       

In the right panel of Fig.~\ref{fig:T01}, $T_{01}$ is plotted against the \HH\ column density for various different samples. $T_{01}$, in general, seem to show a large scatter for lower values of $N(\HH)$. On the contrary, for $\log~N(\HH) >$~19.0 all systems show $T_{01}<$~120~K, irrespective of which samples they come from. This can be understood in terms of efficient shielding of radiation in higher $N(\HH)$ systems. In columns 10 and 11 of Table~\ref{H2tab} we present $T_{02}$ and $T_{13}$, respectively, estimated using Eqn.~3 of \citet{Srianand05}. These values are typically higher than the corresponding $T_{01}$ measurements. Such elevated $T_{02}$ and/or $T_{13}$ values compared to $T_{01}$ have been interpreted as the influence of radiation pumping and/or formation pumping \citep[]{Tumlinson02,Srianand05}. Higher rotational level (i.e. $J \geqslant 4$) populations (e.g. $N_4/N_2$ or $N_5/N_3$ ratios) are even more sensitive probes of radiation/formation pumping \citep[see e.g.][]{Tumlinson02}. None of the low-$z$ \HH\ systems show detectable absorption from $J \geqslant 4$ levels. This clearly suggests that the radiation/formation pumping is not important for the \HH\ systems in our sample.

\begin{figure} 
\centerline{\vbox{
\centerline{\hbox{ 
\includegraphics[height=7.4cm,width=8.6cm,angle=00]{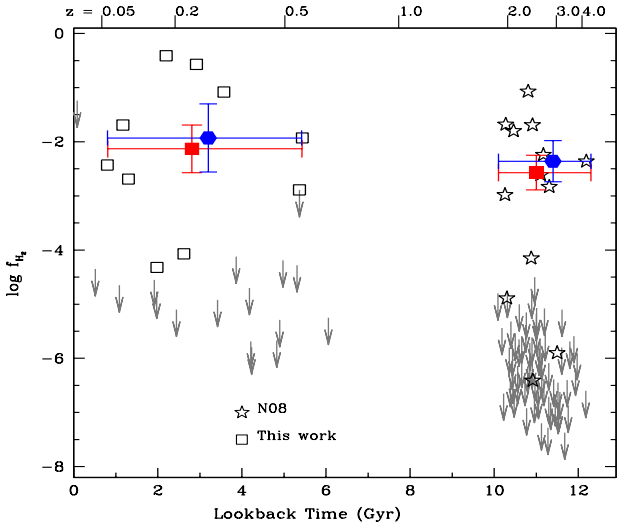}    
}}
}}
\caption{Cosmic evolution of the molecular fraction in DLAs/sub-DLAs. Open squares and stars represent 
measurements from our low-$z$ and the high-$z$ sample of N08, respectively. Filled (red) squares and  
(blue) diamonds, respectively, represent the mean and median of $\log f_{\hh}$, for the systems in 
which \HH\ is detected with a $\log f_{\hh} > -4.9$ in each sample. The error-bars in mean and median 
$\log f_{\hh}$ are computed from bootstrapping. The error-bars along $x$-axis are just to show the 
spread in look-back time for each sample.        
}      
\label{zevol}  
\end{figure}   

\section{Discussions} 
\label{sec:diss}

\subsection{Evolution of $f_{\hh}$}   

In equilibrium between formation and photo-dissociation, the molecular fraction can be written as:  
\begin{equation}           
f_{\hh} = \frac{2Rn_{\hi}}{D}~,      
\end{equation}   
where, $R$ (in cm$^{3}$ s$^{-1}$) is the formation rate coefficient and $D$ (in s$^{-1}$) is the photo-dissociation rate. In a cold and neutral gas phase, \HH\ formation occurs on the surface of dust grains. Thus, $R$ is proportional to the dust-to-gas ratio ($\kappa$). Note that $\kappa$ depends on the metallicity of the gas. The dissociation rate, $D$, on the other hand, depends crucially on the ionizing radiation. There are several indications that  $f_{\hh}$ should increase with cosmic time: \\ 
(a) The global star-formation rate density (or luminosity density) of galaxies decreases with time \citep[see e.g.][]{Bouwens11}. This suggests that if \HH\ systems are related to star forming disks at all epochs, then low-$z$ systems will experience a much weaker ionizing radiation field compared to high-$z$ systems. This will lead to a decrease in the photo-dissociation rate. \\     
(b) The extra-galactic UV background radiation gets fainter by about an order of magnitude at low-$z$ compared to high-$z$ ($z>2$) as a consequence of the decrease in the global star-formation rate density \citep[]{Haardt96,Haardt12}. This implies that if \HH\ systems are related to halo gas, where extra-galactic UV background radiation dominates over the radiation field of the host-galaxy, then again low-$z$ systems will experience a much weaker ionizing radiation compared to high-$z$ systems. \\  
(c) An increase in the cosmic mean metallicity of DLAs with time \citep[]{Prochaska03Zz,Rafelski12}, suggests that the \HH\ formation rate will be higher at low-$z$, leading to an increase in $f_{\hh}$. Furthermore, we note that sub-DLAs tend to show higher metallicities and faster metallicity evolution compared to DLAs \citep[]{Kulkarni07,Som13}. Nevertheless, we find that the median metallicity of our \HH\ systems (listed in column 12 of Table~\ref{H2tab}) is only 0.4 dex higher than that of the N08 sample.    

In Fig.~\ref{zevol} we show the evolution of $f_{\hh}$ over the last 12 Gyr of cosmic time. The high- and low-$z$ samples are presented in open stars and open squares, respectively. Upper limits in both the samples are shown by arrows. The mean and median values of $\log f_{\hh}$, for the low-$z$ systems in which \HH\ is detected with a $\log f_{\hh} > -4.9$, are found to be $-2.13\pm0.44$ and $-1.93\pm0.63$, respectively. These values are only $\sim2.7$ times higher than the corresponding values at high-$z$. This could result from the fact that the $N(\HH)$ distributions at high- and low-$z$ are very similar whereas the low-$z$ systems have systematically lower $N(\HI)$ values. Nevertheless, the mean/median values of $\log f_{\hh}$ at high- and low-$z$ are consistent within $1\sigma$ allowed range. Finally, we recall from Section~\ref{sec:search} that the $N(\HH)$ we derive from COS data could be a lower limit if the lines are narrow and unresolved or partially covering the background continuum source.

Next, we have performed the Kendall's $\tau$ correlation test including censored data points between absorption redshift and $\log f_{\hh}$\footnote{We have used the ``{\sl cenken}" function under the ``{\sl NADA}" package in {\sc r}}. No significant correlation is found when we consider only system with $\log f_{\hh} > -4.9$. However, a mild anti-correlation is suggested by the Kendall's $\tau$ test with a $\tau = -0.27$ and with a confidence of 99.99 percent, when we consider all the systems from both the high- and the low-$z$ samples. This mild anti-correlation is essentially the manifestation of the fact that the detection rate of \HH\ is considerably higher at low-$z$.     

\begin{figure*} 
\centerline{\vbox{
\centerline{\hbox{ 
\includegraphics[height=5.8cm,width=5.8cm,angle=00]{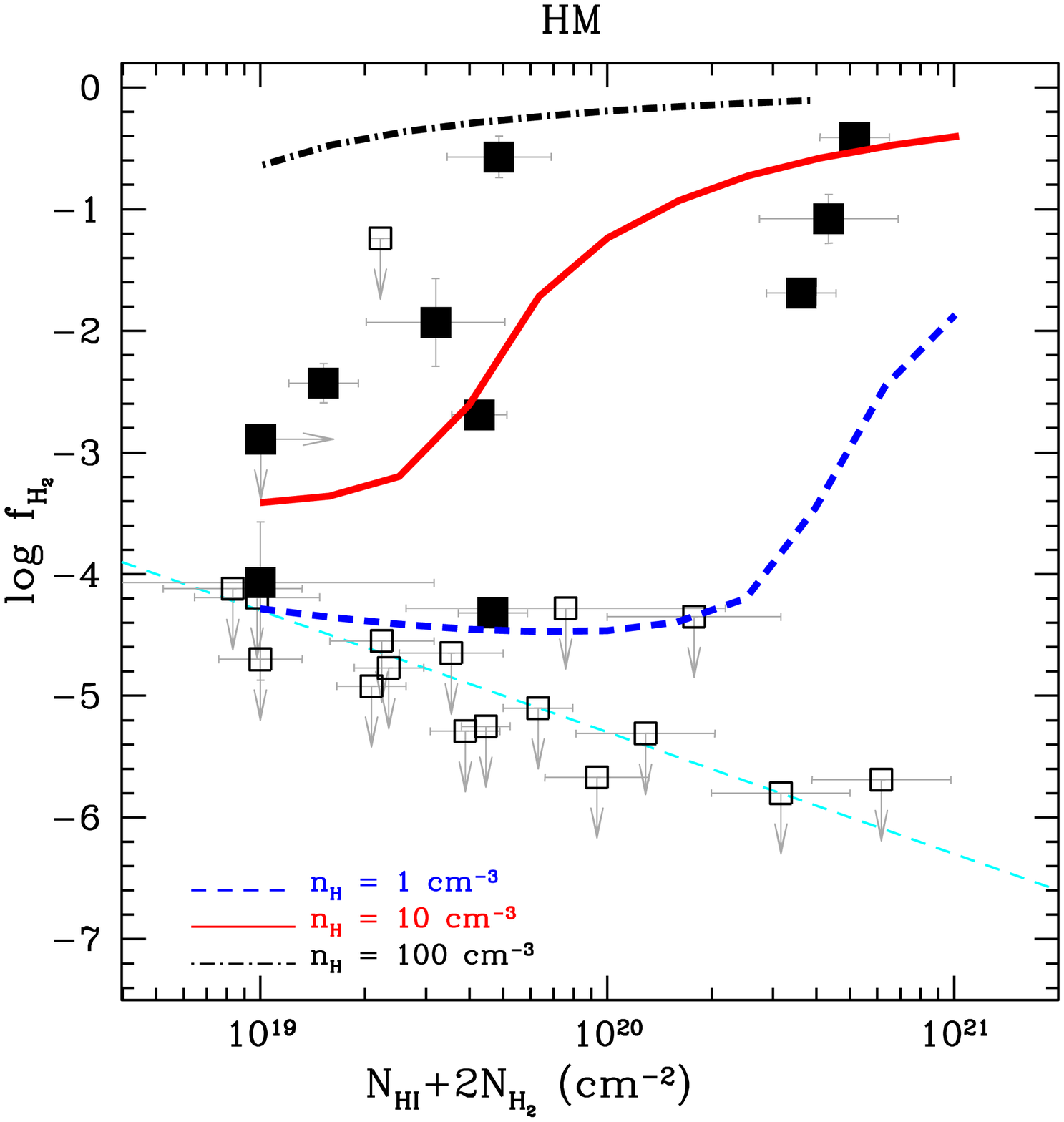}    
\includegraphics[height=5.8cm,width=5.8cm,angle=00]{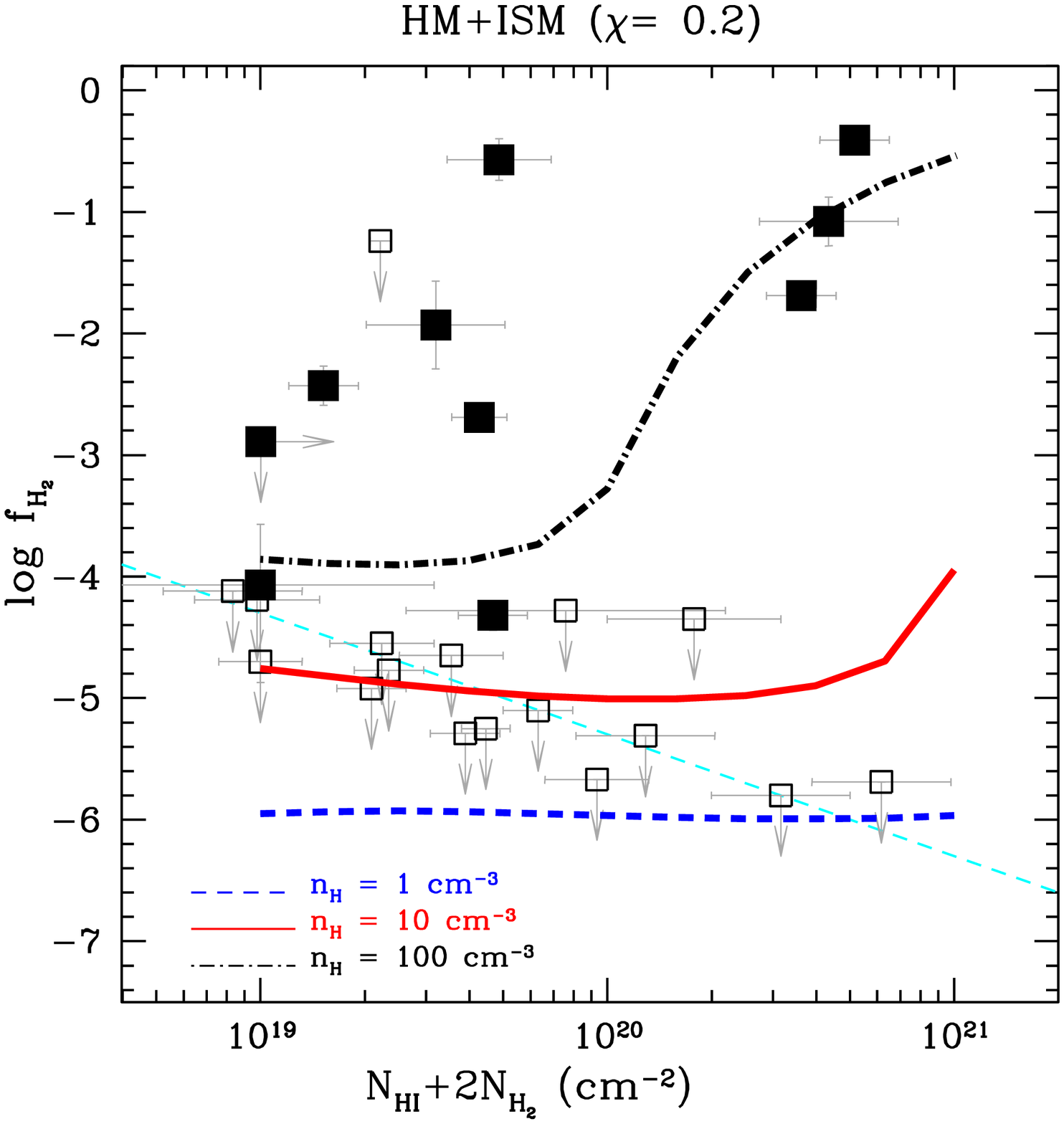}    
\includegraphics[height=5.8cm,width=5.8cm,angle=00]{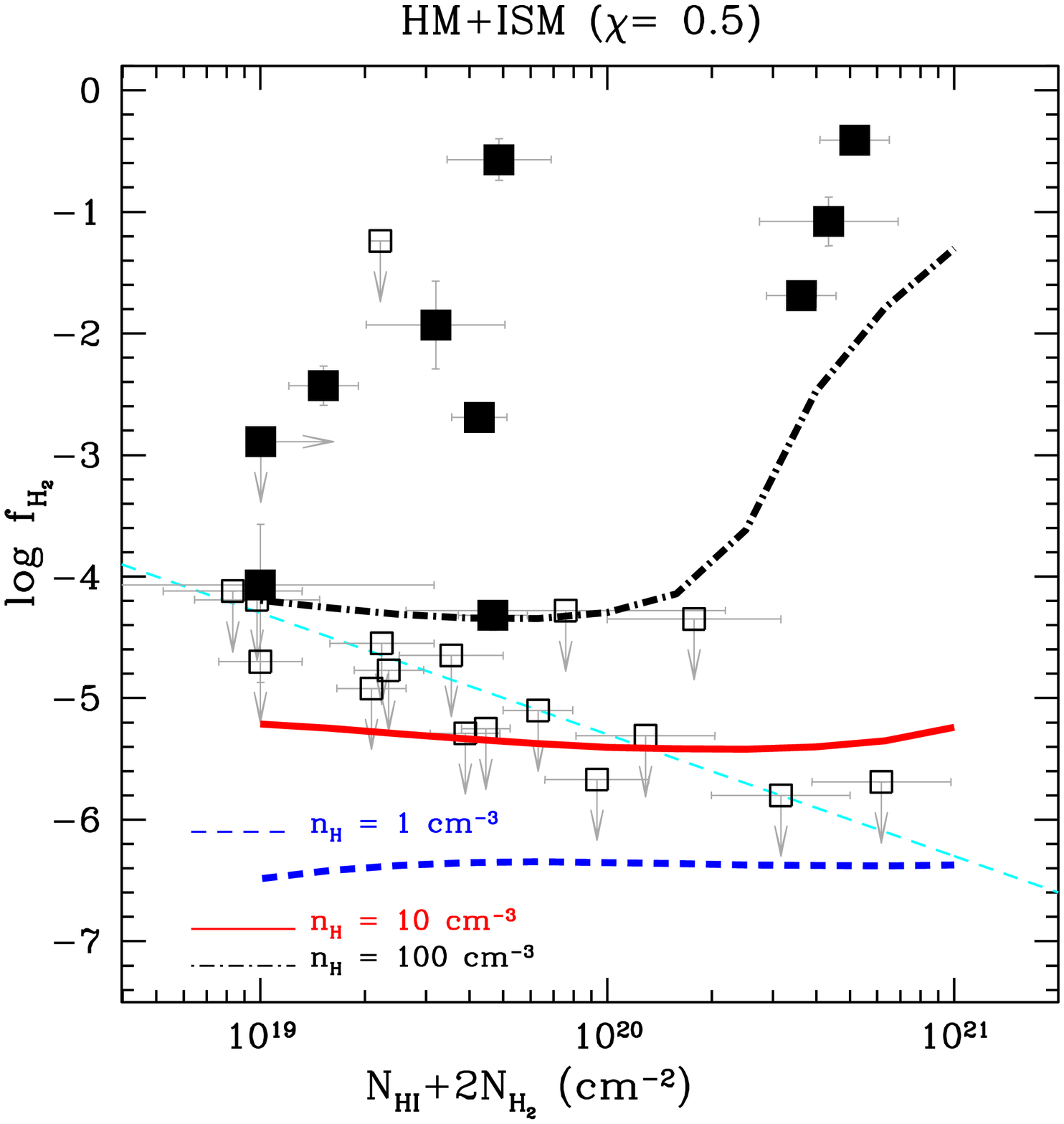}    
}}
}}
\caption{The molecular fraction against the total hydrogen (atomic+molecular) column density 
in our sample. The systems in which \HH\ is detected are plotted as solid squares. The dashed 
straight line corresponds to a $\log N(\HH) = 14.4$. Results of photoionization model predictions 
for three representative densities are shown by smooth curves in each panel. Models of three 
different panels assume three different ionizing radiation (see text). All models assume a 
metallicity of $\log Z/Z_{\odot} = -0.4$ and a dust-to-gas ratio of $\log \kappa = -1.0$.}    
\label{PImodel}    
\end{figure*}   

\subsection{Enhanced detection rate of \HH\ at low-$z$}   
 
We find that the detection rate of \HH\ absorption in our low-$z$ sample is generally higher compared to the high-$z$ sample of N08 for different $N(\HH)$ threshold values up to $\log N(\HH) = 16.5$. In our sample the \HH\ detection rate is found to be $50^{+25}_{-12}$ percent, for systems with $\log N(\HH)> 14.4$, detected in spectra that are sensitive down to $\log N(\HH) = 14.4$. The detection rate is a factor of $\gtrsim 2$ higher compared to the high-$z$ sample of N08, in which it is only $18^{+8}_{-4}$ percent for a similar $N(\HH)$ threshold. The occurrence of enhanced \HH\  detection rate at low-$z$ is unlikely due to sample bias as none of these spectra were obtained in order to study molecular hydrogen. We note that the program ID 12593 (PI: D. Nestor) was proposed to observe low-$z$ DLAs but no \HH\ is detected from that dataset. In addition, program ID 12536 (PI: V. Kulkarni) was proposed to observe low-$z$ sub-DLAs and one of these shows \HH\ absorption. But none of these proposals were focussed on \HH\ that was known beforehand.      

As discussed in the previous section, conditions are more favorable for a molecular gas phase at low-$z$ since the ambient radiation field intensity decreases with cosmic time. In fact, in two systems for which we have conducted detailed ionization modelling we infer a weak radiation field \citep[see e.g.][and Dutta et al. (submitted)]{Srianand14}. Besides, the cosmic mean metallicity of DLAs is known to increase with cosmic time \citep[]{Prochaska03Zz,Rafelski12}. From their best fitting metallicity versus redshift relationship, we expect the metallicity of the low-$z$ DLAs (at $z\lesssim 0.5$) to have $\sim0.5$ dex higher metallicity compared to the DLAs in the sample of N08. Moreover, sub-DLAs have higher metallicities than DLAs at any given epoch and show faster metallicity evolution \citep[e.g.][]{Kulkarni07,Som13}. From Fig.~11 of \citet{Som13} we notice that in the case of sub-DLAs at $z\lesssim0.5$, the enhancement in the $N(\HI)$-weighted mean metallicity could be as high as 0.7 dex compared to sub-DLAs at $z\simeq 2.5$. Both the above mentioned facts suggest that the probability of finding a metal-rich DLA and/or sub-DLA at low-$z$ should be higher than that at high-$z$. Note that the vast majority (22/27) of our low-$z$ systems are sub-DLAs and likely to be metal rich. It has been demonstrated that metal-rich DLAs/sub-DLAs are more likely to have molecules \citep[see e.g.][]{Petitjean06,Noterdaeme08}. Thus a combination of enhanced metallicity and reduced radiation field could possibly explain the higher detection rate of molecular hydrogen at low redshift. We plan to investigate this issue in detail in our future work. In passing, we note that (i) $N(\HI)$ and $N(\HH)$ do not show any significant correlation and (ii) the $N(\HI)$ distributions for systems with and without detected \HH\ are not significantly different. Therefore, the observed difference in the \HH\ detection rate is not possibly due to the difference in the $N(\HI)$ distribution between high- and low-$z$ samples.

\subsection{Overall physical conditions} 
\label{PImod}

To understand the overall physical conditions in these low-$z$ \HH\ systems we run grids of photoionization models using {\sc cloudy} \citep[v13.03, last described by][]{Ferland13}. The models predicted molecular fractions as a function of total hydrogen column density (atomic+molecular) are shown in three different panels of Fig.~\ref{PImodel}. In each panel, models are run for three different densities (i.e. $n_{\rm H} = $~1, 10, and 100~cm$^{-3}$), assuming a metallicity of $\log Z = -0.4$ (the median value for our sample, see column 12 of Table~\ref{H2tab}) and a dust-to-gas ratio of $\log \kappa = -1.0$ (a fiducial value). The ionizing radiation field for the models shown in the left panel is the extra-galactic UV background radiation computed by \citet{Haardt01} at $z = 0.2$. It is apparent from this panel that the model with $n_{\rm H} = $~1 cm$^{-3}$ cannot explain the majority of the systems where \HH\ is detected. The model with $n_{\rm H} = $~10 cm$^{-3}$ seems to be a better choice. Now when we add the mean UV radiation field as seen in the Galactic ISM \citep[]{Black87} but scaled by a factor of $\chi = 0.2$, the model with $n_{\rm H} = $~10 cm$^{-3}$ clearly does not work (see middle panel). A particle density of $n_{\rm H} = $~100 cm$^{-3}$ is preferred instead. If we further increase the contribution of the mean Galactic radiation field by increasing the scaling factor to a $\chi = 0.5$, even the model with $n_{\rm H} = $~100 cm$^{-3}$ fails to reproduce the observed molecular fractions (see right panel).   

At high redshift, the density of the \HH\ bearing components in DLAs found to be in the range 10--200 cm$^{-3}$ with a radiation field of the order of or slightly higher than the mean UV radiation field in the Galactic ISM \citep[i.e. $\chi \gtrsim 1.0$, see][]{Srianand05}. We note that the density of the low-$z$ \HH\ absorbers cannot be much higher than what is seen in the diffuse atomic phase of the ISM (i.e. 10 - 100 cm$^{-3}$), as we do not see HD/CO molecules or the excited fine structure lines of \CI/\CII\ from most of these systems. Weak HD absorption is detected only in one case \citep[see][]{Oliveira14}. If the low-$z$ \HH\ systems originate from a density range 10 -- 100 cm$^{-3}$ then the UV radiation field prevailing in these systems has to be much weaker compared to the Galactic mean field (i.e. $\chi \lesssim 0.5$) and/or the inferred radiation field in high-$z$ DLAs with detected \HH. Therefore, we speculate that the low-$z$ \HH\ systems are possibly not related to star-forming disks, but trace regions that are farther away from the luminous part of a galaxy. The large impact parameters of host-galaxy candidates, as discussed below, further promote such an idea.

\subsection{Connection to galaxies}   

\begin{figure} 
\centerline{\vbox{
\centerline{\hbox{ 
\includegraphics[height=6.4cm,width=8.6cm,angle=00]{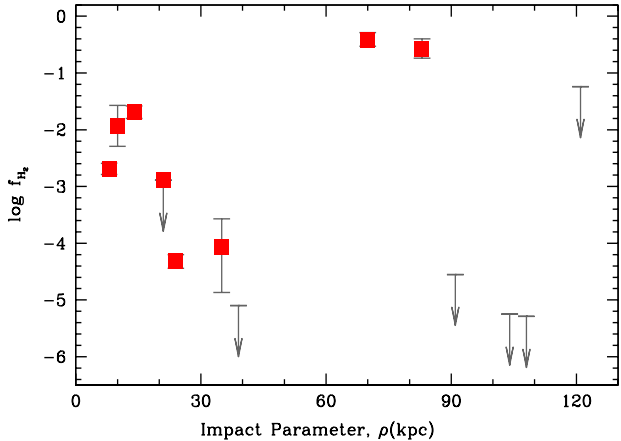}    
}}
}}
\caption{The molecular fractions against the host-galaxy impact parameters   
(see column 10 of Table~\ref{sample1}). The arrows without squares represent 
systems where \HH\ is not detected.}       
\label{fH2_rho}  
\end{figure}

For eight out of the ten low-$z$ \HH\ systems, we have host-galaxy information from the literature (see column 10 of Table~\ref{sample1}). When multiple possible host-galaxy candidates are reported for a given system, the true host is considered to be the galaxy with the lowest impact parameter. In Fig.~\ref{fH2_rho} we show the molecular fractions against the impact parameters of the nearest possible host-galaxy candidate for these systems. There are five more systems where \HH\ absorption is not detected but the impact parameters of the nearest possible host-galaxy candidates are available. These are shown  just by arrows in the figure. Note that all these five systems have $\rho > 30$~kpc. In this limited sample, there is no obvious correlation between the $f_{\hh}$ and the impact parameter. However, the \HH\ detection rate seems to be higher when the impact parameter is $\rho < 30$~kpc.

We notice that three (five) of the eight \HH\ systems show $\rho < 20$~kpc (30~kpc). These are roughly consistent with the median values of impact parameters for the DLA/sub-DLA host-galaxies found by \cite{Rao11}. Nonetheless, the presence of diffuse molecular gas at impact parameters of $>15$~kpc is intriguing. We therefore put forward a scenario where molecular hydrogen detected in DLAs/sub-DLAs is not related to star-forming disks but stems from halo gas. However, the density range of halo gas as suggested by the low ionization metal lines \citep[e.g.][]{Muzahid14,Werk14}, is too low for {\sl in situ} \HH\ production. We thus speculate that the \HH\ molecular gas is tidally stripped or ejected disk-material that retained a molecular phase partially due to self-shielding and the lower ambient UV radiation as suggested from simple photoionization models. The detections of molecular hydrogen in high velocity clouds \citep[]{Richter01} and in the leading arm of the Magellanic Stream \citep[]{Sembach01} reinforce such a scenario \citep[see also][]{Crighton13}. 

Radio observations of the \HI\ 21-cm line is another powerful tool for studying cold and neutral gas \citep[]{Gupta07,Gupta12} and diagnoses physical conditions in host-galaxies when detected \citep[]{Gupta13,Borthakur14}. The only system with $\rho < 10$~kpc (i.e. \zabs\ = 0.10115 towards Q~0439$-$433) shows a tentative detection of 21-cm absorption \citep{Kanekar01}. Even in this case the QSO sight line is well beyond the stellar disk of the galaxy \citep[]{Petitjean96b}. A detailed analysis of this system with follow up VLBA observations are presented in Dutta et al. (submitted).       

It is apparent from Fig.~\ref{fH2_rho} that the two systems showing the highest molecular fractions in our sample are detected at the largest impact parameters, i.e. $\rho \sim $~80~kpc. While \HH\ producing cold gas is expected at low impact parameters (e.g. $\rho \lesssim$~15 kpc), systems at $\rho \sim$~80~kpc with a molecular fraction of $\sim 1/10$ are extremely surprising. The impact parameters for these two systems come from \citet[towards B0120-28]{Oliveira14} and \citet[towards J0925+4004]{Werk13}. In addition to \HH, a detection of the HD molecule has been reported in the former system. We do not have the image of this field accessible to us to make any comment on the host-galaxy. However, we note that the molecular hydrogen is detected in four different components spread over $\sim$~180 \kms, implying a high volume filling factor of \HH\ clouds in this absorber. Naively, one would expect such a case when a line-of-sight is passing through the stellar (or \HI) disk of a galaxy, requiring a much lower impact parameter. In fact, for the two other systems in our sample showing multiple \HH\ absorption components (i.e. \zabs\ = 0.10115 towards Q~0439--433 and \zabs\ = 0.55733 towards Q~0107--0232) show impact parameters of $\sim10$~kpc. Therefore, this system is of particular interest for future deep observations. The later system is a part of the ``COS-Halos" sample \citep[]{Tumlinson11,Werk13}, that were observed to investigate the circum-galactic medium of isolated $L_{\ast}$ galaxies at $z\sim 0.2$. Nonetheless, we note that \citet[]{Werk12} have found two candidate galaxies in the QSO J~0925+4004 field at the redshift of the absorber with impact parameters of 81 and 92 kpc respectively. This possibly could mean that the \HH\ absorption is originating from a group environment and perhaps is not related to a  halo of a single galaxy. We also note that two very faint photometric objects at $\rho \lesssim 20$~kpc, at the absorber's redshift, are seen in the SDSS image. Therefore, we can not rule out the possibility of one these objects being the true host-galaxy candidate. Deep images and determination of spectroscopic redshifts of all the photometric objects in this field are essential for understanding the true origin of molecular hydrogen in this system.

\section{Summary}  
\label{sec:summ} 

We have conducted a systematic search for molecular hydrogen (\HH) in low redshift ($z<$~0.7) DLAs and sub-DLAs using the medium resolution $HST/$COS spectra that were available in the public $HST$ archive before March, 2014. This is the first-ever systematic search for \HH\ below the atmospheric cutoff. In total we found 33 DLAs/sub-DLAs with $\log N(\HI) \gtrsim 19$, of which \HH\ information is available for 27 systems. \HH\ absorption from different rotational levels (up to $J=3$) is seen in a total of 10/27 system with a $N(\HH)$ of $>10^{14.4}$~cm$^{-2}$, 3/5 in DLAs and 7/22 in sub-DLAs. Three \HH\ systems were reported previously by \citet[]{Crighton13}, \citet[]{Oliveira14}, and \citet[]{Srianand14}. The main findings of our analyses are summarized below:   

\begin{itemize}  

\item The \HH\ detection rate, for $N(\HH) > 10^{14.4}$~cm$^{-2}$, in low-$z$ DLAs/sub-DLAs is $50^{+25}_{-12}$ percent. This is a factor of \linebreak $\gtrsim 2$ higher than that (i.e. $18^{+8}_{-4}$ percent) of the high-$z$ sample of \citet{Noterdaeme08}. We argue that the increase of the cosmic mean metallicity of DLAs/sub-DLAs and the dimming of the ambient radiation field intensity due to the decrease of the cosmic star-formation rate density, are responsible for the enhanced detection rate of molecular hydrogen at low-$z$.

\item  The median value of $N(\HI)$ for our sample is $10^{19.6}$~cm$^{-2}$. This is a factor of 10 lower than that of the N08 sample. Importantly, even with systematically lower $N(\HI)$ values, most of the low-$z$ \HH\ systems show a considerably large molecular fraction (i.e. $\log f_{\hh} \gtrsim -2.0$). This implies that either (a) the density is extremely high or (b) the radiation field is fairly weak in these \HH\ absorbers. However, the absence of HD (and/or CO) molecules and the \CI$^{\ast}$, \CII$^{\ast}$ absorption lines in the majority of the systems suggests that the density cannot be much higher than 10--100 cm$^{-3}$, as seen in the diffuse atomic phase of the Galactic ISM.

\item Using simple photoionization models we show that the prevailing radiation field in the low-$z$ \HH\ systems is much weaker than the mean Galactic UV field for particle density in the range 10--100 cm$^{-3}$. The absence of higher-order rotational lines (i.e. $J\geqslant4$) in all our \HH\ systems provides independent evidence for the radiation field being weak. This indeed suggests that the \HH\ bearing gas must be located far away from the star-forming luminous disk of the host galaxy.

\item Eight out of 10 \HH\ systems in our sample have host-galaxy information available from the literature. The majority of them (7/8) show impact parameters $\rho > 10$~kpc, suggesting that they perhaps do not originate from the luminous disks of the host-galaxies. We notice that even for the lowest impact parameter system (i.e. $\rho =$~7.6~kpc for the \zabs = 0.10115 towards Q~0439--433), the line of sight is outside the luminous stellar disk. Nonetheless, the possibility that these \HH\ absorbers are actually originating in faint dwarf galaxies at much lower impact parameters, cannot be ruled out. Therefore, deep photometric searches for sub-$L_{\ast}$ galaxies close to QSO lines of sight and follow up spectroscopic observations are indispensable for understanding the possible origin(s) of \HH\ absorption in low-$z$ DLAs/sub-DLAs.

\item Low-$z$ \HH\ systems occupy a unique region in the $f_{\hh} - N_{\rm H}$ plane (i.e. $\log f_{\hh} \gtrsim -4.5$ and $\log N(\HH) \lesssim 20.7$). Only a handful of \HH\ systems from the high-$z$ and the Galactic disk/Magellanic Cloud samples fall in this range of parameter-space. Interestingly, a significant fraction (i.e. $25/40$) of the Galactic halo systems, with detected \HH, show similar range in $f_{\hh}$ and $N_{\rm H}$. This indicates that the physical conditions of molecular gas in low-z DLAs/sub-DLAs are, presumably, similar to those of the Milky Way halo gas.

\item The $f_{\hh}$ distribution for our low-$z$ sample is significantly different compared to those of the Galactic disk/halo and the Magellanic Clouds (LMC/SMC) samples. Only a mild difference is noticed between the $f_{\hh}$ distributions at low- and high-$z$ for the entire sample. However, no difference is suggested when we consider only systems with $\log f_{\hh} > -4.9$. The mild difference inferred for the entire sample is the manifestation of the fact that the \HH\ detection rate is considerably higher at low-$z$ than at high-$z$.

\item For the \HH\ components with a total $N(\HH) >10^{16.5}$~cm$^{-2}$, the median rotational excitation temperature is found to be $T_{01} = 133\pm55$~K. This is consistent with what has been seen in the ``high-$z$ DLAs" (e.g. 143$\pm$19~K). The inferred $T_{01}$ are, however, slightly higher than those derived for the Galactic disk \citep[i.e. 77$\pm$17 (rms)~K,][]{Savage77}, the Galactic halo \citep[i.e. 115$\pm$13~K,][]{Gillmon06}, and the Magellanic Clouds \citep[i.e. 82$\pm$21 (rms)~K,][]{Tumlinson02} systems.     

\item  For further insight into the origin and nature of low-$z$ DLAs/sub-DLAs, analyses based on observed metal lines and detailed ionization models will be presented elsewhere.

\end{itemize}  

This work is based on observations made with the NASA/ESA Hubble Space Telescope, obtained from the data archive at the Space Telescope Science Institute, which is operated by the Association of Universities for Research in Astronomy, Inc., under NASA contract NAS 5-26555. We thank Patrick Petitjean and the anonymous reviewer for constructive comments. SM thankfully acknowledge Dr. Eric Feigelson for useful discussions on various statistical tests, survival analysis in particular.

\def\aj{AJ}%
\def\actaa{Acta Astron.}%
\def\araa{ARA\&A}%
\def\apj{ApJ}%
\def\apjl{ApJ}%
\def\apjs{ApJS}%
\def\ao{Appl.~Opt.}%
\def\apss{Ap\&SS}%
\def\aap{A\&A}%
\def\aapr{A\&A~Rev.}%
\def\aaps{A\&AS}%
\def\azh{AZh}%
\def\baas{BAAS}%
\def\bac{Bull. astr. Inst. Czechosl.}%
\def\caa{Chinese Astron. Astrophys.}%
\def\cjaa{Chinese J. Astron. Astrophys.}%
\def\icarus{Icarus}%
\def\jcap{J. Cosmology Astropart. Phys.}%
\def\jrasc{JRASC}%
\def\mnras{MNRAS}%
\def\memras{MmRAS}%
\def\na{New A}%
\def\nar{New A Rev.}%
\def\pasa{PASA}%
\def\pra{Phys.~Rev.~A}%
\def\prb{Phys.~Rev.~B}%
\def\prc{Phys.~Rev.~C}%
\def\prd{Phys.~Rev.~D}%
\def\pre{Phys.~Rev.~E}%
\def\prl{Phys.~Rev.~Lett.}%
\def\pasp{PASP}%
\def\pasj{PASJ}%
\def\qjras{QJRAS}%
\def\rmxaa{Rev. Mexicana Astron. Astrofis.}%
\def\skytel{S\&T}%
\def\solphys{Sol.~Phys.}%
\def\sovast{Soviet~Ast.}%
\def\ssr{Space~Sci.~Rev.}%
\def\zap{ZAp}%
\def\nat{Nature}%
\def\iaucirc{IAU~Circ.}%
\def\aplett{Astrophys.~Lett.}%
\def\apspr{Astrophys.~Space~Phys.~Res.}%
\def\bain{Bull.~Astron.~Inst.~Netherlands}%
\def\fcp{Fund.~Cosmic~Phys.}%
\def\gca{Geochim.~Cosmochim.~Acta}%
\def\grl{Geophys.~Res.~Lett.}%
\def\jcp{J.~Chem.~Phys.}%
\def\jgr{J.~Geophys.~Res.}%
\def\jqsrt{J.~Quant.~Spec.~Radiat.~Transf.}%
\def\memsai{Mem.~Soc.~Astron.~Italiana}%
\def\nphysa{Nucl.~Phys.~A}%
\def\physrep{Phys.~Rep.}%
\def\physscr{Phys.~Scr}%
\def\planss{Planet.~Space~Sci.}%
\def\procspie{Proc.~SPIE}%
\let\astap=\aap
\let\apjlett=\apjl
\let\apjsupp=\apjs
\let\applopt=\ao
\bibliographystyle{mn}
\bibliography{mybib}
\newpage 
\appendix 

\begin{figure*} 
\section{Voigt profile fit to \HI\ absorption lines for systems listed in Table~1}   
\label{HIfits1}     
\centerline{
\vbox{
\centerline{\hbox{ 
\includegraphics[height=6.0cm,width=12.0cm,angle=00]{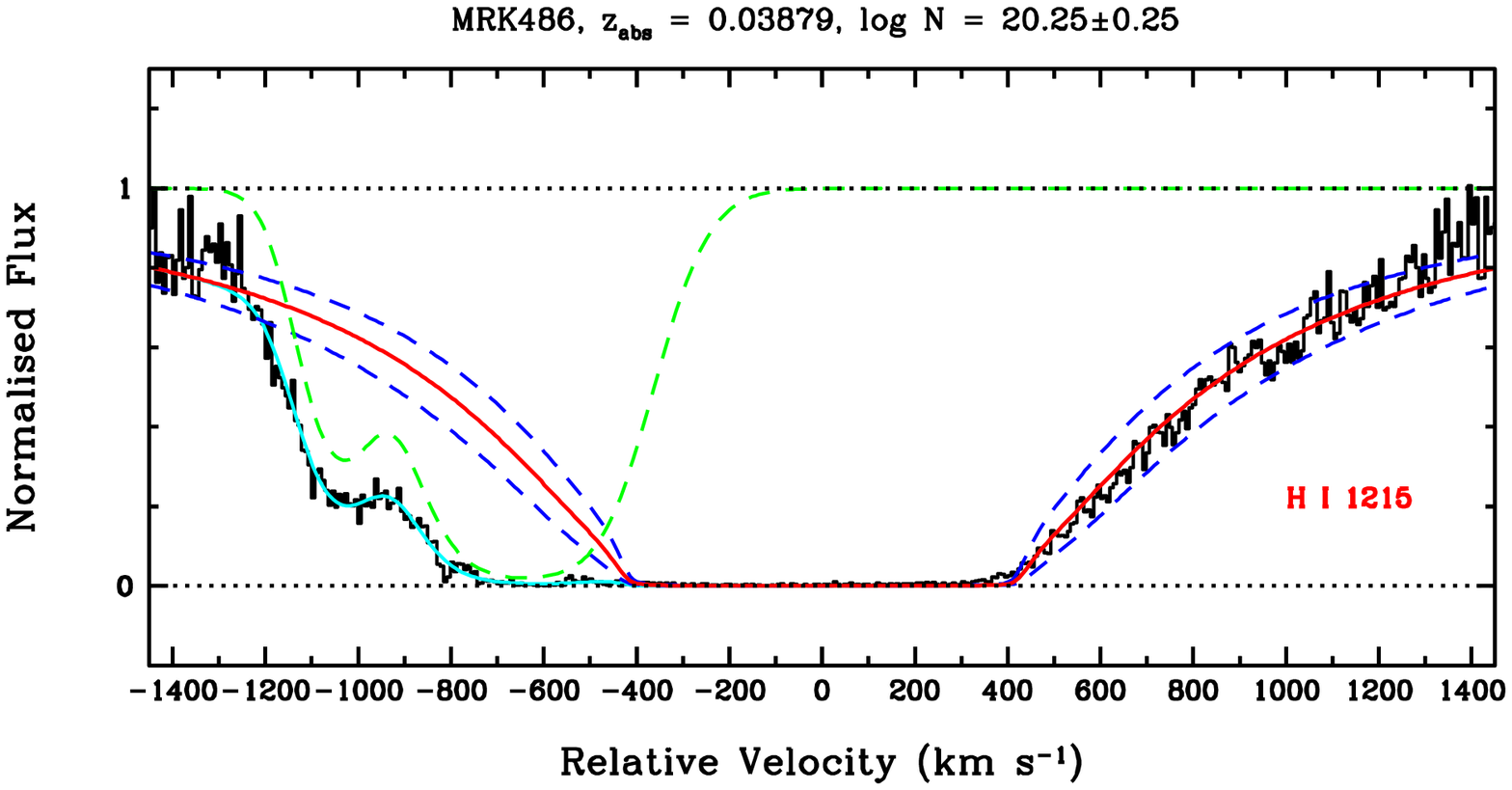}    
}}
}}
\caption{Damped \lya\ absorption from \zabs\ = 0.03879 towards MRK486 system. The red smooth and  
blue dashed curves plotted on the data (black histogram) are the best fitting Voigt profiles for 
the sub-DLA component and its $1\sigma$ uncertainty. Unrelated absorption is usually fitted 
assuming \lya. The green dashed curve at $\sim800$~\kms\ shows an unrelated absorption. Cyan 
curve represents the total (sub-DLA + unrelated) absorption model.}        
\label{HI_MRK486_03879} 
\end{figure*} 

\begin{figure*} 
\centerline{
\vbox{
\centerline{\hbox{ 
\includegraphics[height=6.0cm,width=12.0cm,angle=00]{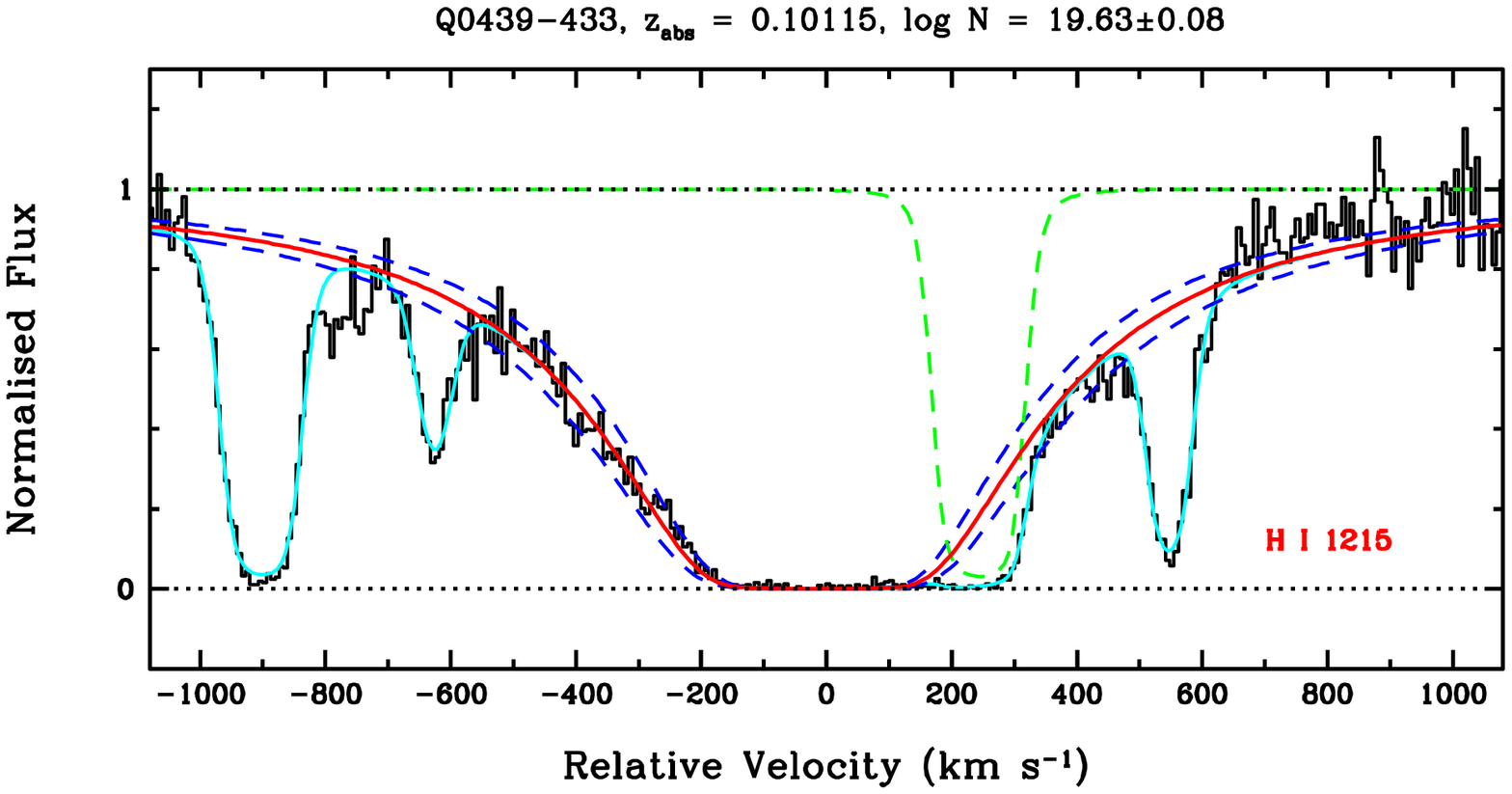}    
}}
}}
\caption{Sub-damped \lya\ absorption from \zabs\ = 0.10115 towards Q~0439-433 system. 
Various curves are as described in Fig.~\ref{HI_MRK486_03879}.}    
\label{HI_Q0439_10110} 
\end{figure*} 

\begin{figure*} 
\centerline{
\vbox{
\centerline{\hbox{ 
\includegraphics[height=6.0cm,width=12.0cm,angle=00]{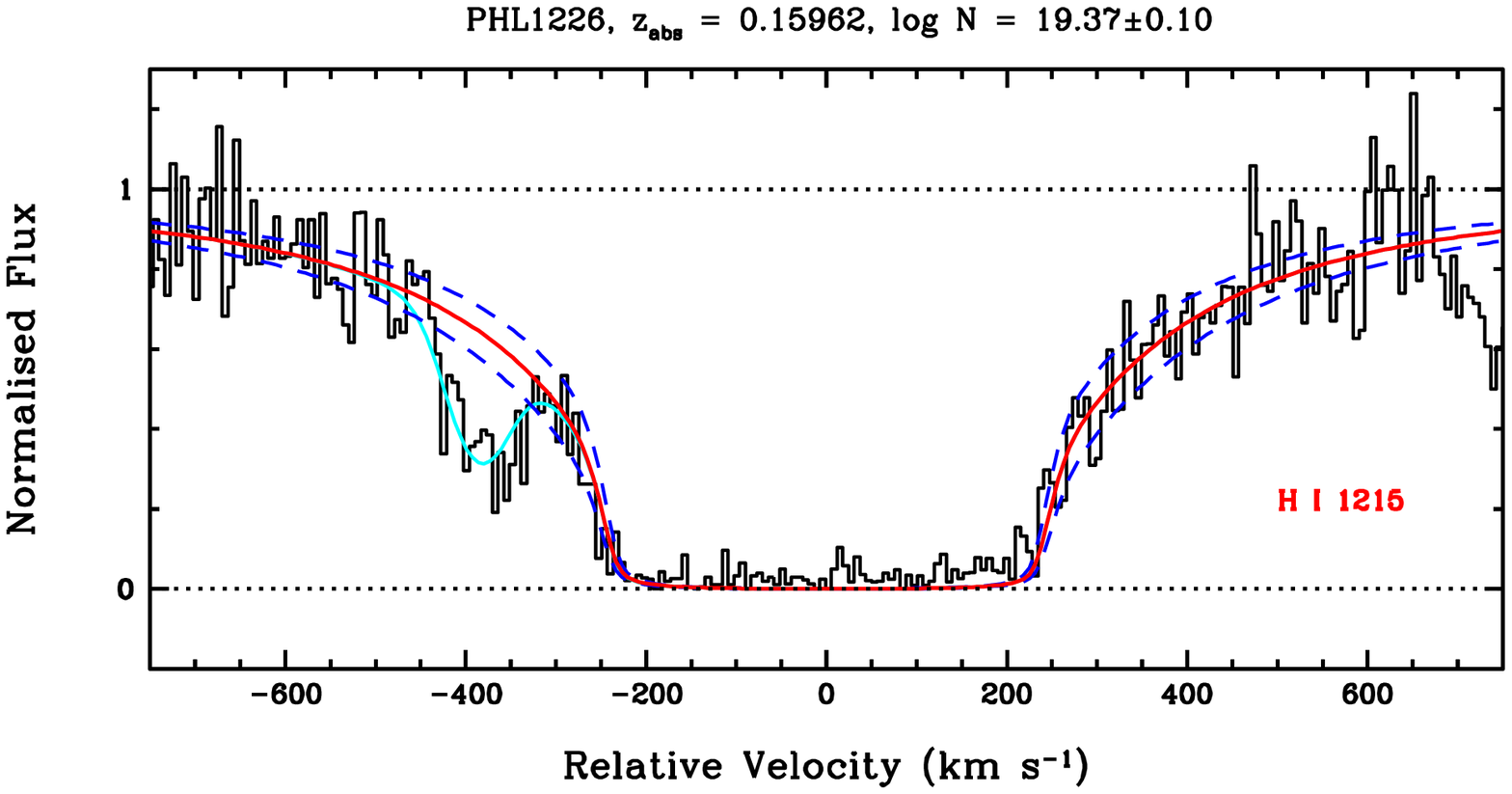}    
}}
}}
\caption{Sub-damped \lya\ absorption from \zabs\ = 0.15962 towards PHL1226 system. 
Various curves are as described in Fig.~\ref{HI_MRK486_03879}.}     
\label{HI_PHL1226_15962} 
\end{figure*} 

\begin{figure*} 
\centerline{
\vbox{
\centerline{\hbox{ 
\includegraphics[height=12.0cm,width=12.0cm,angle=00]{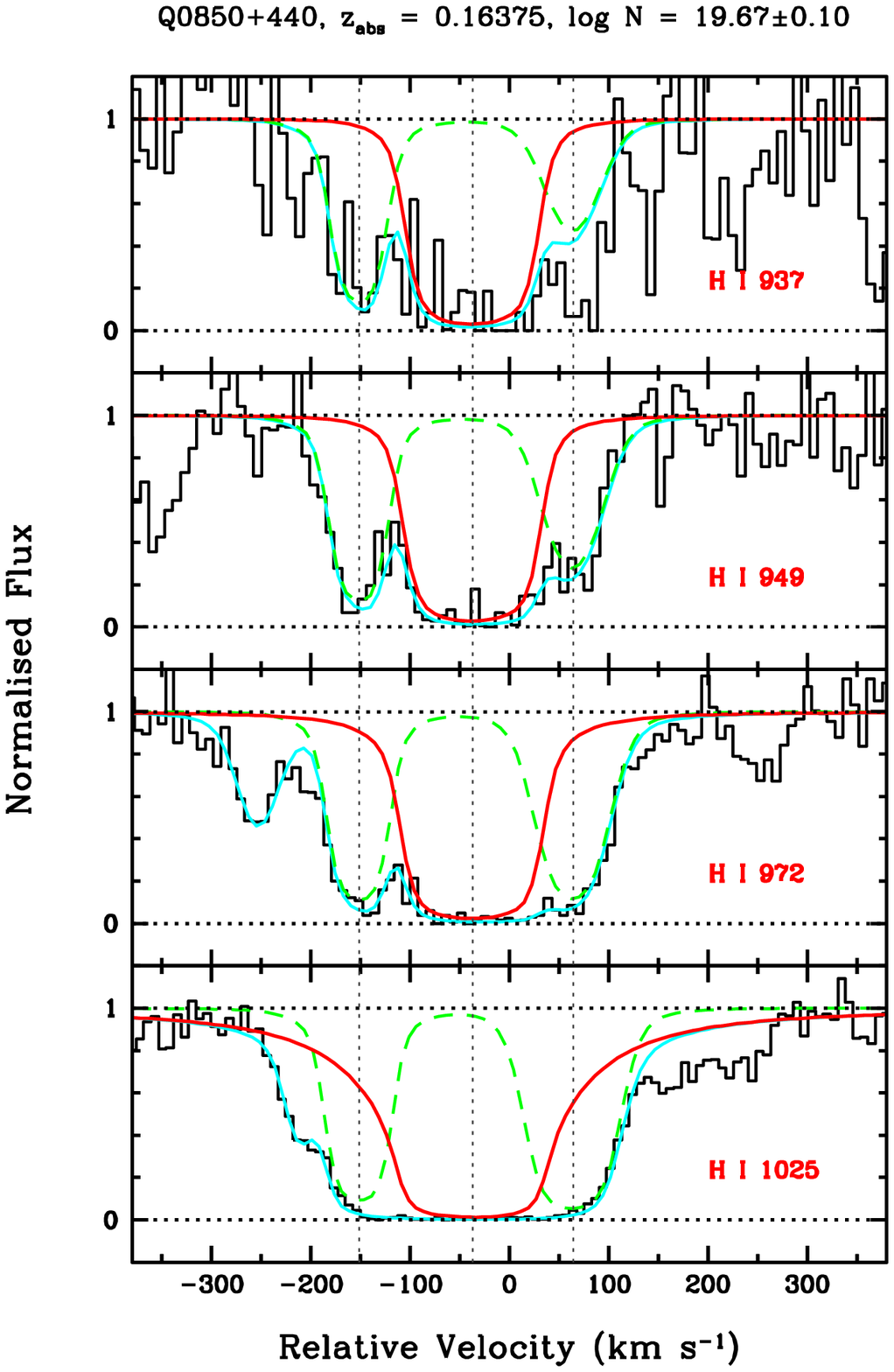} 
}}
}}
\caption{Lyman series absorption from \zabs\ = 0.16375 towards Q~0850+440 system. 
Various curves are as described in Fig.~\ref{HI_MRK486_03879}.} 
\label{HI_Q0850_16375} 
\end{figure*} 

\begin{figure*} 
\centerline{
\vbox{
\centerline{\hbox{ 
\includegraphics[height=6.0cm,width=12.0cm,angle=00]{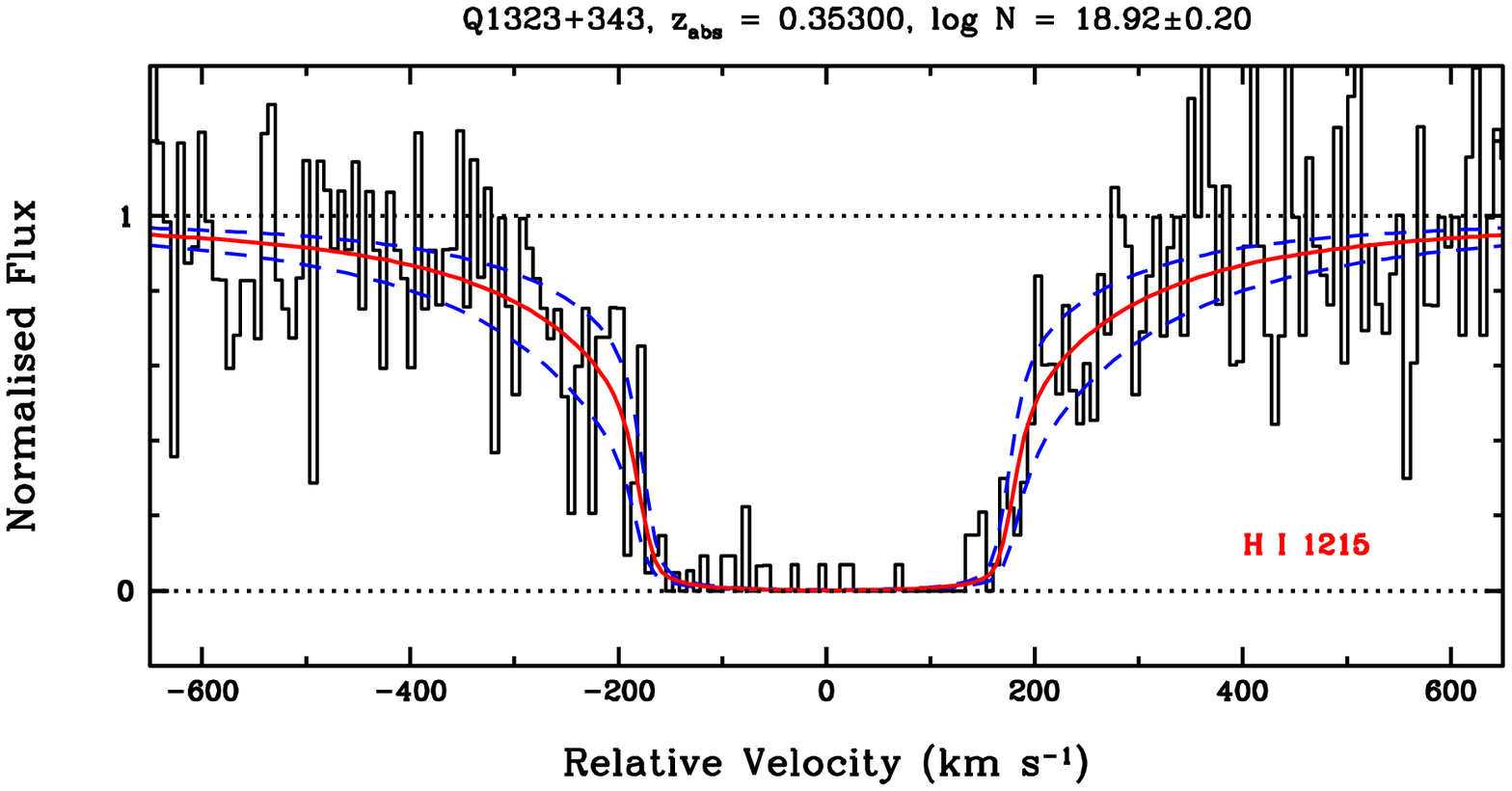} 
}}
}}
\caption{Sub-damped \lya\ absorption from \zabs\ = 0.35300 towards Q1323+343 system. 
Various curves are as described in Fig.~\ref{HI_MRK486_03879}.}   
\label{HI_Q1323_35300} 
\end{figure*} 

\begin{figure*} 
\centerline{
\vbox{
\centerline{\hbox{ 
\includegraphics[height=6.0cm,width=12.0cm,angle=00]{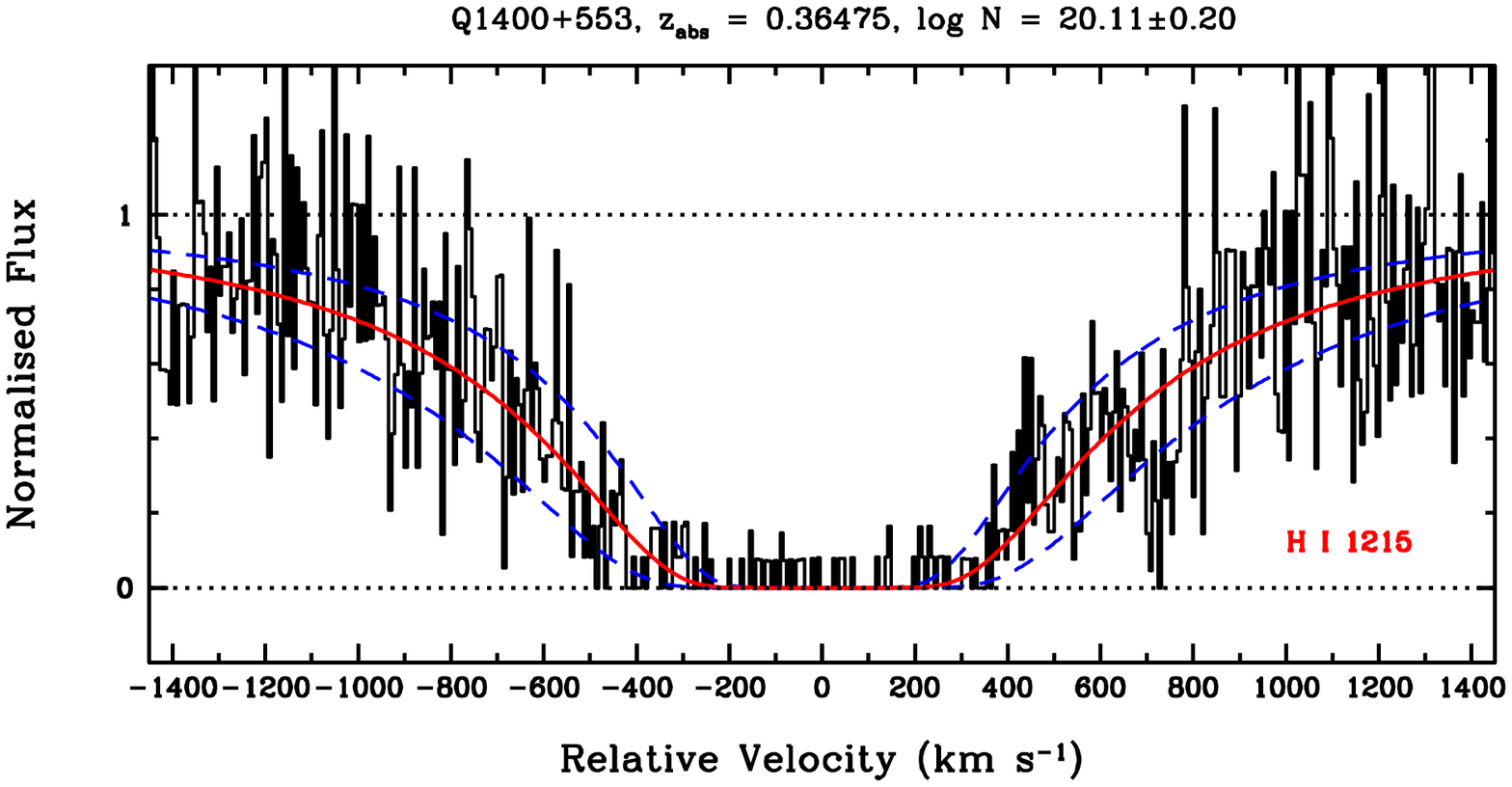} 
}}
}}
\caption{Damped \lya\ absorption from \zabs\ = 0.36475 towards Q1400+553 system. 
Various curves are as described in Fig.~\ref{HI_MRK486_03879}.}   
\label{HI_Q1400_36475} 
\end{figure*} 

\begin{figure*} 
\centerline{
\vbox{
\centerline{\hbox{ 
\includegraphics[height=12.0cm,width=12.0cm,angle=00]{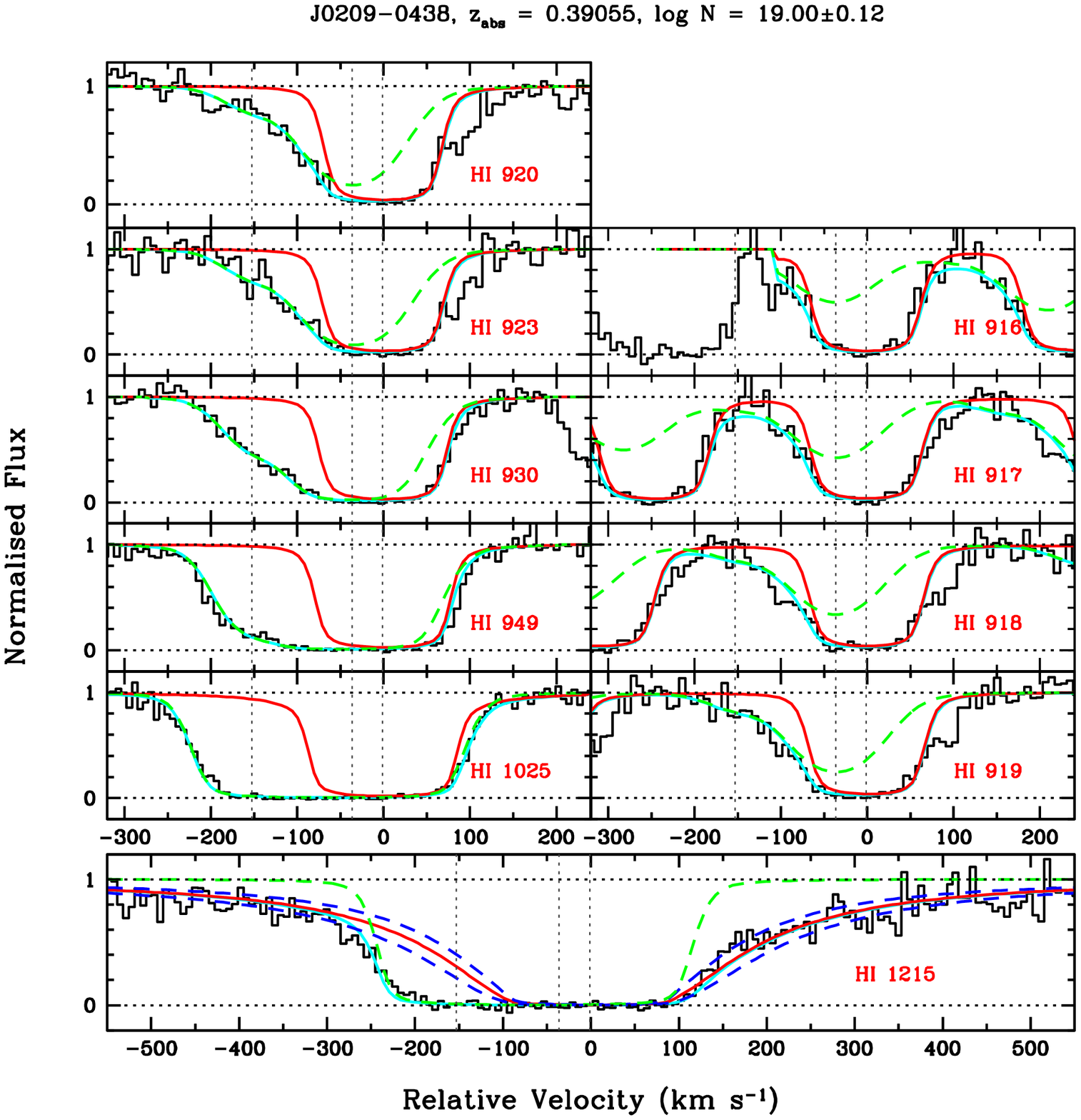}    
}}
}}
\caption{Lyman series absorption from \zabs\ = 0.39055 towards J0209-0438 system.
Various curves are as described in Fig.~\ref{HI_MRK486_03879}.}      
\label{HI_J0209_39055} 
\end{figure*} 

\begin{figure*} 
\centerline{
\vbox{
\centerline{\hbox{ 
\includegraphics[height=11.0cm,width=11.0cm,angle=00]{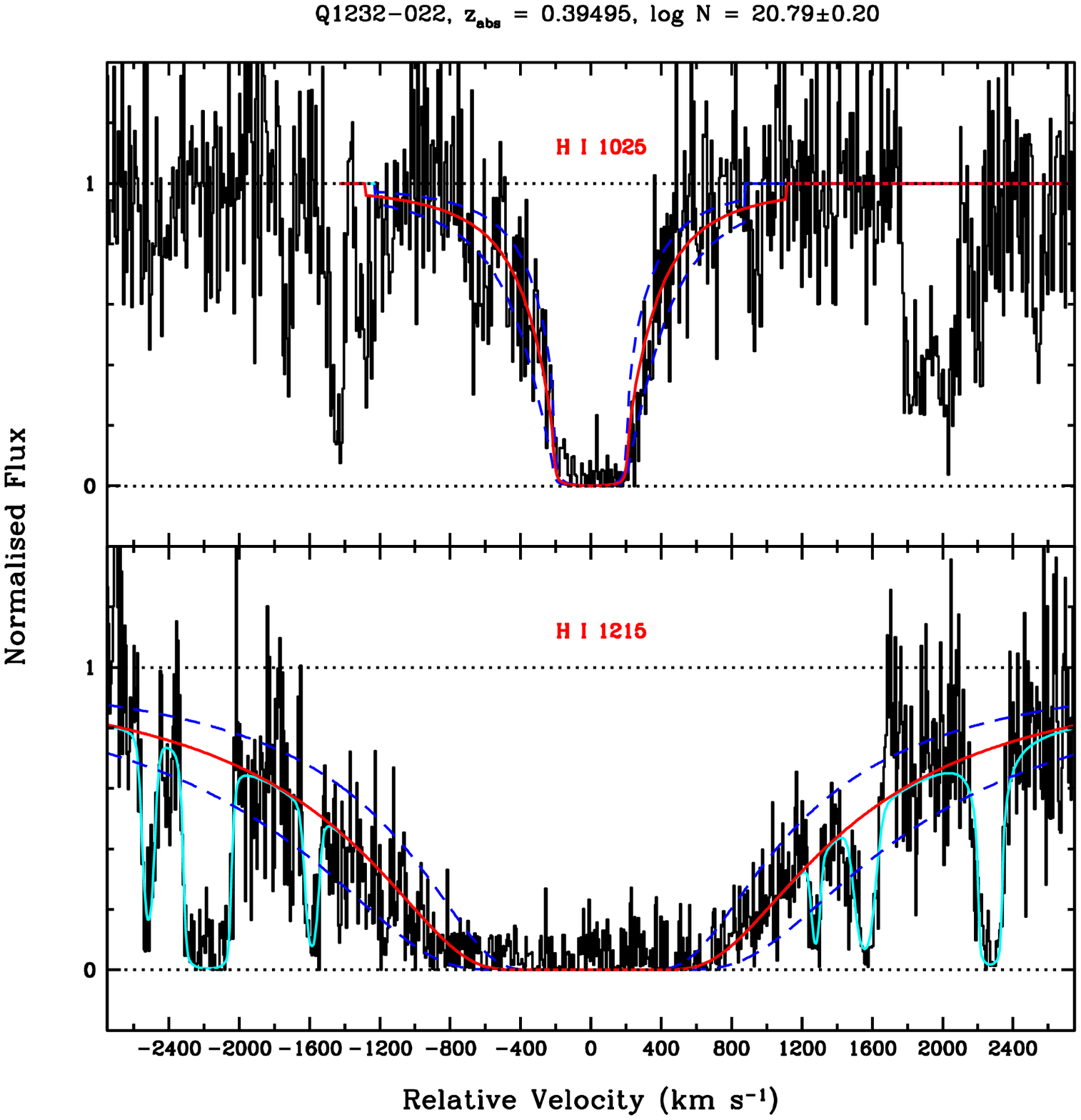} 
}}
}}
\vskip-0.5cm 
\caption{Damped \lya\ and \lyb\ absorption from \zabs\ = 0.39495 towards Q1232$-$022 system. 
Various curves are as described in Fig.~\ref{HI_MRK486_03879}.}    
\label{HI_Q1232_39495} 
\end{figure*} 

\begin{figure*} 
\centerline{
\vbox{
\centerline{\hbox{ 
\includegraphics[height=11.0cm,width=11.0cm,angle=00]{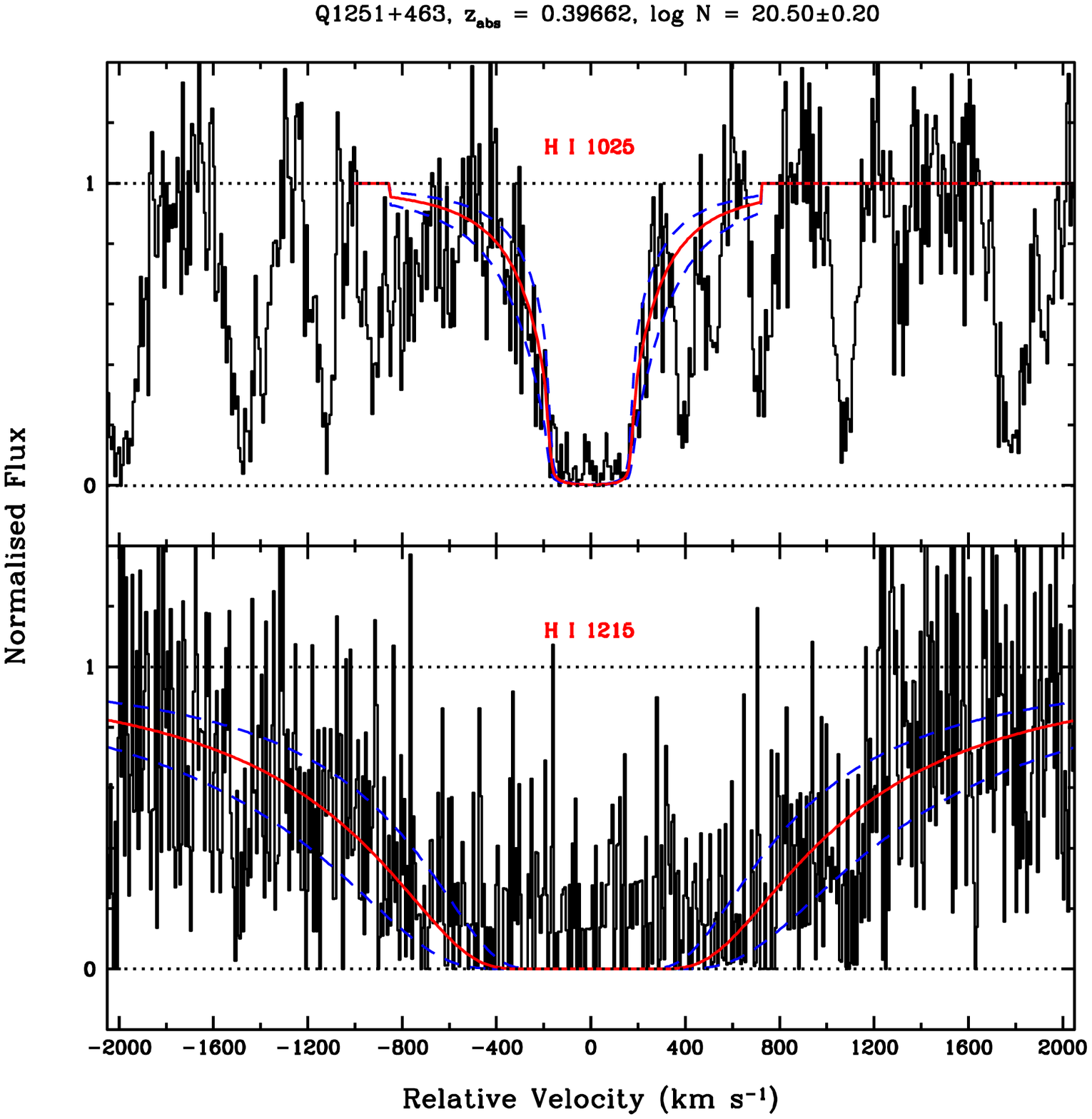} 
}}
}}
\vskip-0.5cm 
\caption{Damped \lya\ and \lyb\ absorption from \zabs\ = 0.39662 towards Q1251+463 system. 
Various curves are as described in Fig.~\ref{HI_MRK486_03879}.}   
\label{HI_Q1251_0.39662} 
\end{figure*} 

\begin{figure*} 
\centerline{
\vbox{
\centerline{\hbox{ 
\includegraphics[height=12.0cm,width=12.0cm,angle=00]{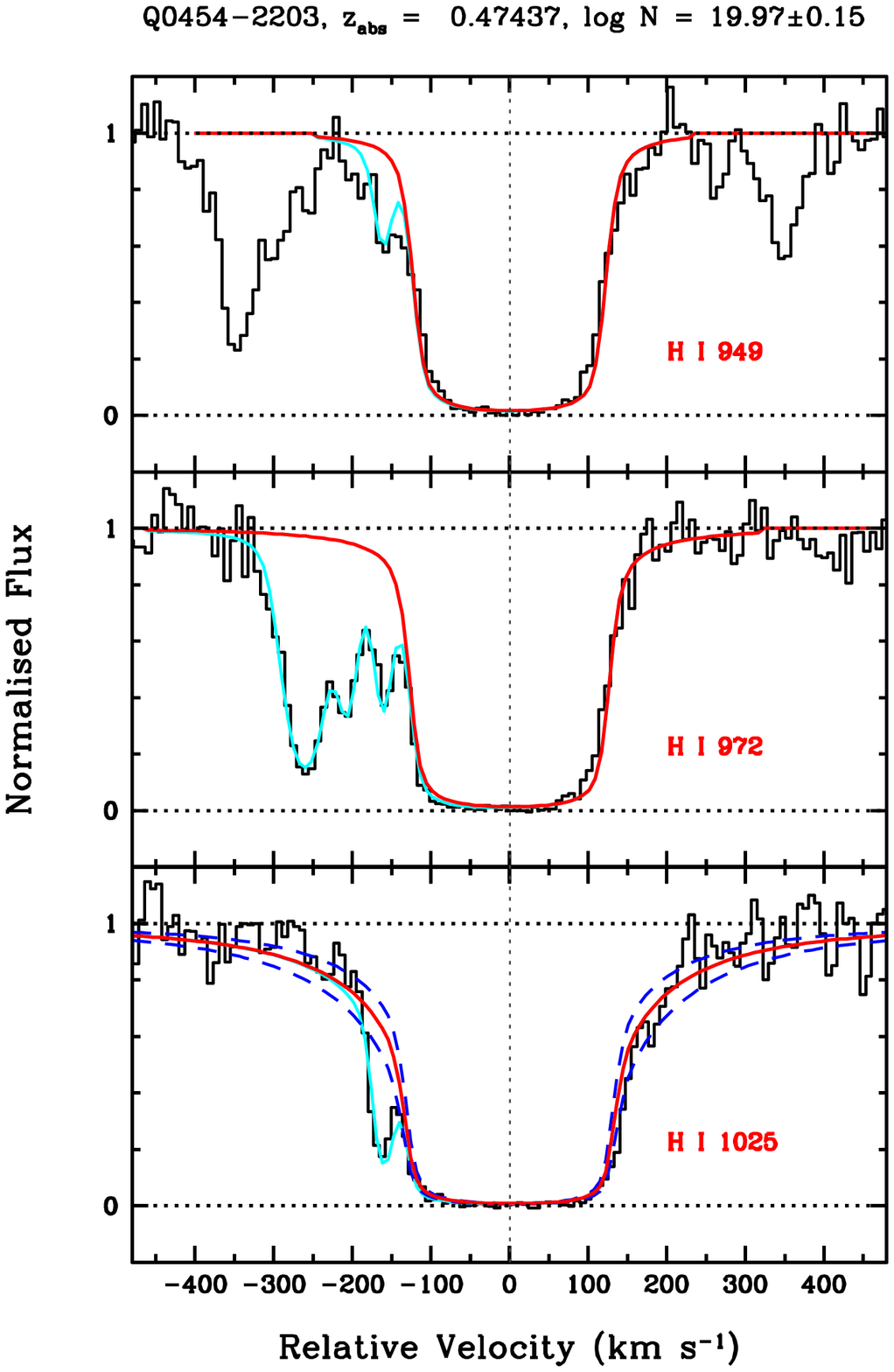}  
}}
}}
\caption{Lyman series absorption from \zabs\ = 0.47437 towards Q0454-2203 system. 
Various curves are as described in Fig.~\ref{HI_MRK486_03879}.}   
\label{HI_Q0454_47437} 
\end{figure*} 

\begin{figure*} 
\centerline{
\vbox{
\centerline{\hbox{ 
\includegraphics[height=12.0cm,width=12.0cm,angle=00]{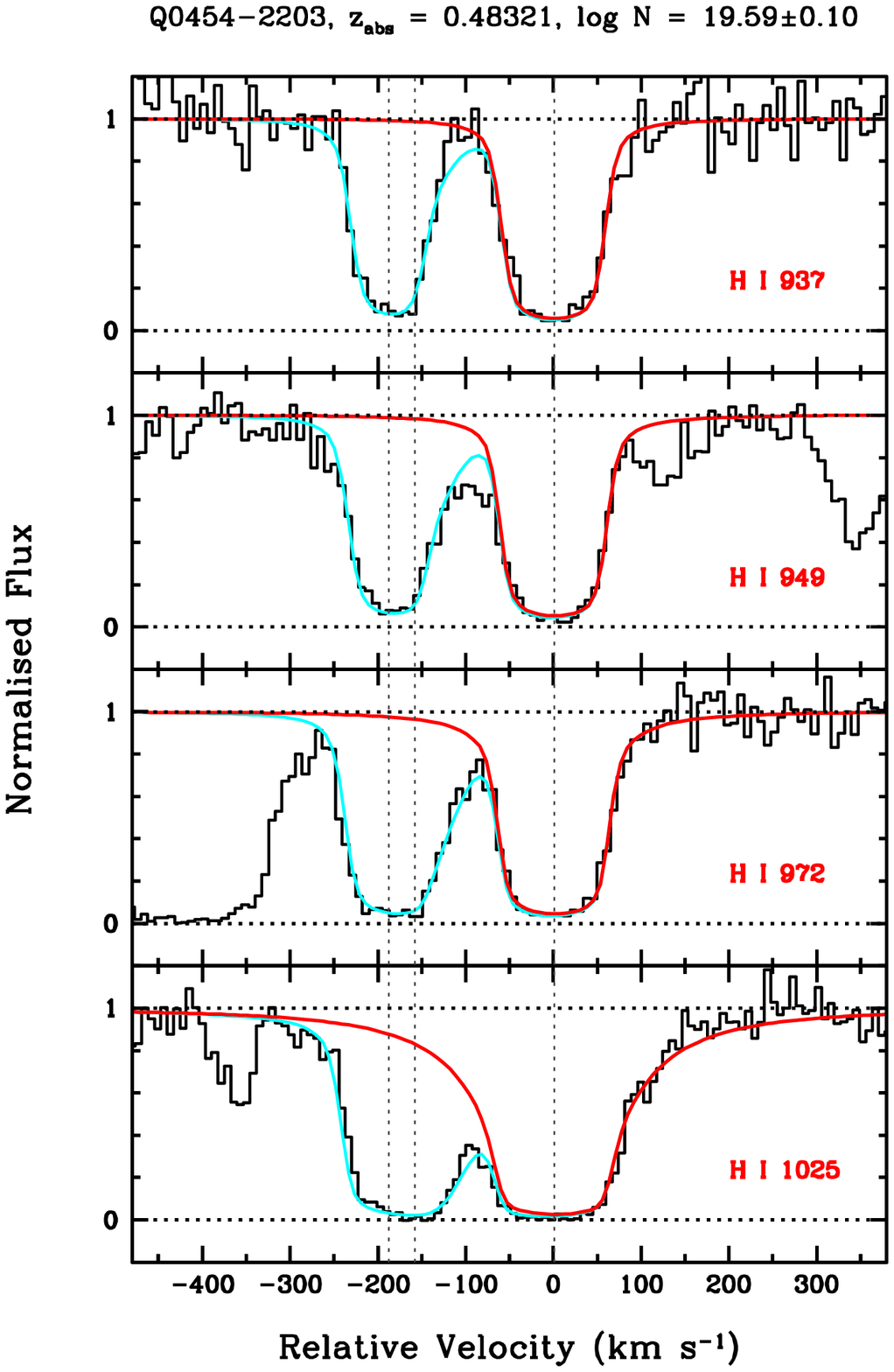} 
}}
}}
\caption{Lyman series absorption from \zabs\ = 0.48321 towards Q0454-2203 system. 
Various curves are as described in Fig.~\ref{HI_MRK486_03879}.}   
\label{HI_Q0454_48321} 
\end{figure*} 

\begin{figure*} 
\centerline{
\vbox{
\centerline{\hbox{ 
\includegraphics[height=12.0cm,width=12.0cm,angle=00]{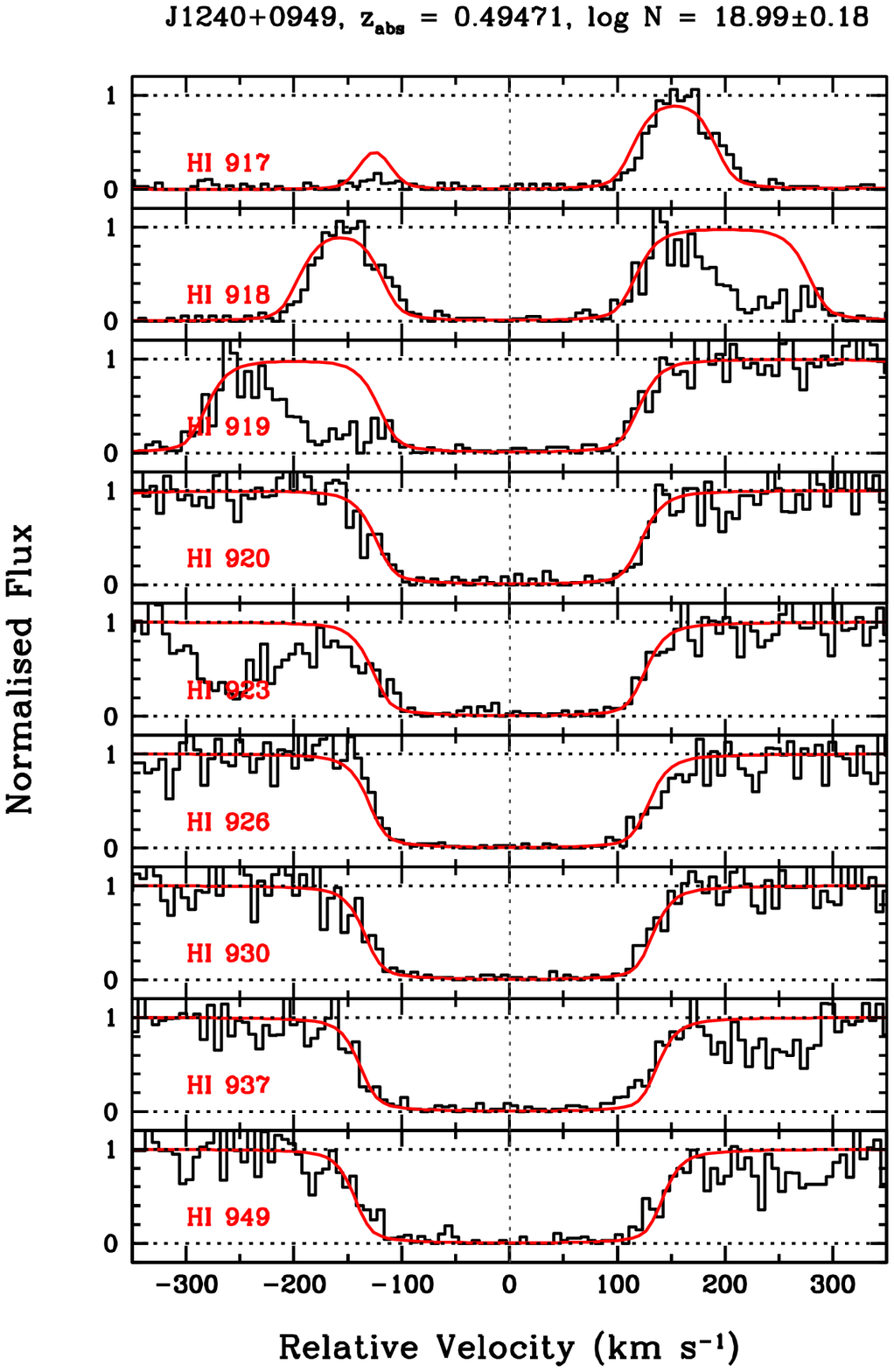}    
}}
}}
\caption{Lyman series absorption from \zabs\ = 0.49471 towards J1240+0949 system. 
Various curves are as described in Fig.~\ref{HI_MRK486_03879}.}     
\label{HI_J1240_49471} 
\end{figure*} 

\begin{figure*} 
\centerline{
\vbox{
\centerline{\hbox{ 
\includegraphics[height=12.0cm,width=12.0cm,angle=00]{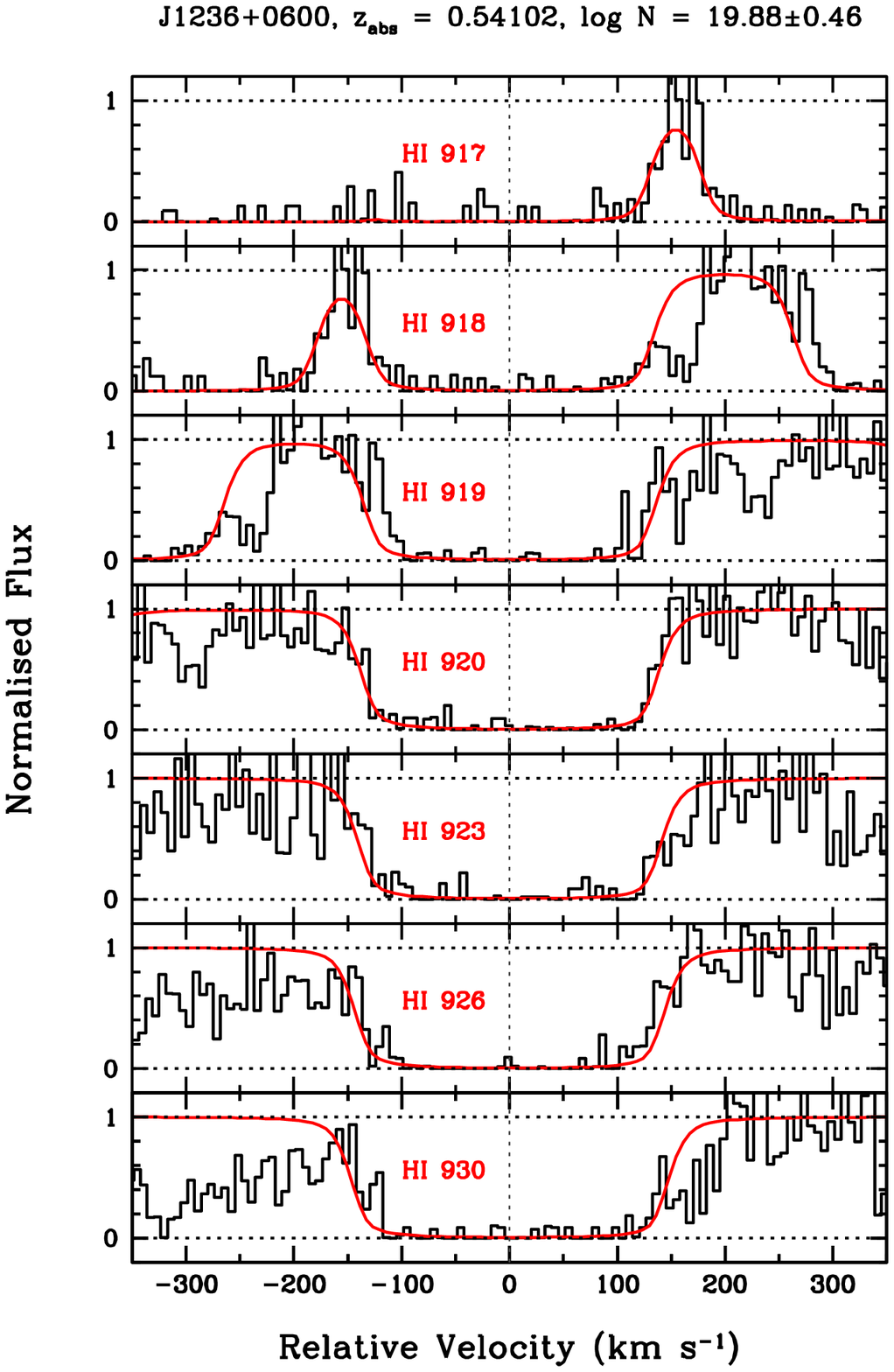}    
}}
}}
\caption{Lyman series absorption from \zabs\ = 0.54102 towards J1236+0600 system. 
Various curves are as described in Fig.~\ref{HI_MRK486_03879}.}    
\label{HI_J1236_54102} 
\end{figure*} 

\begin{figure*} 
\centerline{
\vbox{
\centerline{\hbox{ 
\includegraphics[height=12.0cm,width=12.0cm,angle=00]{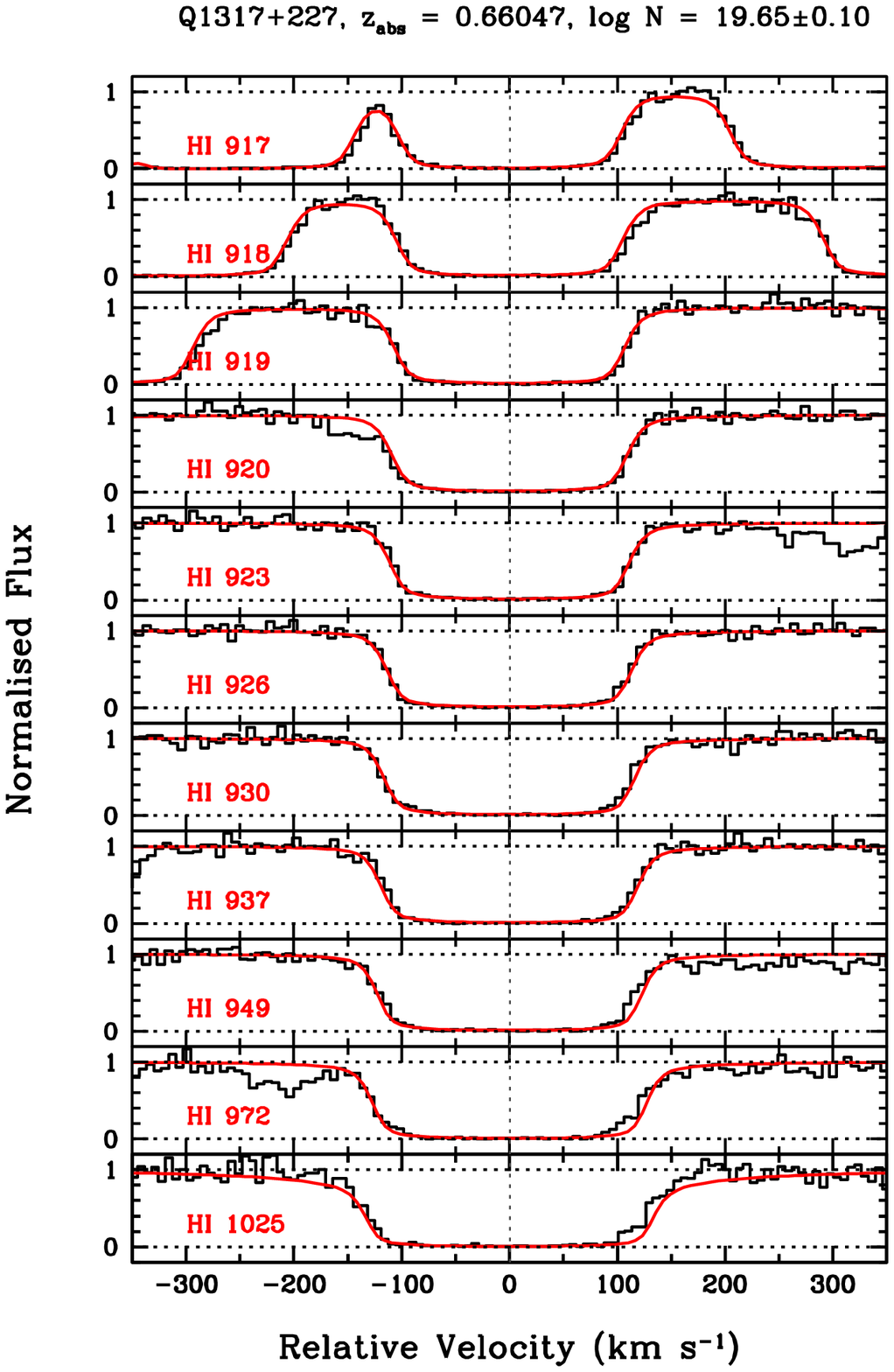}  
}}
}}
\caption{Lyman series absorption from \zabs\ = 0.66047 towards Q1317+227 system. 
Various curves are as described in Fig.~\ref{HI_MRK486_03879}.}    
\label{HI_Q1317_66047} 
\end{figure*} 

\begin{figure*} 
\section{Voigt profile fit to \HI\ absorption lines for systems listed in Table~2}    
\label{HIfits2}     
\centerline{
\vbox{
\centerline{\hbox{ 
\includegraphics[height=6.0cm,width=12.0cm,angle=00]{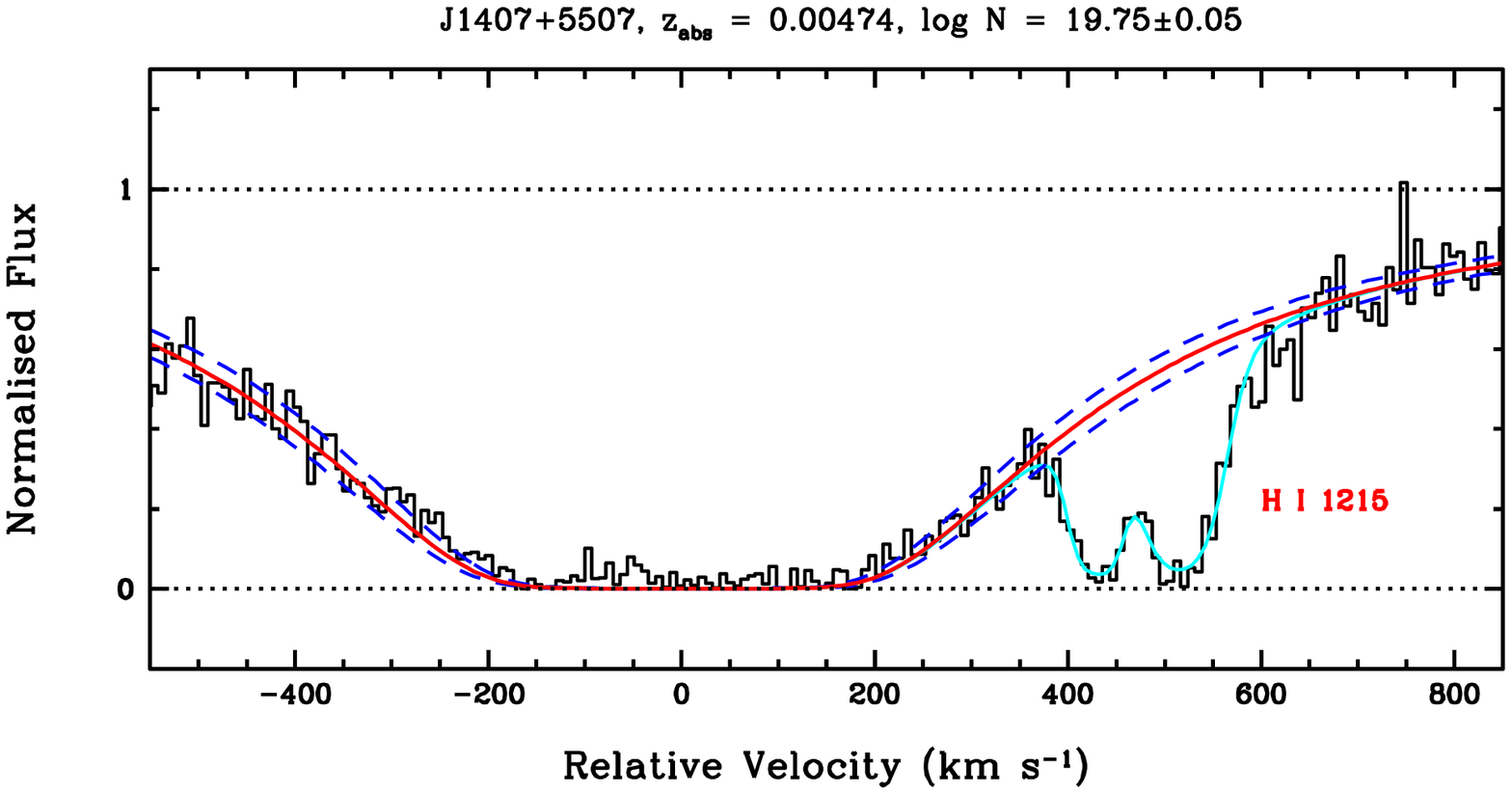}    
}}
}}
\caption{Sub-damped \lya\ absorption from \zabs\ = 0.00474 towards J1407+5507 system. 
The blue wing is affected by the Galactic \lya\ absorption. 
Various curves are as described in Fig.~\ref{HI_MRK486_03879}.}    
\label{HI_J1407_00474} 
\end{figure*} 

\begin{figure*} 
\centerline{
\vbox{
\centerline{\hbox{ 
\includegraphics[height=6.0cm,width=12.0cm,angle=00]{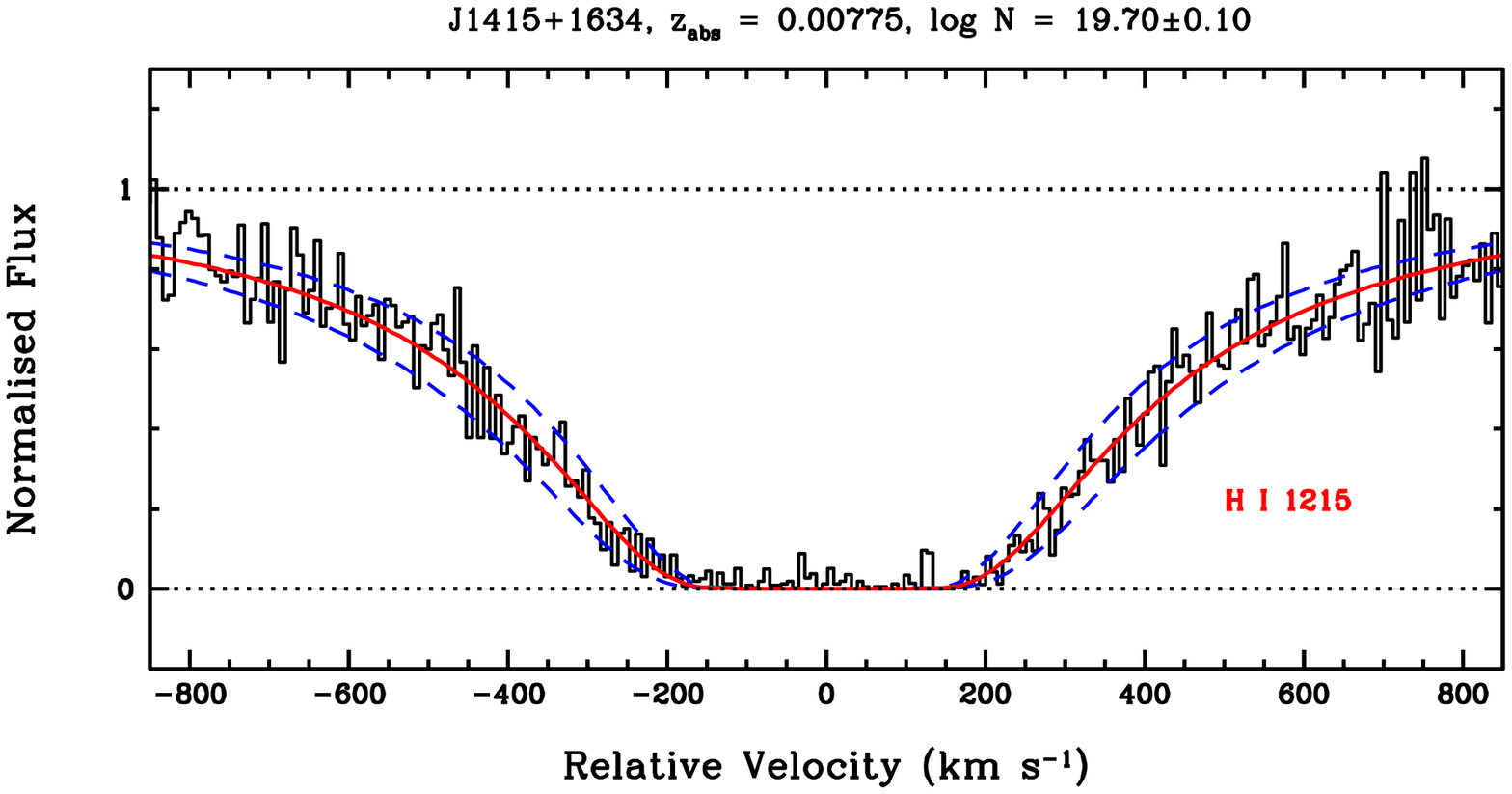}    
}}
}}
\caption{Sub-damped \lya\ absorption from \zabs\ = 0.00775 towards J1415+1634 system. 
Various curves are as described in Fig.~\ref{HI_MRK486_03879}.}      
\label{HI_J1415_00775} 
\end{figure*} 

\begin{figure*} 
\centerline{
\vbox{
\centerline{\hbox{ 
\includegraphics[height=6.0cm,width=12.0cm,angle=00]{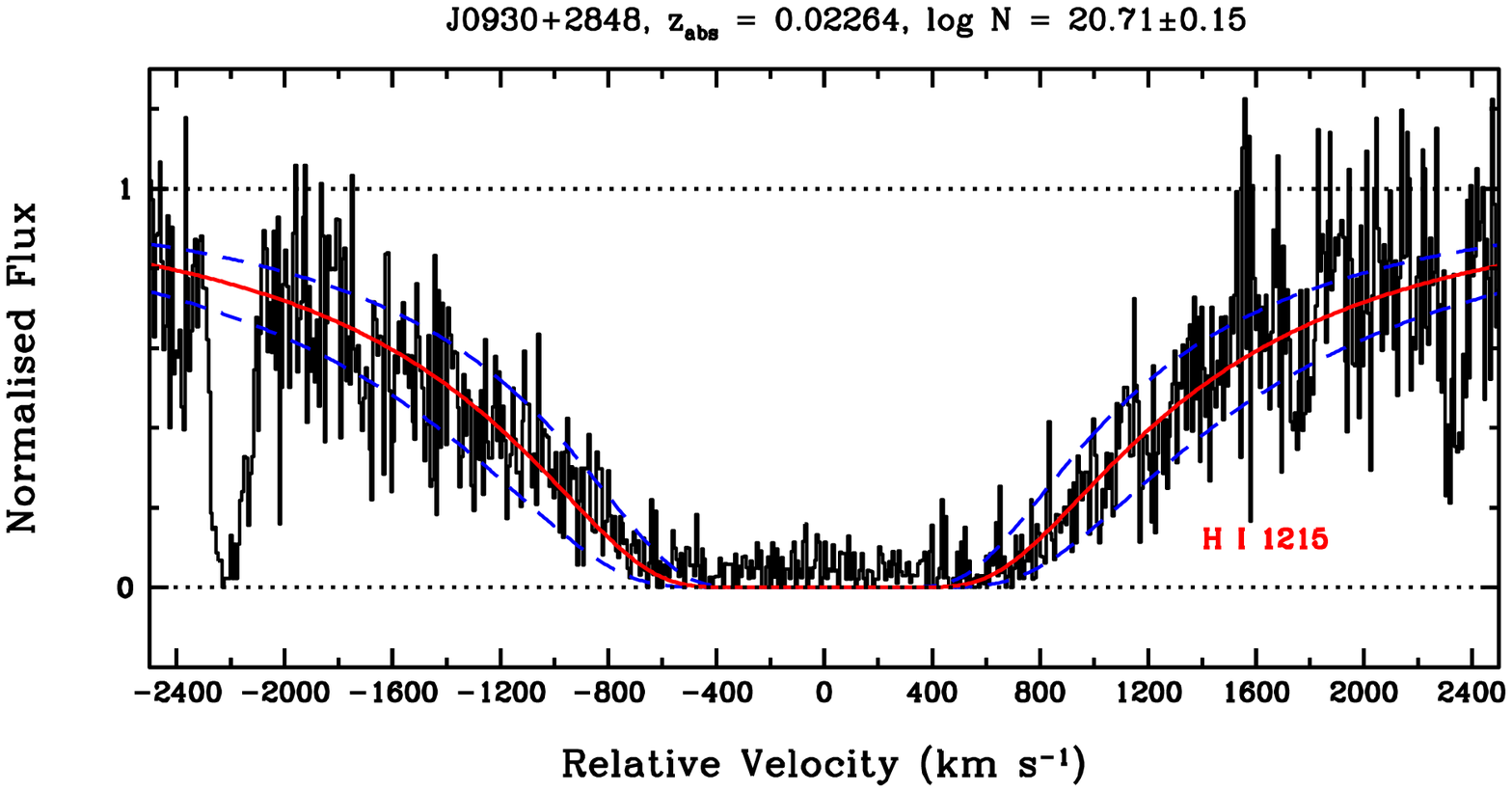}  
}}
}}
\caption{Damped \lya\ absorption from \zabs\ = 0.02264 towards J0930+2848 system. 
Various curves are as described in Fig.~\ref{HI_MRK486_03879}.}   
\label{HI_J0930_02264} 
\end{figure*} 

\begin{figure*} 
\centerline{
\vbox{
\centerline{\hbox{ 
\includegraphics[height=6.0cm,width=12.0cm,angle=00]{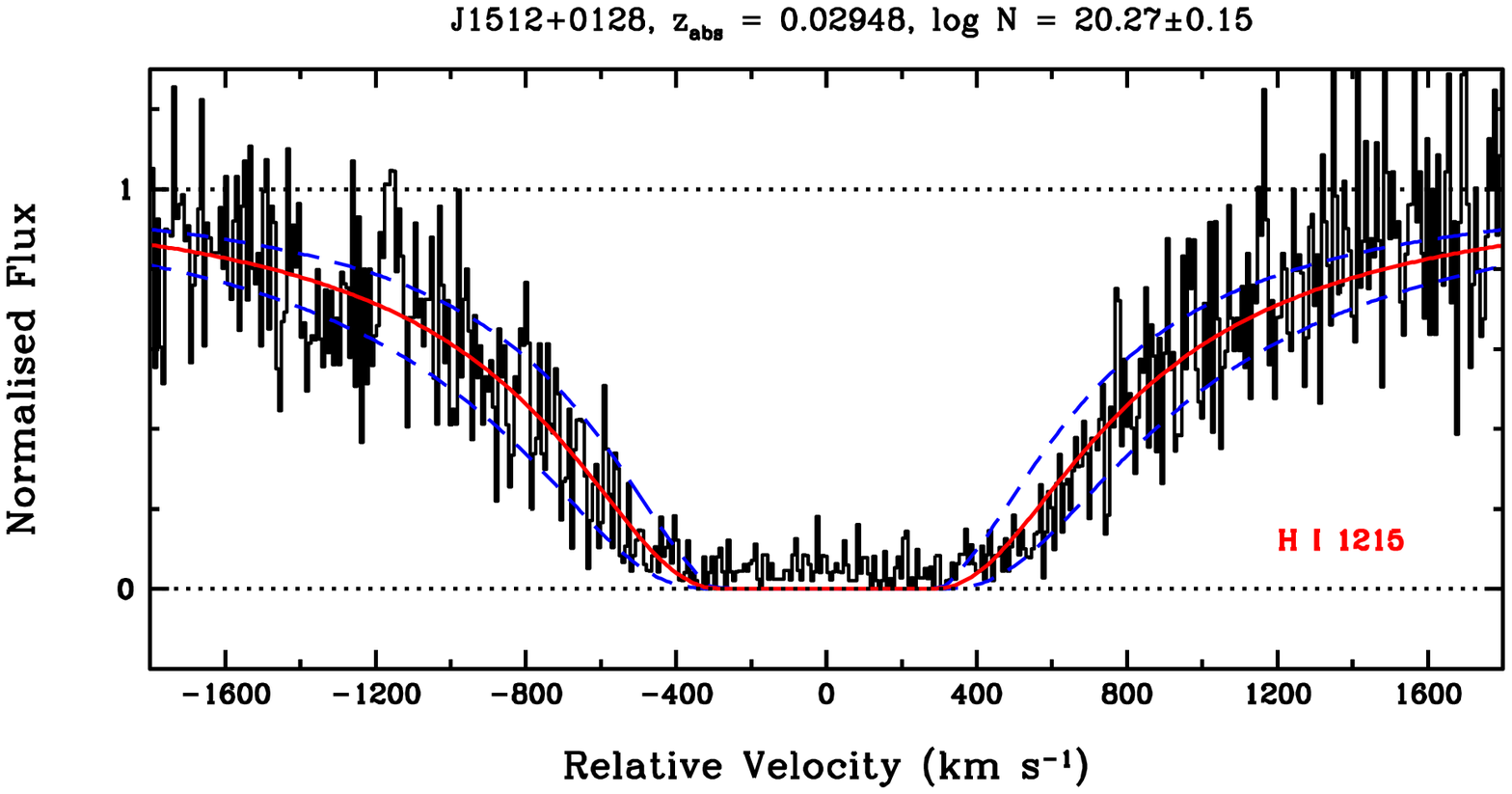}   
}}
}}
\caption{Damped \lya\ absorption from \zabs\ = 0.02948 towards J1512+0128 system. 
Various curves are as described in Fig.~\ref{HI_MRK486_03879}.}      
\label{HI_J1512_02948} 
\end{figure*} 

\begin{figure*} 
\centerline{
\vbox{
\centerline{\hbox{ 
\includegraphics[height=6.0cm,width=12.0cm,angle=00]{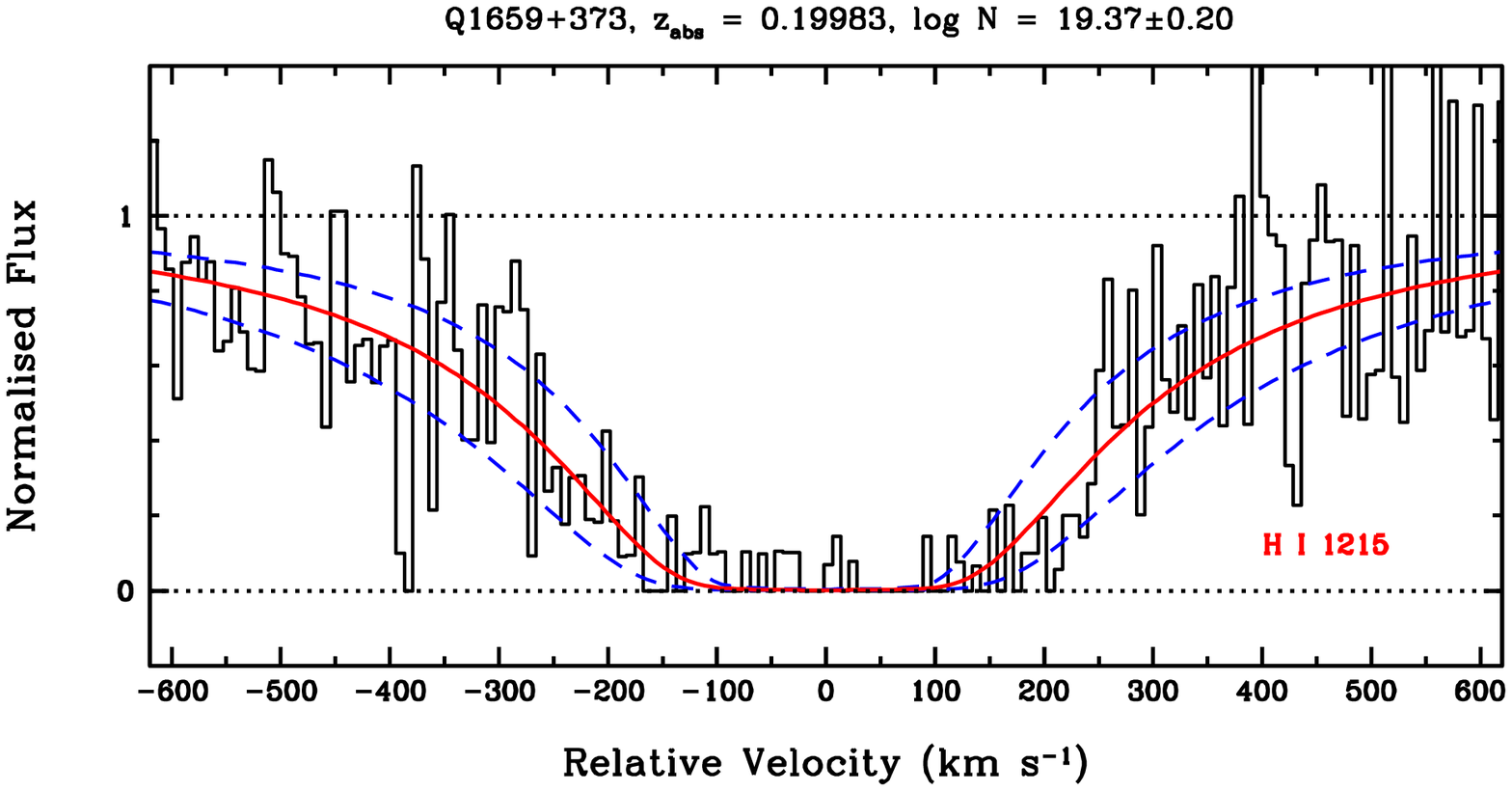} 
}}
}}
\caption{Sub-damped \lya\ absorption from \zabs\ = 0.19983 towards Q1659+373 system. 
Various curves are as described in Fig.~\ref{HI_MRK486_03879}.}     
\label{HI_Q1659_19983} 
\end{figure*} 


\newpage

\setcounter{figure}{0}   
\begin{figure*} 
\section{Voigt profile fit to molecular hydrogen absorption}     
\label{H2fits}      
\centerline{
\vbox{
\centerline{\hbox{ 
\includegraphics[height=12.0cm,width=12.0cm,angle=00]{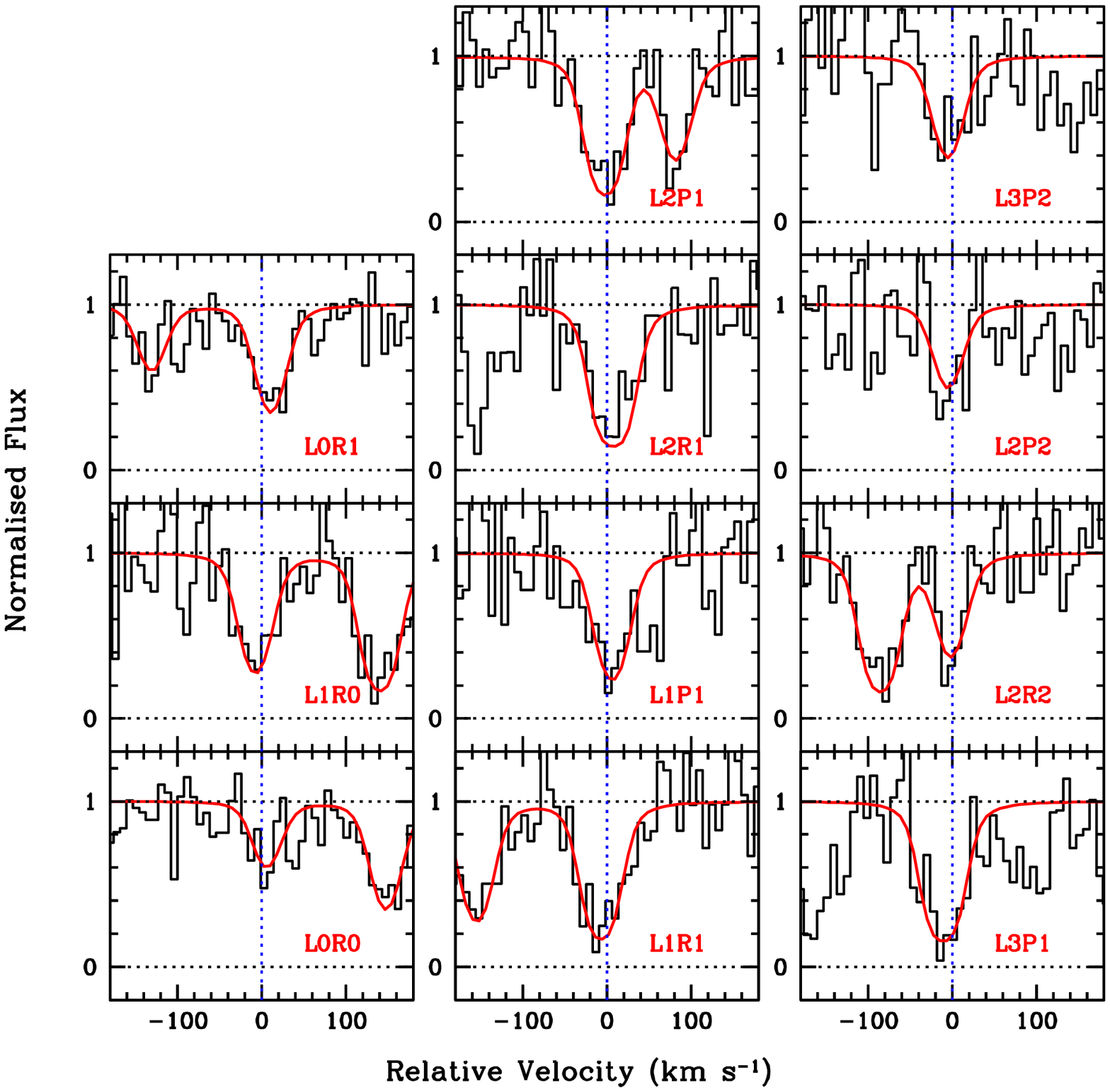} 
}}  
}}
\vskip-0.5cm 
\caption{Molecular hydrogen absorption from the \zabs\ = 0.06650 towards J1241+2852 system. The 
red smooth curves are the best fitting Voigt profiles to the data (in black histogram). The 
vertical dotted line marks the median line centroid.}      
\label{H2_Q0439_10115}  
\end{figure*} 

\setcounter{figure}{1}  
\begin{figure*} 
\centerline{
\vbox{
\centerline{\hbox{ 
\includegraphics[height=12.0cm,width=12.0cm,angle=00]{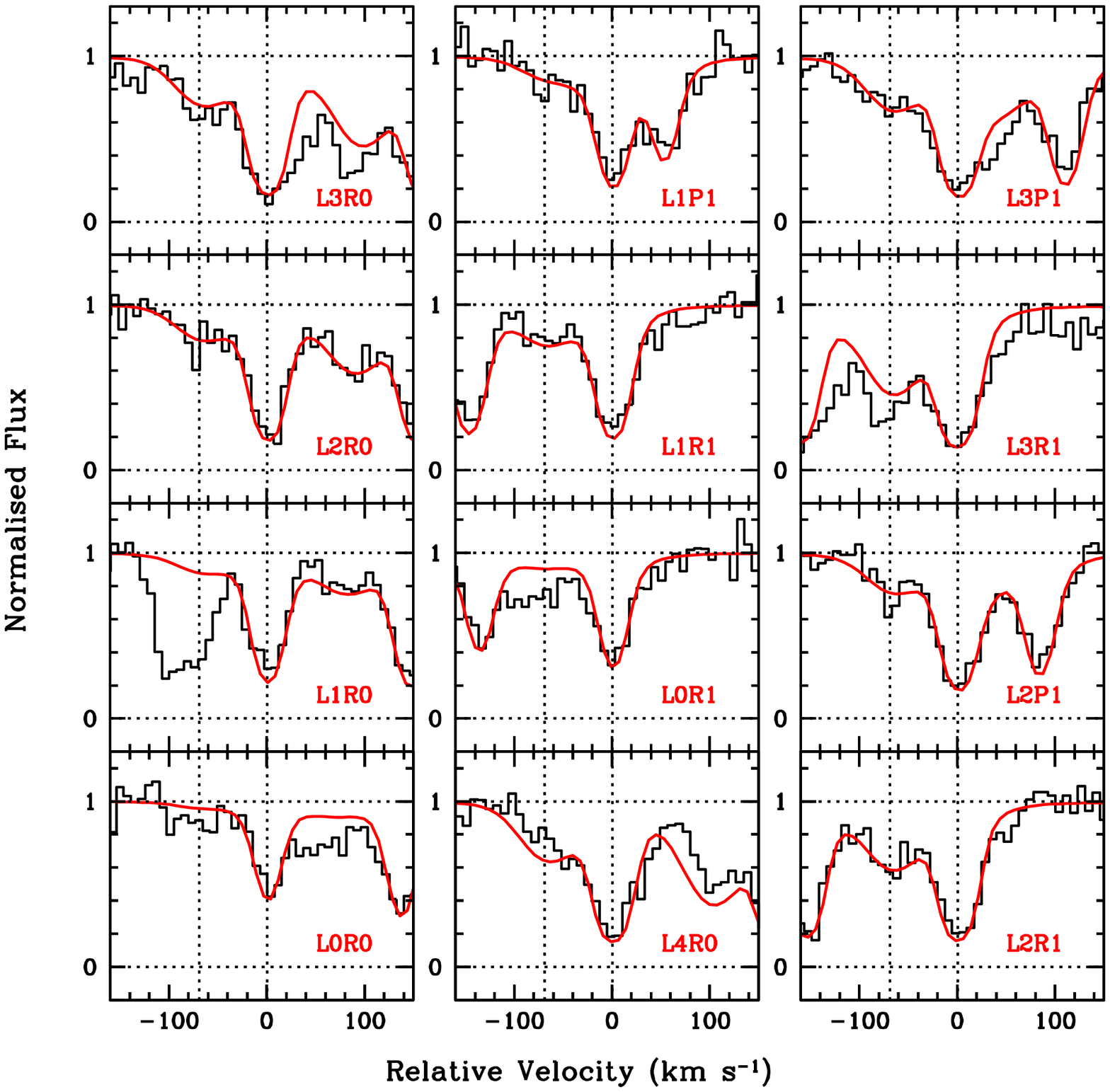} 
}}  
\centerline{\hbox{ 
\includegraphics[height=12.0cm,width=12.0cm,angle=00]{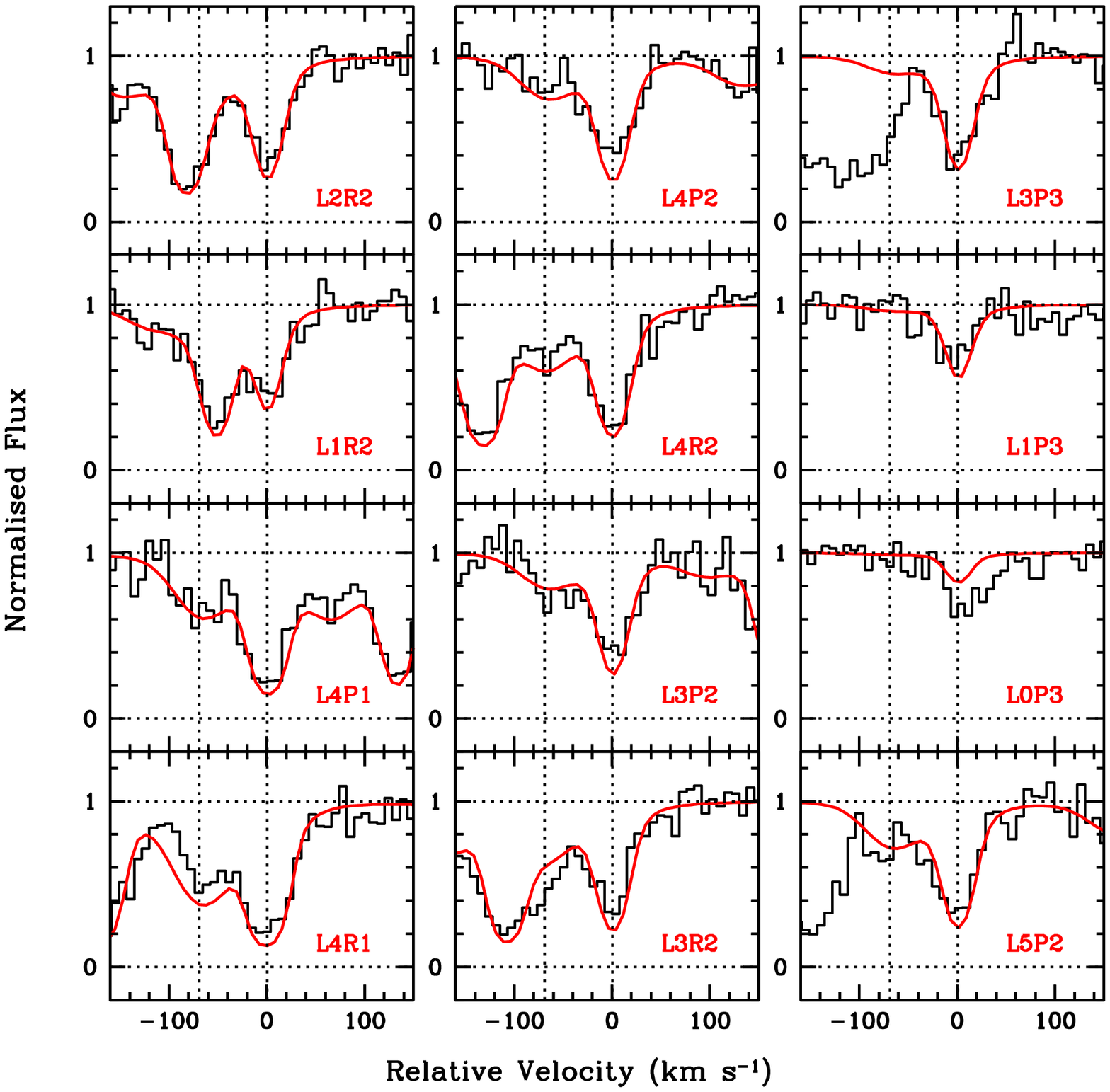} 
}}
}}
\vskip-0.5cm 
\caption{Molecular hydrogen absorption from the \zabs\ = 0.10115 towards Q0439--433 system. A weak 
component is possibly detected at $\sim70$~\kms.} 
\label{H2_Q0439_10115}  
\end{figure*} 

\setcounter{figure}{1}  
\begin{figure*} 
\centerline{
\vbox{
\centerline{\hbox{ 
\includegraphics[height=7.0cm,width=12.0cm,angle=00]{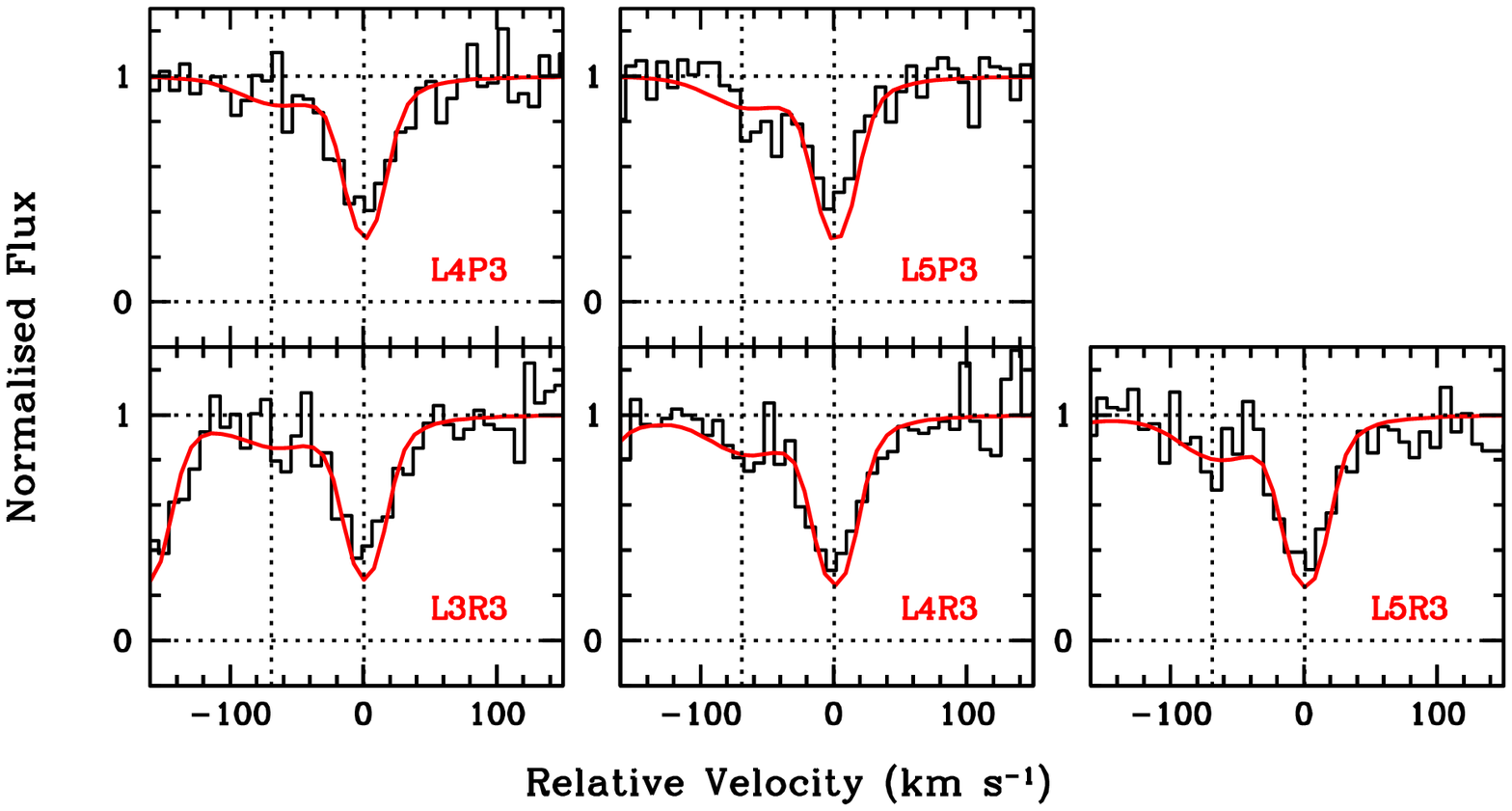} 
}}  
}}
\vskip-0.5cm 
\caption{Continued.} 
\label{H2_Q0439_10115}  
\end{figure*} 

\setcounter{figure}{2}  
\begin{figure*} 
\centerline{
\vbox{
\centerline{\hbox{ 
\includegraphics[height=12.0cm,width=12.0cm,angle=00]{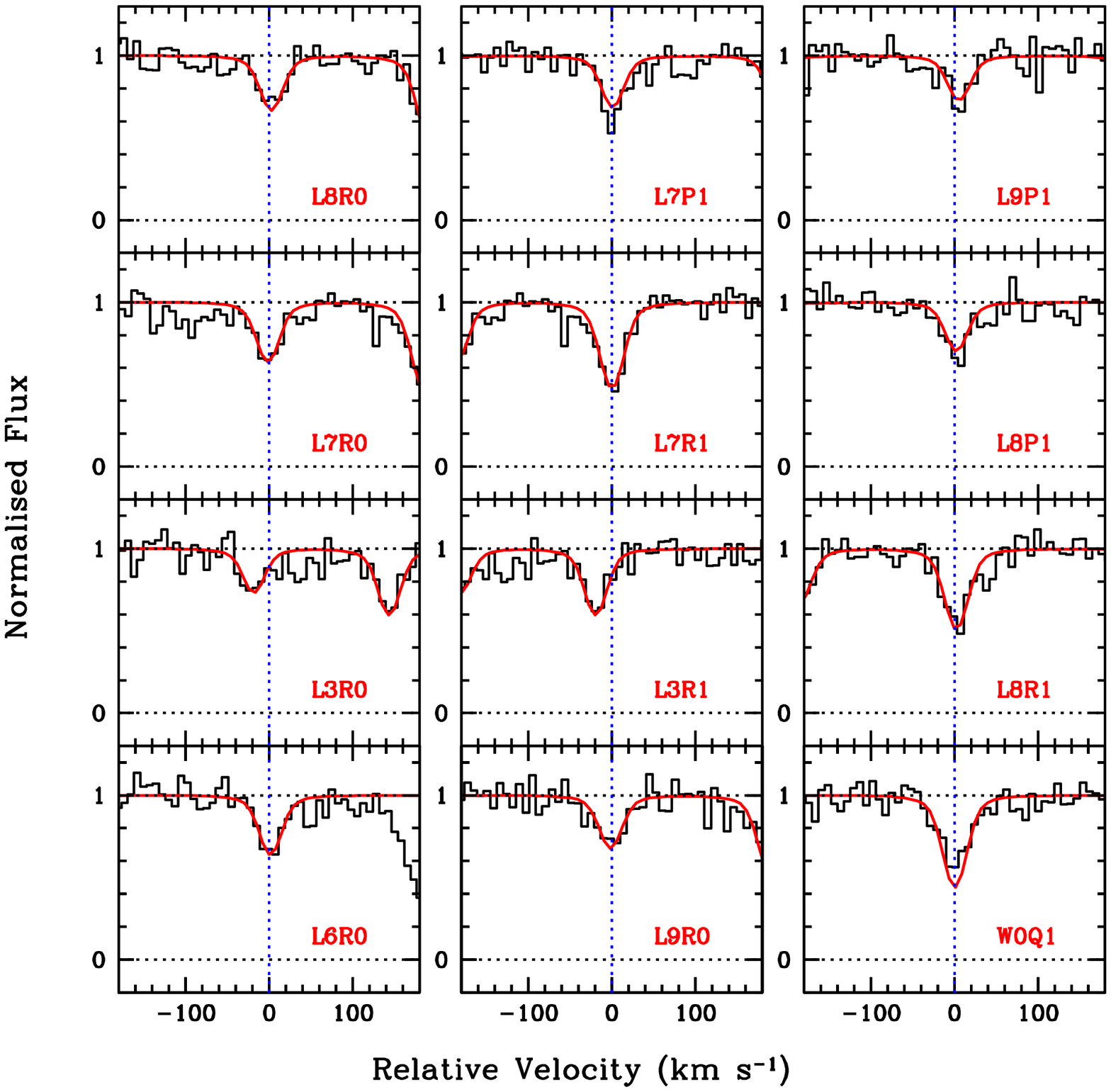} 
}}  
\centerline{\hbox{ 
\includegraphics[height=7.0cm,width=12.0cm,angle=00]{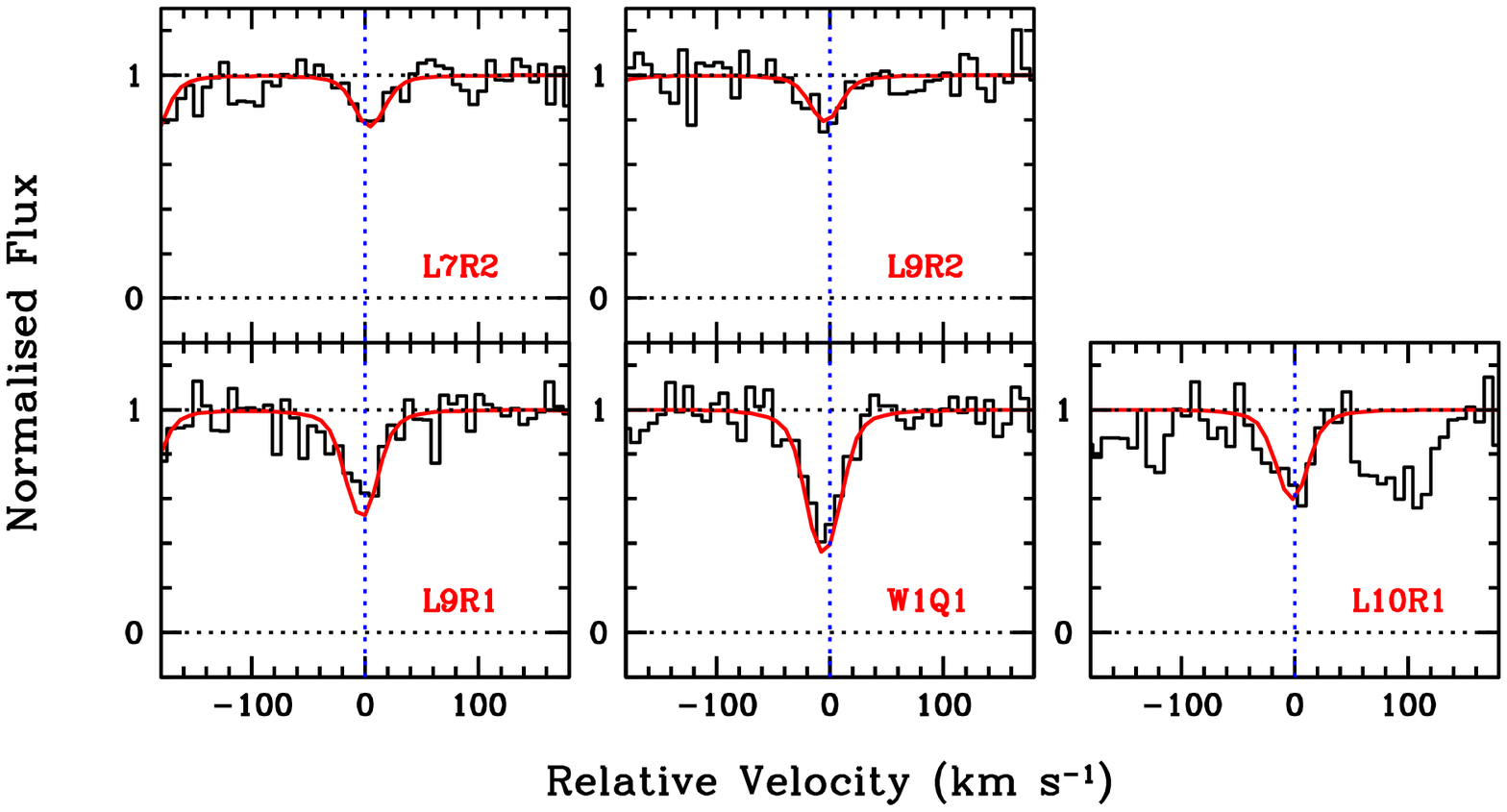} 
}}
}}
\vskip-0.5cm 
\caption{Molecular hydrogen absorption from the \zabs\ = 0.16375 towards Q0850+440 system. 
 Note that both the L3R0~$\lambda1062.8$ and L3R1~$\lambda 1063.4$ lines fall at the 
edge of the detector segment ``B" and are shifted towards blue by slightly more than 20 \kms.}     
\label{H2_Q0850_16375}  
\end{figure*} 

\setcounter{figure}{3}  
\begin{figure*} 
\centerline{
\vbox{
\centerline{\hbox{ 
\includegraphics[height=7.0cm,width=12.0cm,angle=00]{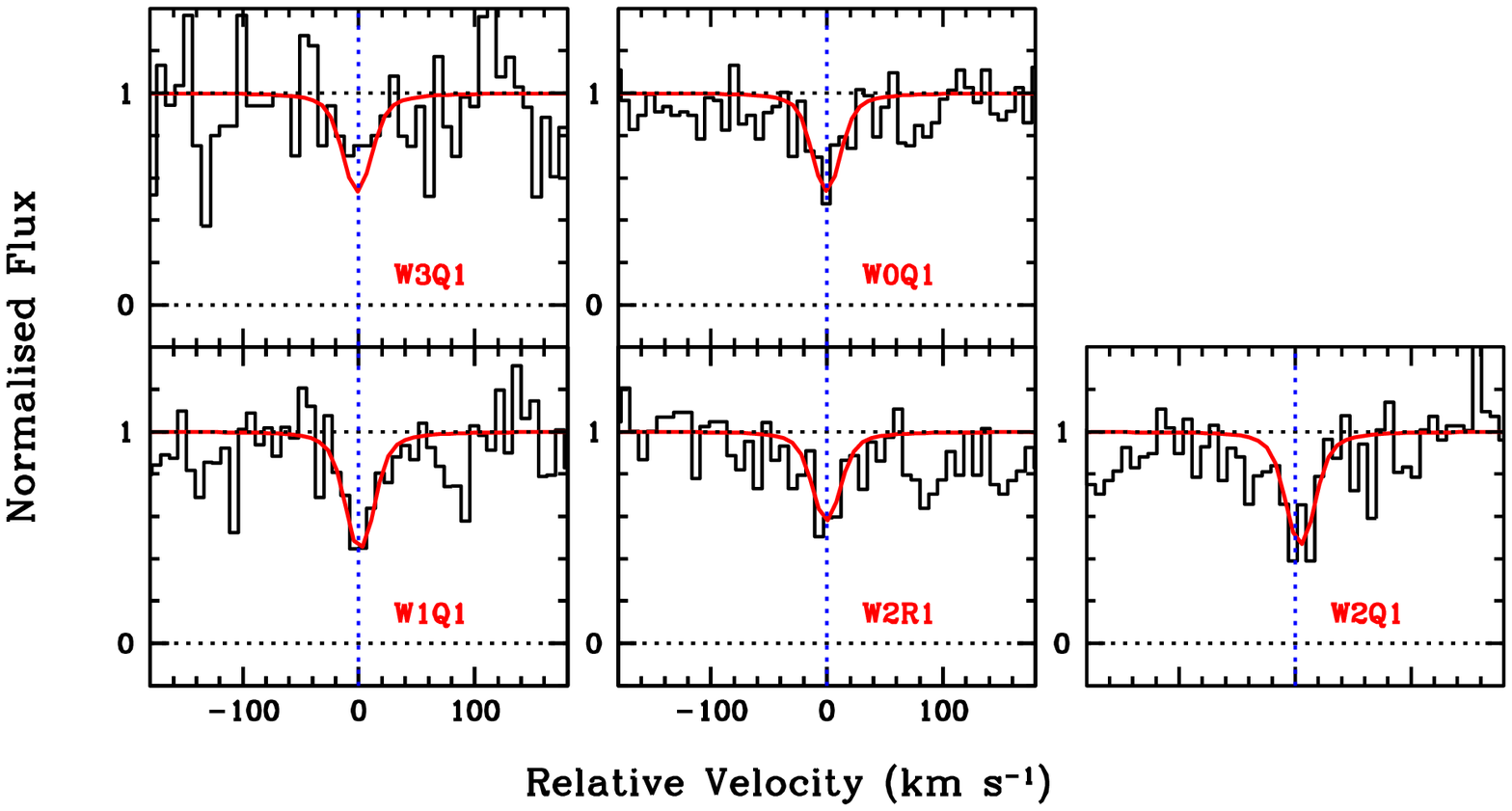} 
}}  
}}
\vskip-0.5cm 
\caption{Molecular hydrogen absorption from the \zabs\ = 0.22711 towards J1342--0053 system.}  
\label{H2_J1342_22711}  
\end{figure*} 

\setcounter{figure}{4}  
\begin{figure*} 
\centerline{
\vbox{
\centerline{\hbox{ 
\includegraphics[height=12.0cm,width=12.0cm,angle=00]{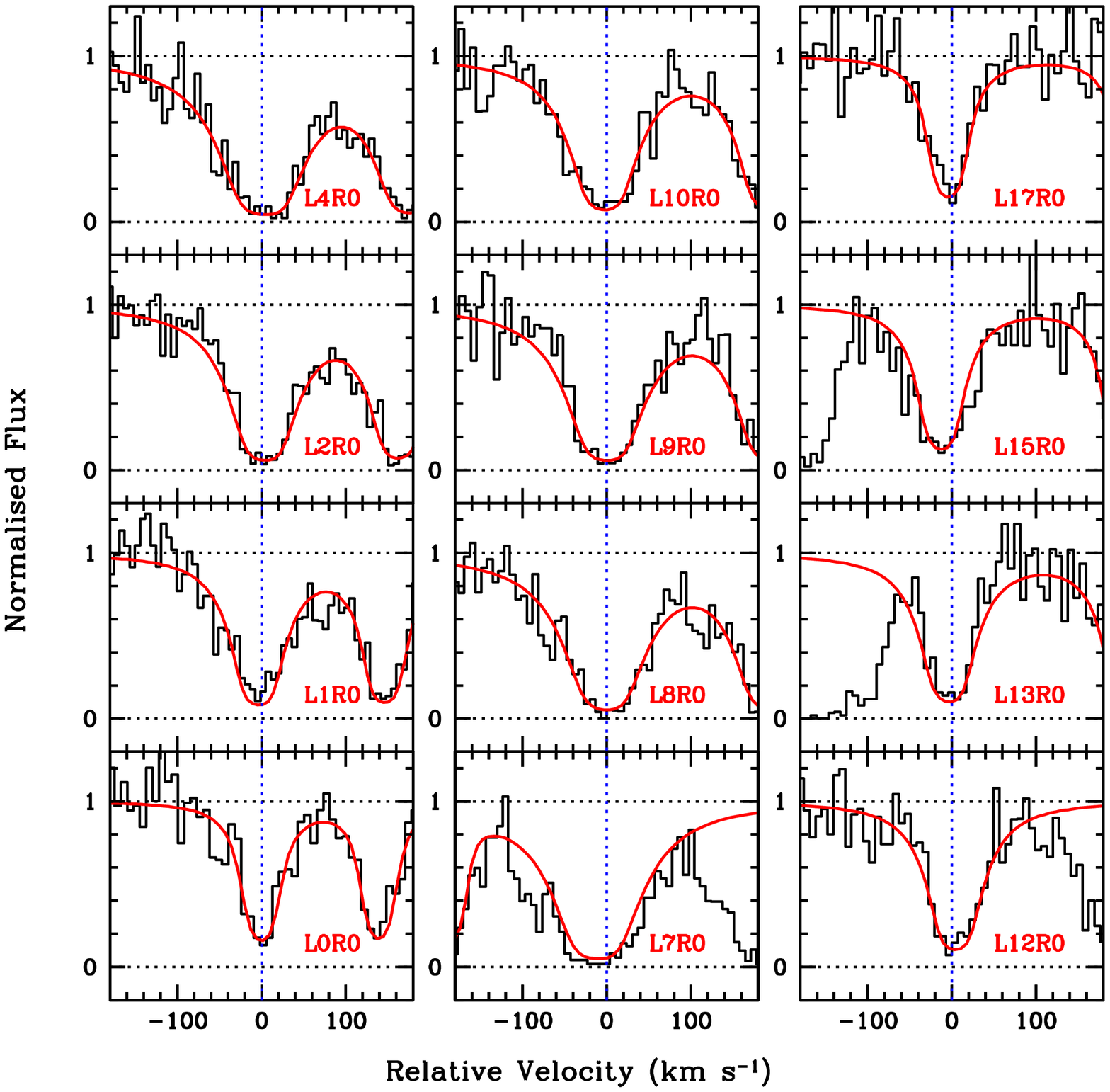} 
}}  
\vskip-0.5cm 
\centerline{\hbox{ 
\includegraphics[height=12.0cm,width=12.0cm,angle=00]{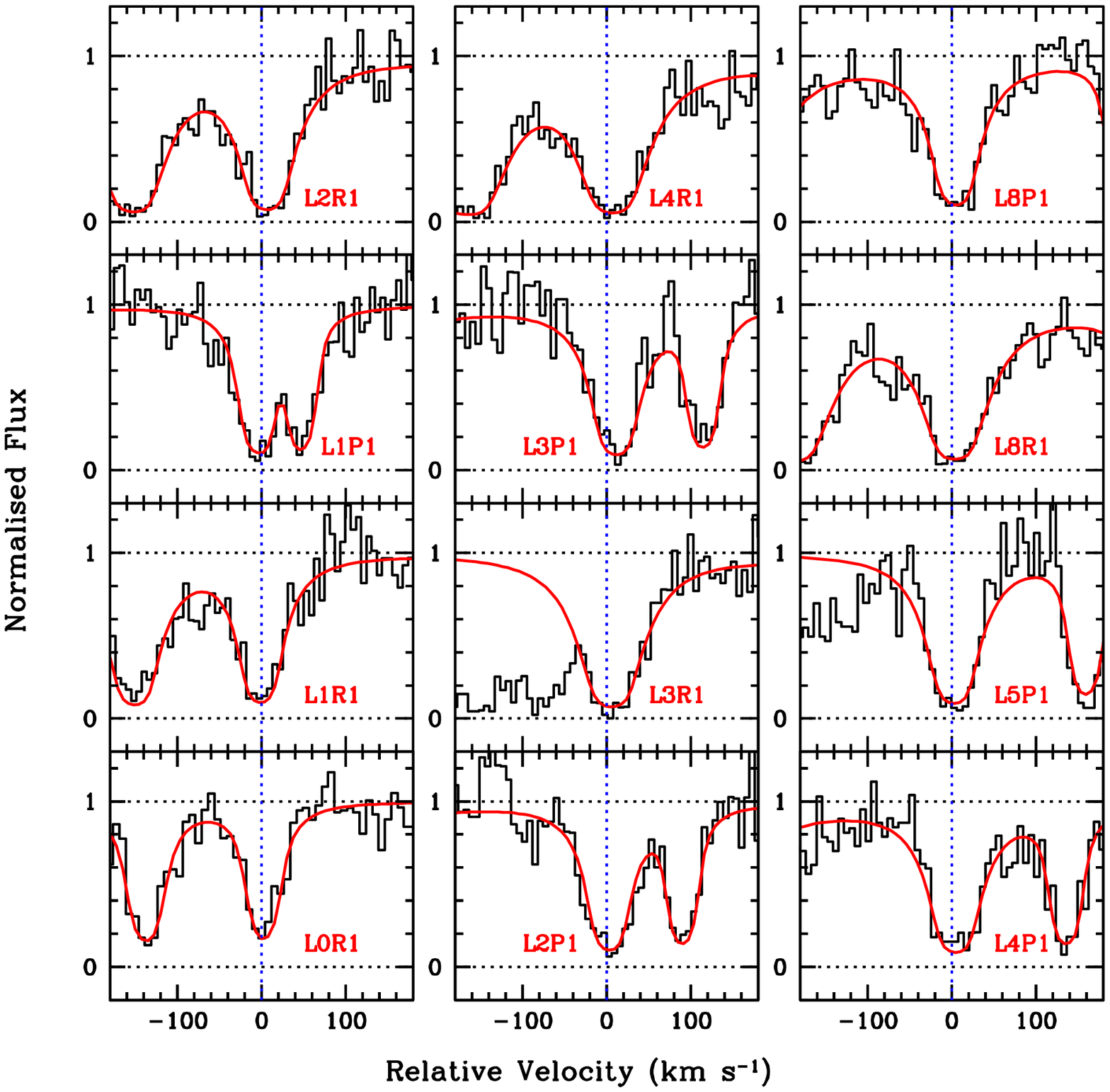} 
}}
}}
\vskip-0.5cm 
\caption{Molecular hydrogen absorption from the \zabs\ = 0.32110 towards J1616+4154 system.} 
\label{H2_J1616_32110}  
\end{figure*} 

\setcounter{figure}{4}  
\begin{figure*} 
\centerline{
\vbox{
\centerline{\hbox{ 
\includegraphics[height=12.0cm,width=12.0cm,angle=00]{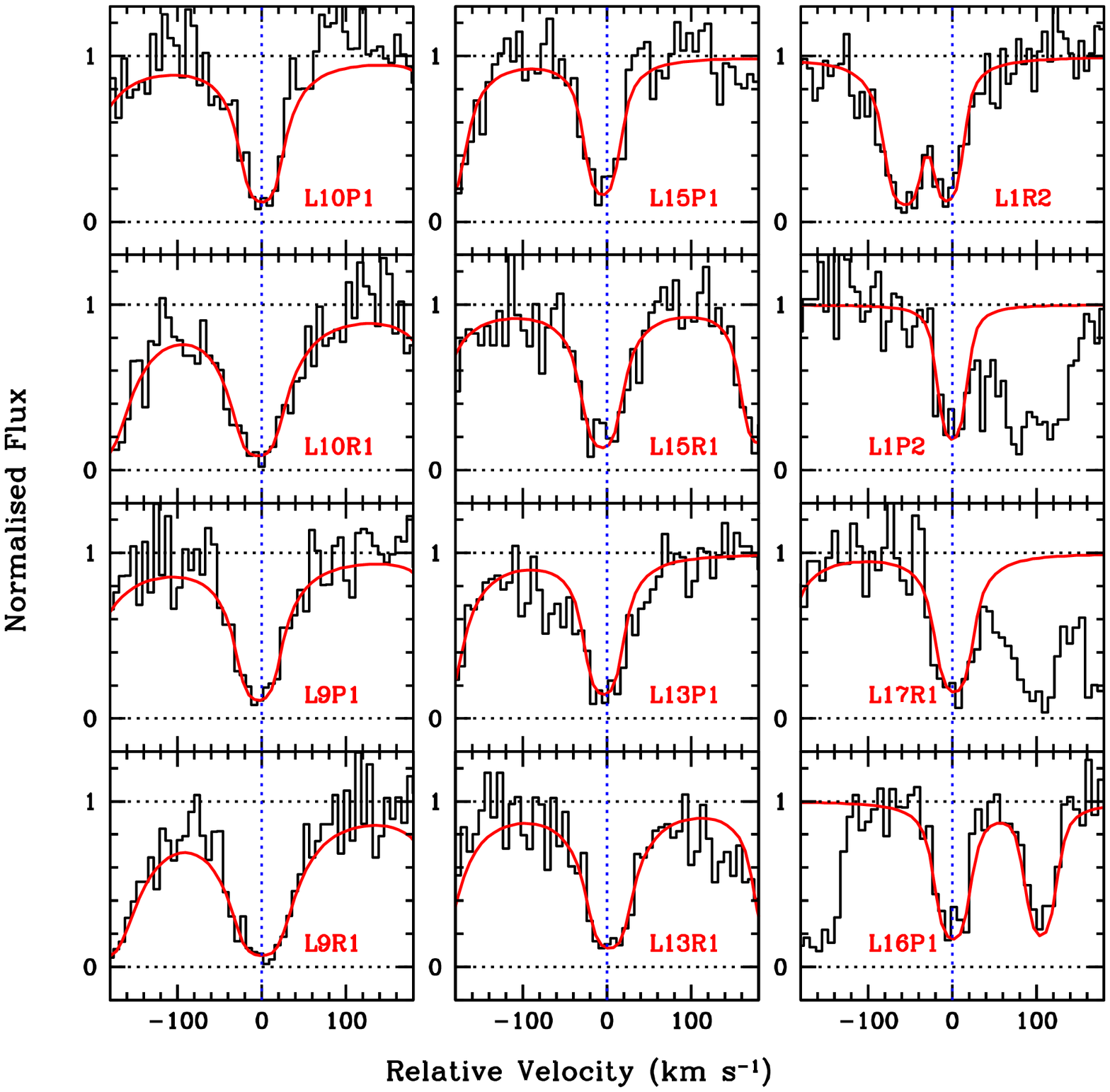} 
}}  
\centerline{\hbox{ 
\includegraphics[height=12.0cm,width=12.0cm,angle=00]{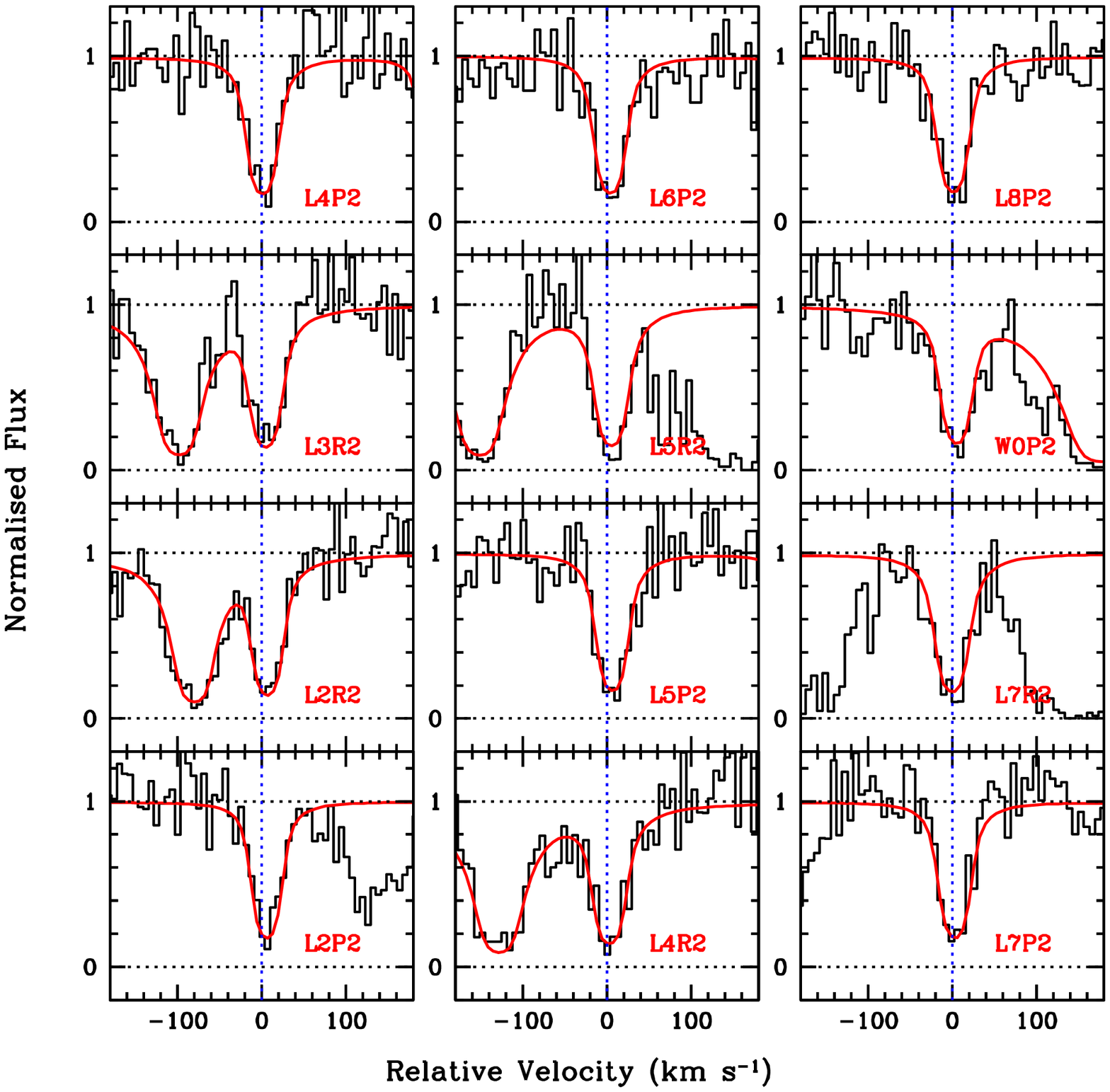} 
}}
}}
\vskip-0.5cm 
\caption{Continued.}  
\label{H2_J1616_32110}  
\end{figure*} 

\setcounter{figure}{4}  
\begin{figure*} 
\centerline{
\vbox{
\centerline{\hbox{ 
\includegraphics[height=12.0cm,width=12.0cm,angle=00]{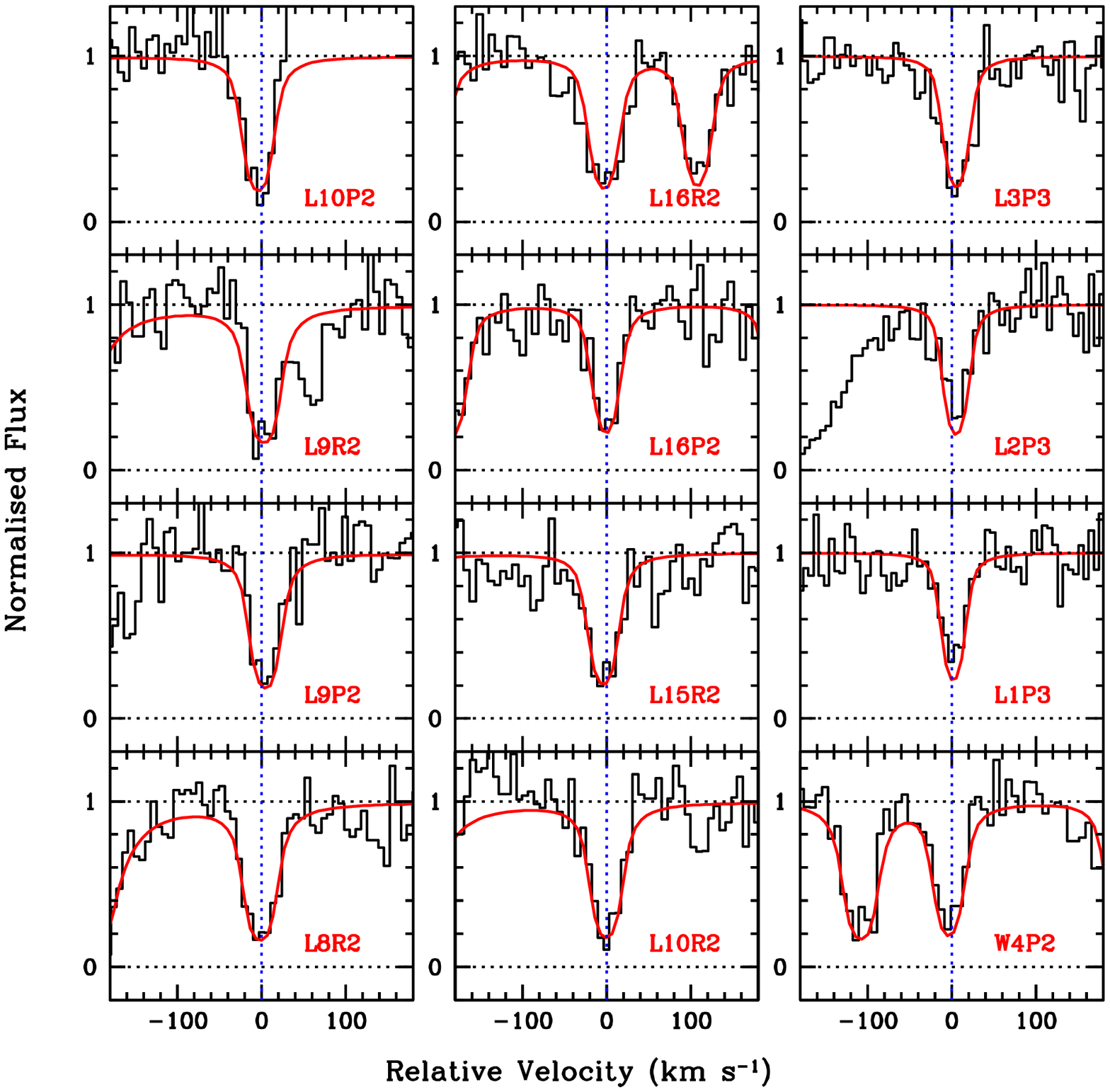} 
}}  
\centerline{\hbox{ 
\includegraphics[height=12.0cm,width=12.0cm,angle=00]{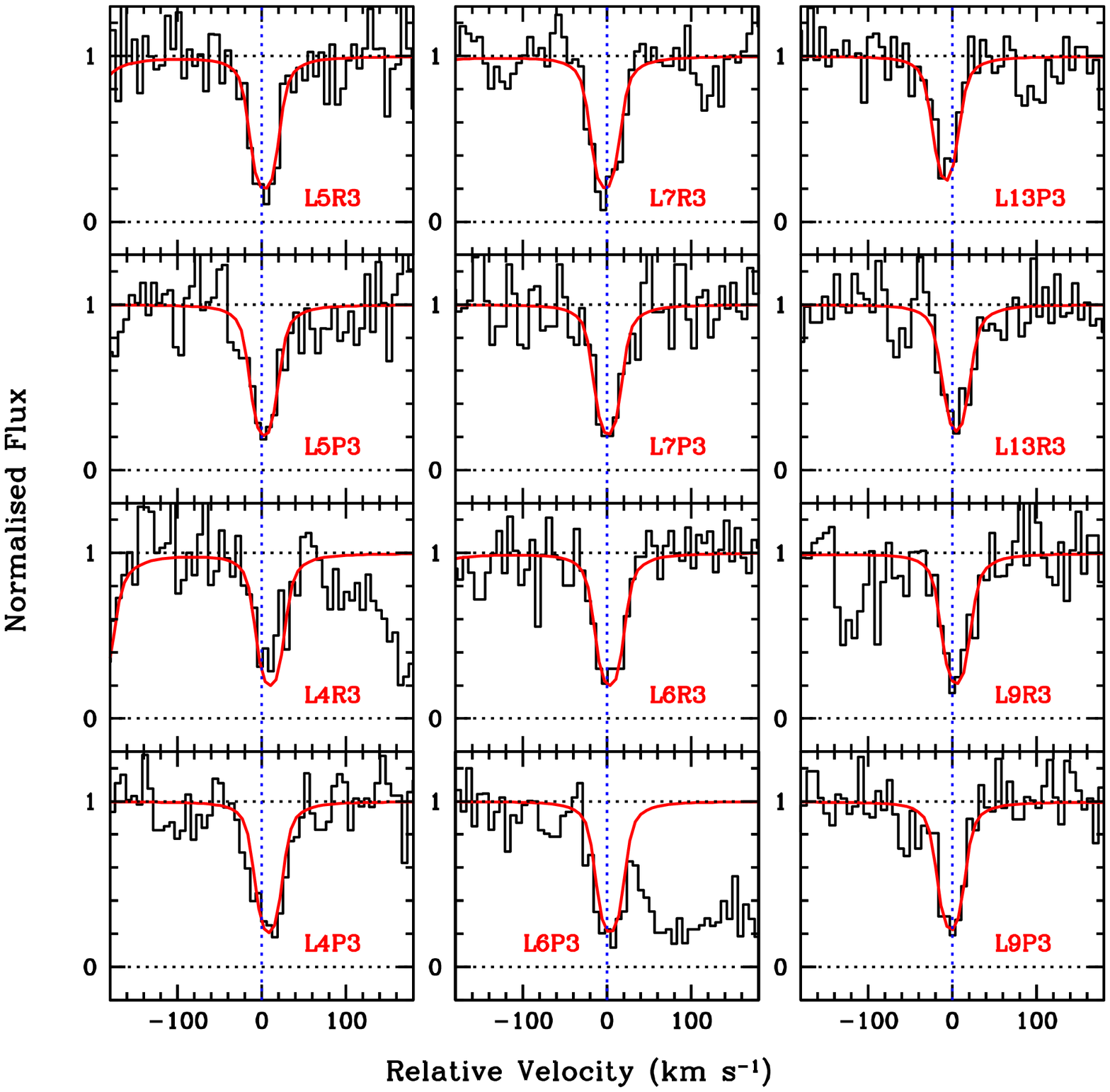} 
}}
}}
\vskip-0.5cm 
\caption{Continued.} 
\label{H2_J1616_32110}  
\end{figure*} 
 
\setcounter{figure}{4}  
\begin{figure*} 
\centerline{
\vbox{
\centerline{\hbox{ 
\includegraphics[height=4.0cm,width=12.0cm,angle=00]{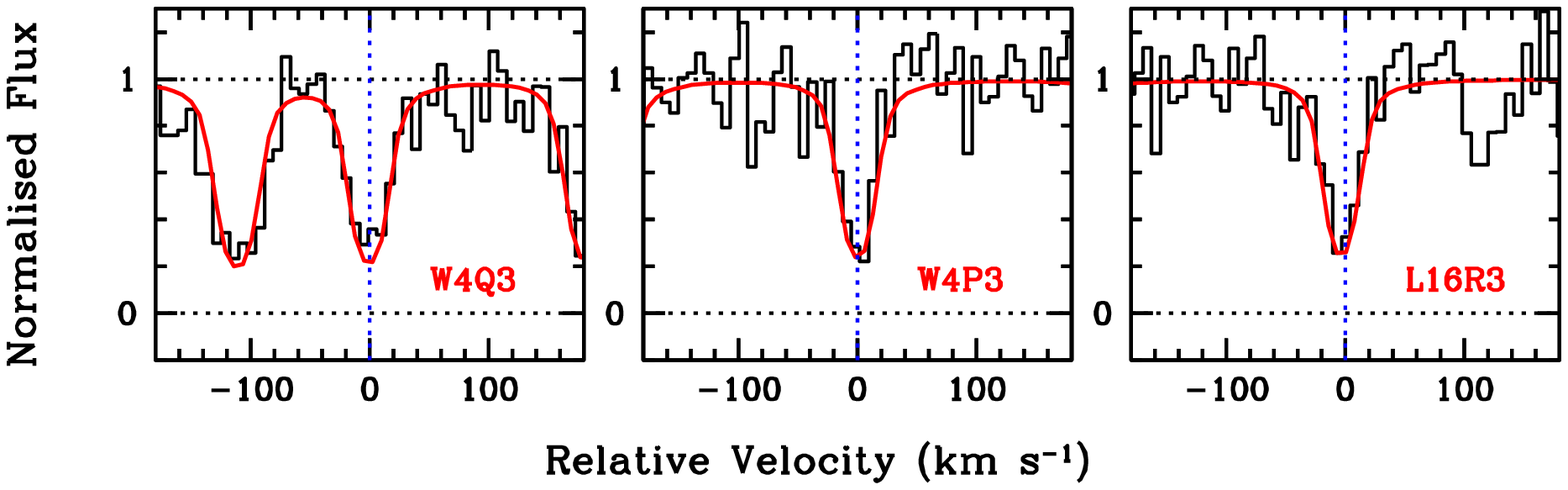} 
}}  
}}
\vskip-0.5cm 
\caption{Continued.} 
\label{H2_J1616_32110}  
\end{figure*} 

\setcounter{figure}{5}  
\begin{figure*} 
\centerline{
\vbox{
\centerline{\hbox{ 
\includegraphics[height=12.0cm,width=12.0cm,angle=00]{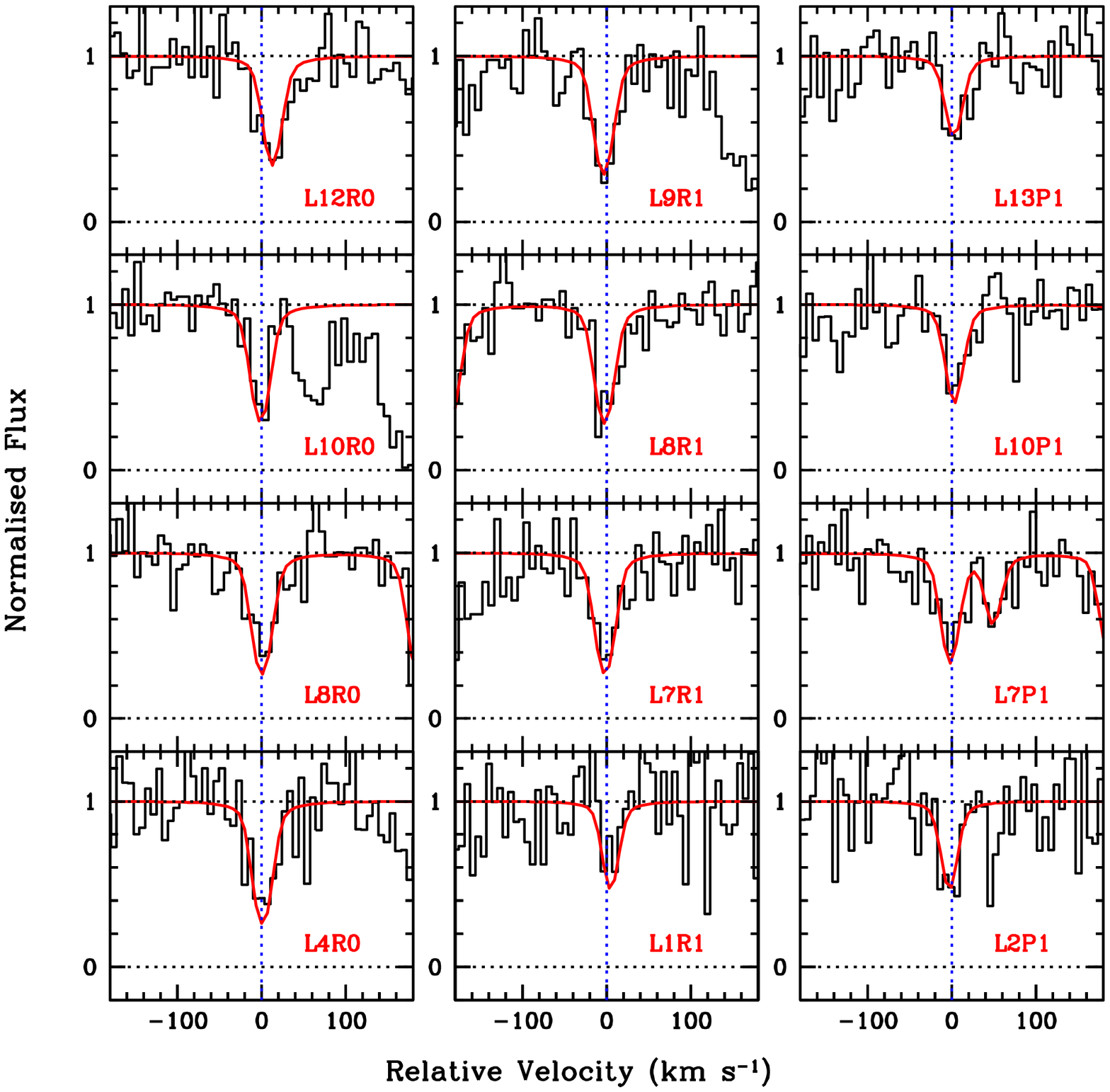} 
}}  
\centerline{\hbox{ 
\includegraphics[height=12.0cm,width=12.0cm,angle=00]{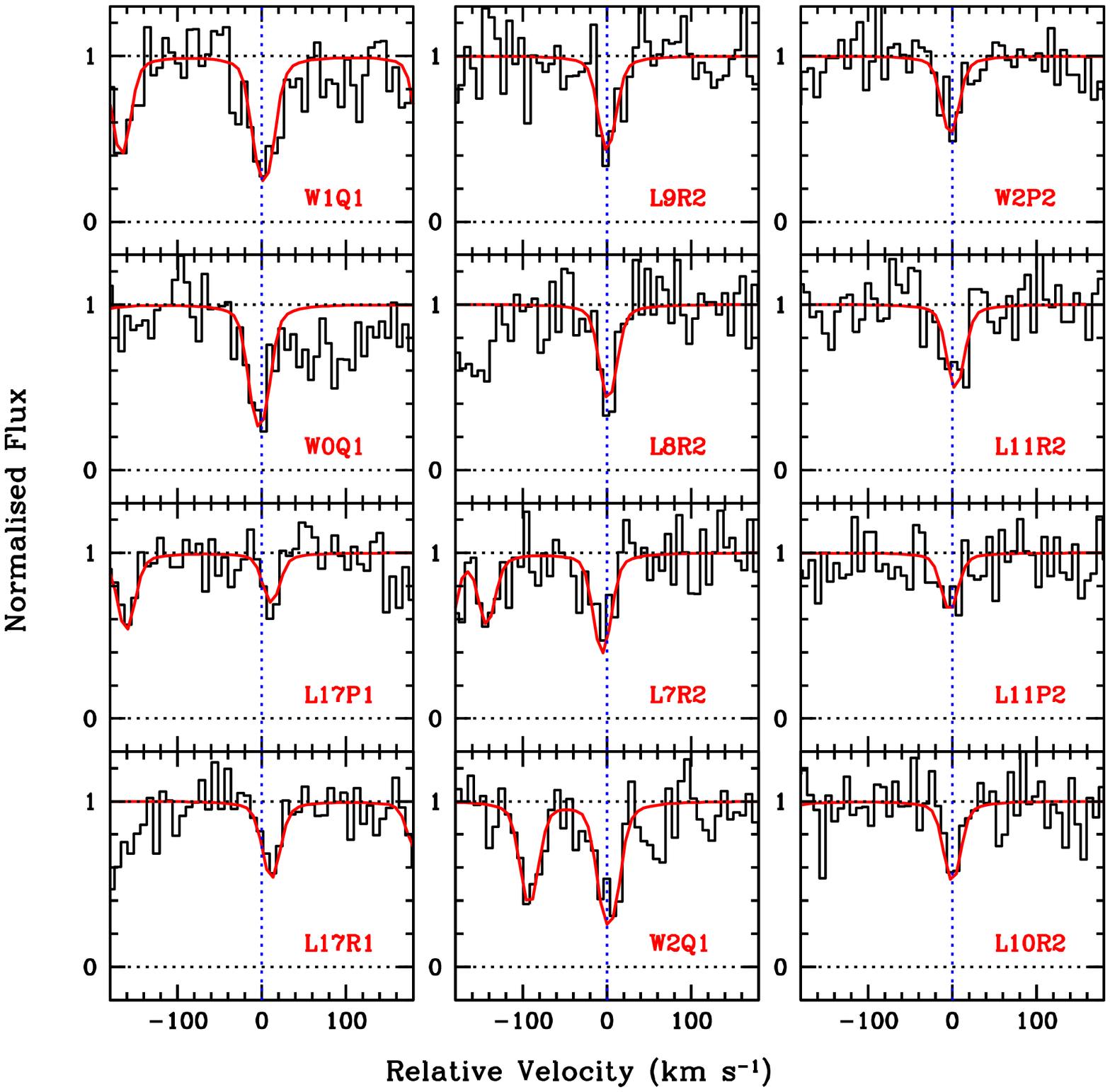} 
}}
}}
\vskip-0.5cm 
\caption{Molecular hydrogen absorption from the \zabs\ = 0.55048 towards Q1241+176 system.}  
\label{H2_Q1241_55048}  
\end{figure*} 

\setcounter{figure}{5}  
\begin{figure*} 
\centerline{
\vbox{
\centerline{\hbox{ 
\includegraphics[height=10.0cm,width=12.0cm,angle=00]{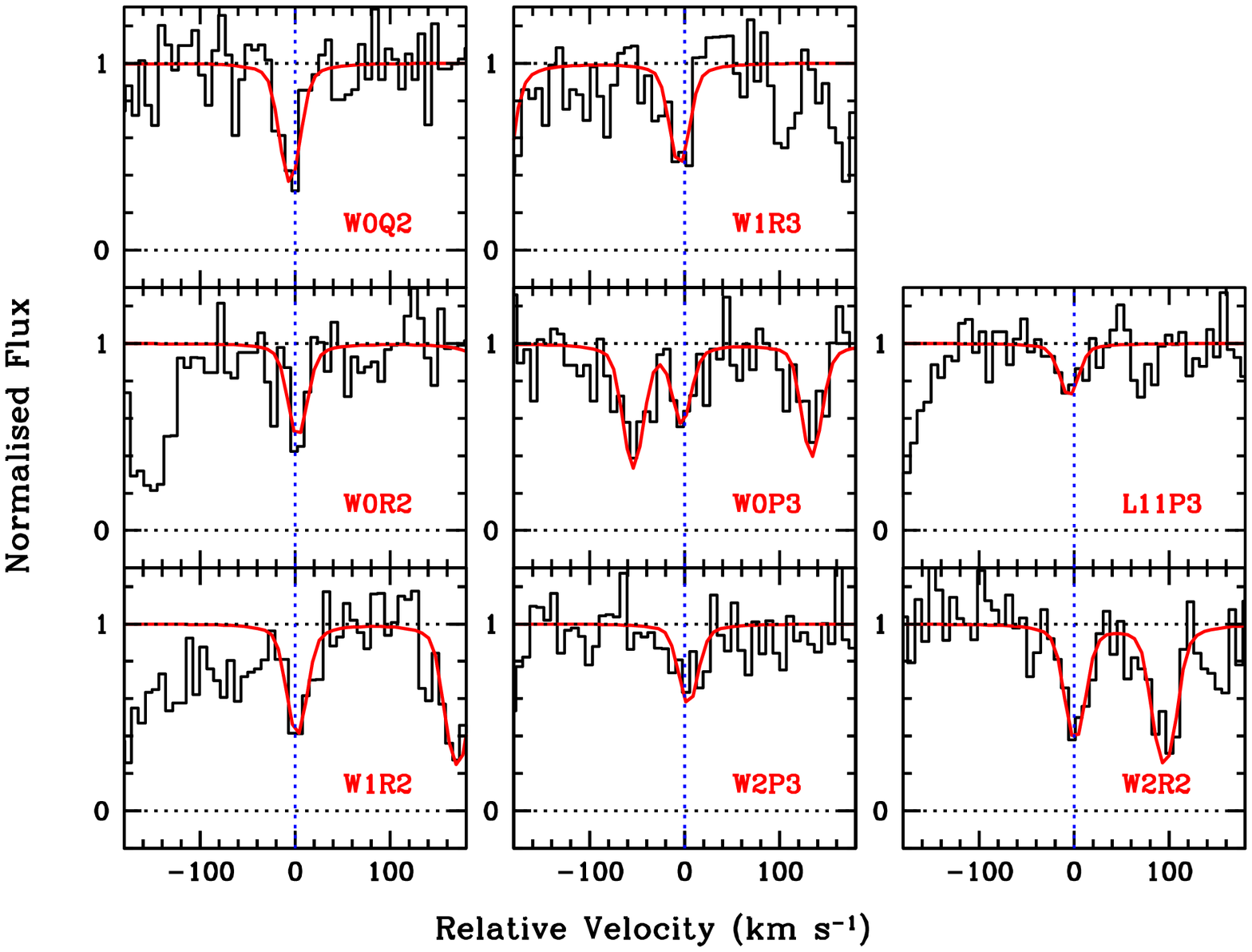} 
}}  
}}
\vskip-0.5cm 
\caption{Continued.}  
\label{H2_Q1241_55048}  
\end{figure*} 

\end{document}

%% file: tab1.tex
\begin{table*} 
\caption{Core sample of low-$z$ DLAs/sub-DLAs for which \HH\ information is available.}      
\begin{tabular}{lcclcm{1.5cm}m{1.2cm}m{1.5cm}rl} 
\hline \hline 
QSO$^{1}$    & \zem$^{2}$    & PID$^{3}$	  &     PI$^{4}$      &   \zabs$^{5}$    &   $\log N(\HI)^{6}$  &    $\log N(\HH)^{7}$     &   $\log N({\HH})^{8}$  &  $\log f_{\hh}^{9}$       &   $\rho^{10}$ (kpc)     \\ 
       &         &        &             &             &                  &    Sensitivity  &      &      &      \\   
(1)    &  (2)    & (3)    &     (4)     &    (5)      &    (6)           &     (7)   &         (8)   &      (9)            &    (10)          \\ 
\hline  
PG1216+069	& 0.331 & 12025  &   J. Green	     &  0.00635   &    19.32$\pm$0.03$^{a}$      & 17.8 &   $<$17.8$^{a}$           &   $<-$1.24$^{a}$             & 121$^{a}$        \\ 
MRK486	        & 0.039 & 12276  &   B. Wakker       &  0.03879   &    20.25$\pm$0.25 	         & 15.6 &     $<$15.6               &   $<-$4.35	       \\ 
J1241+2852	& 0.589 & 12603  &   T. Heckman      &  0.06650   &    19.18$\pm$0.10 	         & 14.9 &    16.45$\pm$0.12         &   $-$2.43$\pm$0.16   \\ 	
J1553+3548	& 0.723 & 11598  &   J. Tumlinson    &  0.08300   &    19.55$\pm$0.15$^{b}$      & 14.6 &    $<$14.6      	     &   $<-$4.65	       \\ 
J1619+3342	& 0.472 & 11598  &   J. Tumlinson    &  0.09630   &    20.55$\pm$0.10$^{b}$      & 14.4 &    18.57$\pm$0.06$^{c}$   &   $-$1.69$\pm$0.11$^{c}$     & 14$^{c}$         \\  	
Q0439-433	& 0.594 & 12536  &   V. Kulkarni     &  0.10115   &    19.63$\pm$0.08 	         & 14.2 &    16.64$\pm$0.05         &   $-$2.69$\pm$0.09           & 8 (Pet96)        \\ 		
J0928+6025	& 0.296 & 11598  &   J. Tumlinson    &  0.15380   &    19.35$\pm$0.15$^{b}$      & 14.5 &    $<$14.5      	     &   $<-$4.55                   & 91 (Wer13)       \\ 
PHL1226  	& 0.404 & 12536  &   V. Kulkarni     &  0.15962   &    19.37$\pm$0.10 	         & 14.3 &     $<$14.3               &   $<-$4.77	   \\ 		
Q0850+440	& 0.515 & 13398  &   C. Churchill    &  0.16375   &    19.67$\pm$0.10$^{d}$      & 13.9 &    15.05$\pm$0.07         &   $-$4.32$\pm$0.12           & 24 (Kac11)       \\    	 
B0120-28	& 0.434 & 12204  &   C. Thom	     &  0.18562   &    20.50$\pm$0.10$^{e}$      & 13.9 & 20.00$\pm$0.10$^{e}$      &   $-$0.41$\pm$0.12$^{e}$     & 70$^{e}$         \\ 
J1435+3604	& 0.430 & 11598  &   J. Tumlinson    &  0.20260   &    19.80$\pm$0.10$^{b}$      & 14.4 &    $<$14.4    	     &   $<-$5.10	               & 39 (Wer13)     \\ 	
J1342-0053	& 0.326 & 11598  &   J. Tumlinson    &  0.22711   &    19.0$^{+0.5}_{-0.8}$$^{f}$& 14.2 &  14.63$\pm$0.06           &   $-$4.07$^{+0.50}_{-0.80}$  &  35 (Wer13)     \\ 
J0925+4004	& 0.472 & 11598  &   J. Tumlinson    &  0.24788   &    19.55$\pm$0.15$^{b}$      & 14.6 &    18.82$\pm$0.13         &   $-$0.57$\pm$0.17           &  83 (Wer13)    \\ 	
J1001+5944	& 0.747 & 12248  &   J. Tumlinson    &  0.30350   &    19.32$\pm$0.10$^{g}$      & 14.1 &     $<$14.1               &   $<-$4.92 	   \\ 
J1616+4154	& 0.441 & 11598  &   J. Tumlinson    &  0.32110   &    20.60$\pm$0.20$^{b}$      & 14.2 &    19.26$\pm$0.09         &   $-$1.08$\pm$0.20           & Not found$^{g}$  \\   	
Q1323+343	& 0.443 & 12593  &   D. Nestor	     &  0.35300   &    18.92$\pm$0.20 	         & 14.5 &     $<$14.5               &   $<-$4.12	   \\ 
Q1400+553	& 0.840 & 12593  &   D. Nestor	     &  0.36475   &    20.11$\pm$0.20            & 14.5 &     $<$14.5               &   $<-$5.31	   \\ 
J0209-0438	& 1.132 & 12264  &   S. Morris	     &  0.39055   &    19.00$\pm$0.12$^{h}$      & 14.0 &     $<$14.0               &   $<-$4.70   	   \\ 
Q1232-022	& 1.043 & 12593  &   D. Nestor	     &  0.39495   &    20.79$\pm$0.20 	         & 14.8 &     $<$14.8               &   $<-$5.69	   \\ 	
Q1251+463	& 1.460 & 12593  &   D. Nestor	     &  0.39662   &    20.50$\pm$0.20 	         & 14.4 &     $<$14.4               &   $<-$5.80	   \\ 		
Q0454-2203	& 0.533 & 12466  &   J. Charlton     &  0.47437   &    19.97$\pm$0.15$^{i}$      & 14.0 &     $<$14.0               &   $<-$5.67	   \\		
Q0454-2203	& 0.533 & 12466  &   J. Charlton     &  0.48321   &    19.59$\pm$0.10$^{j,k}$    & 14.0 &     $<$14.0               &   $<-$5.29	               & 108 (Kac11)  \\ 	
J1240+0949      & 1.045 & 11698  &   M. Putman	     &  0.49471   &    18.99$\pm$0.18$^{j}$      & 14.5 &     $<$14.5               &   $<-$4.19        \\  	
J1236+0600	& 1.286 & 11698  &   M. Putman	     &  0.54102   &    19.88$\pm$0.46$^{j}$      & 15.3 &     $<$15.3               &   $<-$4.28        \\ 	
Q1241+176	& 1.273 & 12466  &   J. Charlton     &  0.55048   &    $>$19.00$^{l}$            & 14.3 &    15.81$\pm$0.17         &   $<-$2.89 	               & 21 (Kac11)   \\		
Q0107-0232	& 0.728 & 11585  &   N. Crighton     &	0.55733   &	19.50$\pm$0.20$^{m}$     & 14.0 &	 17.27$\pm$0.30$^{m}$   &   $-$1.93$\pm$0.36$^{m}$     & 10$^{m}$     \\ 
Q1317+227	& 1.022 & 11667  &   C. Churchill    &  0.66047   &    19.65$\pm$0.10$^{j,n}$    & 14.1 &     $<$14.1               &   $<-$5.25	               & 104 (Kac11)  \\
\hline \hline 
\end{tabular}        
\label{sample1} 
~\\ ~\\    
\raggedright   
Notes -- 
$^{1}$QSO name. 
$^{2}$Emission redshift of QSO.  
$^{3}$$HST$ proposal ID. 
$^{4}$Principal investigator.  
$^{5}$Absorption redshift.  
$^{6}$Logarithmic neutral hydrogen column density.      
$^{7}$\HH\ detection sensitivity of the spectrum at 3$\sigma$ level.   
$^{8}$Total \HH\ column density when detected, else a 3$\sigma$ upper limit is given.  
$^{9}$Molecular mass fraction.    
$^{10}$Impact parameter of the nearest possible host-galaxy candidate when available.    
References for $\rho$: Pet96--\citet{Petitjean96b}; Wer13--\citet{Werk13}; Kac11--\citet{Kacprzak11}     
$^{a}$From \citet{Tripp05a}, a 3$\sigma$ upper limit on $N(\HH)$ is obtained from FUSE spectrum.    
$^{b}$From \citet{Meiring11}.    
$^{c}$From \citet{Srianand14}.  
$^{d}$$\log N(\HI)$~=~19.81$\pm$0.04 \citep[][using low resolution data]{Lanzetta97}.      
$^{e}$From \citet{Oliveira14}.  
$^{f}$From \citet{Werk14}.  
$^{g}$From \citet{Battisti12}.  
$^{h}$$\log N(\HI)$~=~18.87$\pm$0.03 \citep[]{Tejos14}.   
$^{i}$$\log N(\HI)$~=~19.45$^{+0.02}_{-0.03}$ \citep[][using low resolution data]{Rao06}.   
$^{j}$$N(\HI)$ is estimated from Voigt profile fitting of higher order Lyman series lines with minimum number of components. Therefore, $N(\HI)$ value should be taken with caution. 
$^{k}$$\log N(\HI)$~=~18.65$\pm$0.02 \citep[][using low resolution data]{Rao06}.  
$^{l}$$N(\HI)$ is not well constrained from the COS spectrum due to blend, so we adopt a conservative lower limit; $\log N(\HI)$~=~18.90$^{+0.07}_{-0.09}$ \citep[][using low resolution data]{Rao06}. 
$^{m}$From \citet{Crighton13}. 
$^{n}$$\log N(\HI)$~=~18.57$\pm$0.02 \citep[][using low resolution data]{Rao06}.     
\end{table*}

%% file: tab2.tex
\begin{table}
\caption{Low-$z$ DLAs/sub-DLAs for which \HH\ information is not available.}      
\begin{tabular}{cclcc} 
\hline \hline 
 QSO    &   PID	  &        PI	      &   \zabs\    &   $\log N(\HI)$ 	    \\ 
\hline  
J1407+5507	&  12486  &   D. Bowen	      &  0.00474$^{a}$         &    19.75$\pm$0.05         \\ 
J1415+1634	&  12486  &   D. Bowen	      &  0.00775$^{a}$         &    19.70$\pm$0.10         \\ 
J0930+2848	&  12603  &   T. Heckman      &  0.02264$^{a}$         &    20.71$\pm$0.15         \\ 
J1512+0128      &  12603  &   T. Heckman      &  0.02948$^{a}$         &    20.27$\pm$0.15         \\ 
J1009+0713	&  11598  &   J. Tumlinson    &  0.11400$^{b}$         &    20.68$\pm$0.10$^{d}$   \\ 
Q1659+373	&  12593  &   D. Nestor	      &  0.19983$^{c}$         &    19.37$\pm$0.20         \\  
\hline \hline 
\end{tabular}        
\label{sample2} 
~\\  
Notes -- $^{a}$No \HH\ coverage.  
$^{b}$The expected wavelength range of \HH\ transitions are covered but there is no flux 
below $\lambda 1241$ \AA\ due to the complete Lyman limit break from \zabs\ = 0.35590 system. 
$^{c}$Only G160M spectrum is available in which \HH\ transitions are not covered.   
$^{d}$From \citet{Meiring11}. 
\end{table}

%% file: tab3.tex
\begin{table*}
\caption{Voigt profile fit parameters and other measured quantities for the \HH\ bearing systems.}  
\begin{tabular}{m{1.6cm}m{0.8cm}m{1.3cm}m{1.3cm}m{1.3cm}m{1.3cm}m{1.0cm}m{1.0cm}ccclc} 
\hline \hline 
QSO  &   \zabs  &          \multicolumn{5}{c}{$\log N(\HH)$ (cm$^{-2}$)}   & $b$(H$_2$)              &  $T_{01}$  &  $T_{02}$  &  $T_{13}$ &  $\log Z/Z_{\odot}$   \\   \cline{3-7}        
     &          &   $J=0$        &   $J=1$        &  $J=2$        &  $J=3$   &  $J=4$  &   (\kms)    &    (K)     &  (K)       &     (K)   &                       \\ 
(1)  &    (2)   &    (3)   &    (4)         &      (5)       &     (6)       &     (7)       &    (8)      &    (9)     &  (10)      &    (11)   &        (12)           \\   
\hline  
J1241+2852        & 0.06650  &   15.72$\pm$0.08  &   16.30$\pm$0.06  &  15.45$\pm$0.08  &  $<14.8^{1}$      & $<14.7^{1}$ &   17.1$\pm$1.1  &   197 &  229 &  ...$^{3}$ &  $-$0.62$\pm$0.13$^{a}$  \\ 	
J1619+3342$^{b}$  & 0.09630  &   18.17$\pm$0.04  &   18.36$\pm$0.04  &  15.97$\pm$0.25  &  15.27$\pm$0.29   & $<14.1^{1}$ &    4.1$\pm$0.4  &    97 &   77 &  107 &  $-$0.62$\pm$0.13$^{f}$  \\    
Q0439-433         & 0.10091  &   14.67$\pm$0.05  &   15.21$\pm$0.02  &  14.91$\pm$0.04  &  14.52$\pm$0.07   & $<14.1^{1}$ &   32.7$\pm$2.6  &   178 &  484 &  350 &  $+$0.32$\pm$0.14$^{a}$  \\   	
                  & 0.10115  &   15.98$\pm$0.06  &   16.38$\pm$0.05  &  15.70$\pm$0.04  &  15.56$\pm$0.04   & $<14.1^{1}$ &   12.0$\pm$0.5  &   133 &  227 &  311 &                          \\   	
Q0850+440         & 0.16375  &   14.40$\pm$0.03  &   14.83$\pm$0.02  &  14.33$\pm$0.06  &  $<13.7^{1}$      & $<13.6^{1}$ &   11.9$\pm$0.8  &   141 &  289 &  ...$^{3}$ &  $<-$1.36$^{c}$          \\  
B0120-28$^{d}$    & 0.18495  &   16.14$\pm$0.14  &   17.23$\pm$0.08  &    $<14.2^{1}$   &  $<14.5^{1}$      & $<13.8^{1}$ &      ...$^{2}$  &   ...$^{3}$ &  ...$^{3}$ &  ...$^{3}$ &  $-$1.19$\pm$0.21        \\    
                  & 0.18524  &   16.80$\pm$0.13  &   17.45$\pm$0.08  & 14.55$^{+0.16}_{-0.08}$ & 14.21$\pm$0.10 & $<13.5^{1}$ &      ...$^{2}$  &   243 &   75 &  103 &                          \\ 
                  & 0.18550  &   16.81$^{+0.87}_{-0.22}$ &  18.91$\pm$0.07 &  18.32$\pm$0.05 &  17.73$\pm$0.05  & $<13.9^{1}$ &      ...$^{2}$  &    87 &  ...$^{3}$ &  239 &                          \\        
                  & 0.18568  &   19.72$\pm$0.02  &   19.53$\pm$0.03  &  18.40$\pm$0.04  &  17.60$\pm$0.05   &  $<13.8^{1}$    &      ...$^{2}$  &    65 &  110 &  161 &                          \\    
J1342-0053        & 0.22711  &   $<13.4^{1}$     &   14.63$\pm$0.06  &  $<13.6^{1}$     &  $<14.0^{1}$      & $<14.1^{1}$ &   10.1$\pm$2.3  &   ...$^{3}$  &  ...$^{3}$ &  ...$^{3}$ &  $-$0.40$^{e}$           \\ 
J0925+4004        & 0.24788  &   18.15$\pm$0.08  &   18.63$\pm$0.04  &  17.90$\pm$0.10  &  16.83$\pm$0.15   & $<14.3^{1}$ &    8.4$\pm$0.5  &   156 &  234 &  171 &  $-$0.29$\pm$0.17$^{f}$  \\	
J1616+4154        & 0.32110  &   18.95$\pm$0.02  &   18.93$\pm$0.02  &  17.83$\pm$0.09  &  16.99$\pm$0.12   & $<14.0^{1}$ &    6.9$\pm$0.5  &    76 &  122 &  160 &  $-$0.38$\pm$0.23$^{f}$  \\ 
Q1241+176         & 0.55048  &   15.35$\pm$0.13  &   15.42$\pm$0.06  &  14.95$\pm$0.05  &  14.80$\pm$0.08   & $<14.0^{1}$ &    7.8$\pm$0.4  &    83 &  202 &  375 &  $<+$0.18$^{a}$          \\   	
Q0107-0232$^{g}$  & 0.55715  &   16.17$\pm$0.25  &   17.05$\pm$0.28  &  16.19$\pm$0.19  &  15.77$\pm$0.12   & $<14.0^{1}$ &    6.7$\pm$0.6  &   997 &  327 &  225 &  $-$0.72$\pm$0.32        \\     
                  & 0.55729  &   15.63$\pm$0.39  &   16.42$\pm$0.40  &  15.65$\pm$0.25  &  15.47$\pm$0.18   & $<14.0^{1}$ &    4.3$\pm$0.7  &   450 &  327 &  281 &                          \\   
\hline \hline 
\end{tabular}        
\label{H2tab}
~\\ 	
\raggedright  
Notes -- 
$^{1}$Not detected, a formal $3\sigma$ upper limit is given.    
$^{2}$Not reported by \citep{Oliveira14}.    
$^{3}$Temperature is undefined.   
$^{a}$Average metallicities ([S/H] or [Si/H]) from this work without ionization correction.   
$^{b}$All values are taken from \citet{Srianand14}.     
$^{c}$From \citet{Lanzetta97}. 
$^{d}$All values are taken from \citet{Oliveira14}.  
$^{e}$From \citet{Werk13}. 
$^{f}$From \citet{Battisti12}. 
$^{g}$All values are taken from \citet{Crighton13} except the excitation temperatures.   
\end{table*}